\documentclass[9pt,journal,twoside]{IEEEtran}
\ifCLASSINFOpdf
\else
\fi
\usepackage{graphicx}
\usepackage{latexsym}
\usepackage{amssymb}
\usepackage{amsmath}
\usepackage{amsmath,empheq}
\usepackage{booktabs}
\usepackage[export]{adjustbox}
\usepackage{subfig}
\usepackage{enumerate}
\usepackage{multicol}
\usepackage{rotating}
\usepackage{pdflscape}
\DeclareMathSizes{12}{8}{7}{4}
\usepackage{nicefrac}
\usepackage{mathtools}
\usepackage{svg}
\usepackage{sidecap}
\usepackage{siunitx}
\usepackage{placeins}
\usepackage{comment}
\usepackage{sidecap}
\usepackage{etoolbox}

\usepackage{xargs}                      

\usepackage[sc]{mathpazo}
\usepackage{algorithm}
\usepackage[noend]{algpseudocode}
\makeatletter
\renewcommand\fs@ruled{%
  \def\@fs@cfont{\itshape}%
  \let\@fs@capt\floatc@plain%
  \def\@fs@pre{\hrule height.8pt depth0pt \kern2pt}
  \def\@fs@post{}
  \def\@fs@mid{\kern2pt\hrule\kern2pt}
  \let\@fs@iftopcapt\iffalse}
\makeatother
\makeatletter
\renewcommand{\ALG@name}{\small\textbf{Algorithm}}
\makeatother
\usepackage{etoolbox}

\usepackage[sort,compress]{cite}
\usepackage{multirow}
\usepackage{makecell}
\usepackage{textcomp}
\usepackage{paralist}
\usepackage{cancel}
\usepackage{color, colortbl}
\usepackage{geometry}
\usepackage{longtable}
\usepackage{graphbox}
\usepackage{cases} 

\usepackage[labelfont=bf,textfont=it,belowskip=0pt,aboveskip=5pt,tableposition=top]{caption}
\usepackage[colorlinks=true,
citecolor=blue,
filecolor=black,
linkcolor=blue,
urlcolor=blue]{hyperref}
\usepackage{lineno}

\graphicspath{{figures/}{tables/}}

\newcounter{runidnum}

\newcommand{\figref}[1]{Fig.~\ref{#1}}
\newcommand{\tabref}[1]{Tab.~\ref{#1}}
\newcommand{\secref}[1]{\S\ref{#1}}
\renewcommand{\algref}[1]{Alg.~\ref{#1}}

\newcommand{\nf}[2] {\nicefrac{#1}{#2}}

\newcolumntype{R}{>{\columncolor{gray!20}}r}
\newcolumntype{C}{>{\columncolor{gray!20}}c}
\newcolumntype{L}{>{\columncolor{gray!20}}l}
\newcolumntype{G}{>{\columncolor{gray!30}}l}

\newcommand{\distCM}{\ensuremath{\mbox{D}_{cm}}}
\newcommand{\distPi}{\ensuremath{\mbox{D}_{\vect{p}}}}

\newcommand{\mpPatientBrain}[1][]{\ifthenelse { \equal {#1} {} }
    { \vect{m}_D }   
    { \vect{m}_{D,#1} } }  

\newcommand{\mpWarpedAtlasBrain}[1][]{\ifthenelse { \equal {#1} {} }
    { \vect{m}_A^{(1,1)} }   
    { \vect{m}_{A,#1}^{(1,1)} }}   

\newcommand{\D}[1]{\ensuremath{\mathcal{#1}}}                           

\newcommand{\ns}[1]{\ensuremath{\mathbf{#1}}}

\newcommand{\idiv}{\ensuremath{\mbox{div}\,}}
\newcommand{\igrad}{\ensuremath{\nabla}}

\renewcommand{\d}[1]{\mathop{}\!\mathrm{d}#1}

\newcommand{\dx}{\d{\vect{x}}}

\newcommand{\p} {\partial}

\newcommand{\vect}[1]{\boldsymbol{#1}} 
\newcommand{\mat}[1]{\boldsymbol{#1}}  

\newcommand{\bipa}{\begin{inparaenum}[(\itshape i\upshape)]}
\newcommand{\eipa}{\end{inparaenum}}

\newcommand{\bipasub}{\begin{inparaenum}[(\itshape a\upshape)]}
\newcommand{\eipasub}{\end{inparaenum}}

\newcommand{\ipoint}[1]{\textit{\textbf{\color{darkgray}#1}}}

\newcommand{\iemph}[1]{\emph{#1}}

\newcommand{\defeq}{\ensuremath{\mathrel{\mathop:}=}}

\definecolor{sred}{cmyk}{0.01,0.98,0,0.2} 
\reversemarginpar
\newcommand{\mmargin}[1]{{\marginpar{\em\tiny #1}}}\renewcommand{\mmargin}[1]{}

\newcommand{\zapspace}{\topsep=0pt\partopsep=0pt\itemsep=0pt\parskip=0pt}

\definecolor{C0}{rgb}{0.000,0.447,0.741}
\definecolor{C1}{rgb}{0.850,0.325,0.098}
\definecolor{C2}{rgb}{0.929,0.694,0.125}
\definecolor{C3}{rgb}{0.494,0.184,0.556}
\definecolor{C4}{rgb}{0.466,0.674,0.188}
\definecolor{C5}{rgb}{0.301,0.745,0.933}
\definecolor{C6}{rgb}{0.635,0.078,0.184}
\definecolor{C7}{rgb}{0.887,0.465,0.758}
\definecolor{C8}{rgb}{0.496,0.496,0.496}

\definecolor{anthrazit}{RGB}{62,68,76}
\definecolor{mittelblau}{RGB}{0,81,158}
\definecolor{helllblau}{RGB}{0,190,255}

\definecolor{mdtRed}{RGB}{180,0,0}

\algrenewcommand{\alglinenumber}[1]{\footnotesize\color{anthrazit}{{#1}}}
\algrenewcommand{\algorithmicrequire}{\textbf{Input:}}
\algrenewcommand{\algorithmicensure}{\textbf{Output:}}

\algrenewcommand{\textproc}{}

\algnewcommand{\Break}{\textbf{break}}
\algnewcommand{\Continue}{\textbf{continue}}
\algnewcommand{\True}{\textbf{true}}
\algnewcommand{\False}{\textbf{false}}
\algnewcommand{\Null}{\textbf{null}}
\algrenewcommand{\algorithmicend}{\textbf{end}}
\algrenewcommand{\algorithmicdo}{\textbf{do}}
\algrenewcommand{\algorithmicwhile}{\textbf{while}}
\algrenewcommand{\algorithmicfor}{\textbf{for}}
\algrenewcommand{\algorithmicforall}{\textbf{for all}}
\algrenewcommand{\algorithmicloop}{\textbf{loop}}
\algrenewcommand{\algorithmicrepeat}{\textbf{repeat}}
\algrenewcommand{\algorithmicuntil}{\textbf{until}}
\algrenewcommand{\algorithmicprocedure}{\textbf{procedure}}
\algrenewcommand{\algorithmicfunction}{\textbf{func}}
\algrenewcommand{\algorithmicif}{\textbf{if}}
\algrenewcommand{\algorithmicthen}{\textbf{then}}
\algrenewcommand{\algorithmicelse}{\textbf{else}}
\algrenewcommand{\algorithmicreturn}{\textbf{return}}
\algnewcommand{\algorithmicforever}{\textbf{for ever}}
\algdef{S}[FOR]{ForEver}{\algorithmicforever\ \algorithmicdo}
\algnewcommand{\algorithmicgoto}{\textbf{go to}}
\algnewcommand{\Goto}[1]{\algorithmicgoto\ \cref*{#1}}
\algnewcommand{\funccall}[1]{\textbf{\color{darkgray}#1}}

\makeatletter
\g@addto@macro\ALG@beginalgorithmic{\small}
\makeatother

\algrenewcomment[2][.3\linewidth]{\hfill\leavevmode\hfill\makebox[#1][l]{$\#$ \emph{\normalfont\color{darkgray}#2}}}

\newcommand\crule[3][black]{\textcolor{#1}{\rule{#2}{#3}}}

\definecolor{pastGcolrA}{RGB}{4,92,155}
\definecolor{pastGcolrB}{RGB}{6,140,117}
\definecolor{pastGcolrC}{RGB}{122,211,227}
\definecolor{pastGcolrD}{RGB}{12,163,171}
\definecolor{pastGcolrE}{RGB}{124,180,212}

\definecolor{colorA}{RGB}{189,201,225}
\definecolor{babyblue}{RGB}{29,162,220}
\definecolor{anthrazit}{RGB}{0,153,229}
\definecolor{red}{RGB}{240,85,97}
\definecolor{colorB}{RGB}{103,169,207}
\definecolor{colorC}{RGB}{ 28,144,153}
\definecolor{colorD}{RGB}{  1,108, 89}
\definecolor{anthrazit}{RGB}{62,68,76}
\definecolor{mittelblau}{RGB}{0,81,158}
\definecolor{hellblau}{RGB}{0,190,255}

\definecolor{c1color1}{RGB}{125,97,234}
\definecolor{c1color2}{RGB}{100,200,214}
\definecolor{c1color3}{RGB}{209,241,154}
\definecolor{c1color4}{RGB}{255,20,20}
\definecolor{c0color1}{RGB}{243,77,250}
\definecolor{c0color2}{RGB}{121,173,252}
\definecolor{dcolor}{RGB}{255,255,228}
\definecolor{segED}{RGB}{64,64,64}
\definecolor{segWM}{RGB}{192,192,192}
\definecolor{segGM}{RGB}{160,160,160}
\definecolor{segCSF}{RGB}{255,255,255}

\definecolor{pastelgreen}{RGB}{60,179,113}


\begin{document}
\bstctlcite{IEEEexample:BSTcontrol}
%

\title{Automatic MRI-Driven Model Calibration for Advanced Brain Tumor Progression Analysis}
%
%
%

\author{Klaudius Scheufele, Shashank Subramanian, George Biros, \IEEEmembership{Member~IEEE}
\thanks{K. Scheufele, Oden Institute, University of Texas at Austin, Texas, USA e-mail: kscheufele@{austin}.utexas.edu.}
\thanks{S. Subramanian, Oden Institute, University of Texas at Austin, Texas, USA e-mail: shashanksubramanian@{utmail}.utexas.edu.}
\thanks{G. Biros, Oden Institute, University of Texas at Austin, Texas, USA e-mail: biros@{oden}.utexas.edu.}
\thanks{\emph{Preprint submitted to IEEE Transactions on Medical Imaging}. Manuscript received Jan 19, 2020.}}

%
%

\markboth{{Preprint submitted to} IEEE Transactions on Medical Imaging}
{Scheufele \MakeLowercase{\textit{et al.}}: Automatic MRI-Driven Model Calibration for Advanced Brain Tumor Progression Analysis}
%



\maketitle

\vspace*{-30pt}

\begin{abstract}
Our objective is the calibration of mathematical tumor growth models from a single multiparametric scan. The target problem is the analysis of preoperative  Glioblastoma (GBM) scans. To this end, we present a fully automatic tumor-growth calibration methodology that integrates a single-species reaction-diffusion partial differential equation (PDE) model for tumor progression with multiparametric Magnetic Resonance Imaging (mpMRI) scans to robustly extract patient specific biomarkers i.e., estimates for \bipa \item the tumor \iemph{cell proliferation rate}, \item the tumor \iemph{cell migration rate}, and \item the original, \iemph{localized} \iemph{site(s) of tumor initiation}.\eipa{}
Our method is based on a sparse reconstruction algorithm for the tumor initial location (TIL). This problem  is particularly challenging due to nonlinearity, ill-posedeness, and ill conditioning. We propose  a coarse-to-fine multi-resolution continuation scheme with parameter decomposition to stabilize the inversion. We demonstrate robustness and practicality of our method by applying the proposed method to clinical data of 206 GBM patients. We analyze the extracted biomarkers and relate tumor origin with patient overall survival by mapping the former into a common atlas space. We present preliminary results that suggest improved accuracy for prediction of patient overall survival when a set of imaging features is augmented with estimated biophysical parameters. All extracted features, tumor initial positions, and biophysical growth parameters are made publicly available for further analysis. To our knowledge, this is the first fully automatic scheme that can handle multifocal tumors and can localize the TIL to a few millimeters.

\end{abstract}


%
\IEEEpeerreviewmaketitle

\vspace{-8pt}
\section{Introduction}
\label{sec:intro}
\IEEEPARstart{G}{ioblastoma} multiforme (GBM) is a highly aggressive primary brain tumor with poor prognosis despite radical treatment.  GBMs are challenging due to their heterogeneity and their infiltration into surrounding healthy tissue. Several research groups are working on developing  image-driven biophysical tumor-growth models for GBMs that can be applied in a clinical setting and assist with  diagnosis,  prognosis, and treatment planning. For example, such models can be used to determine survival, probability of recurrence, molecular subtypes, overall tumor aggressiveness, and epidemiology~\cite{yankeelov-miga13,rockne-roadmap-e19,Lipkova2019a,akbari-e18}.

\begin{figure}
\centering
\includegraphics[width=.495\textwidth, trim=3 0 0 0, clip]{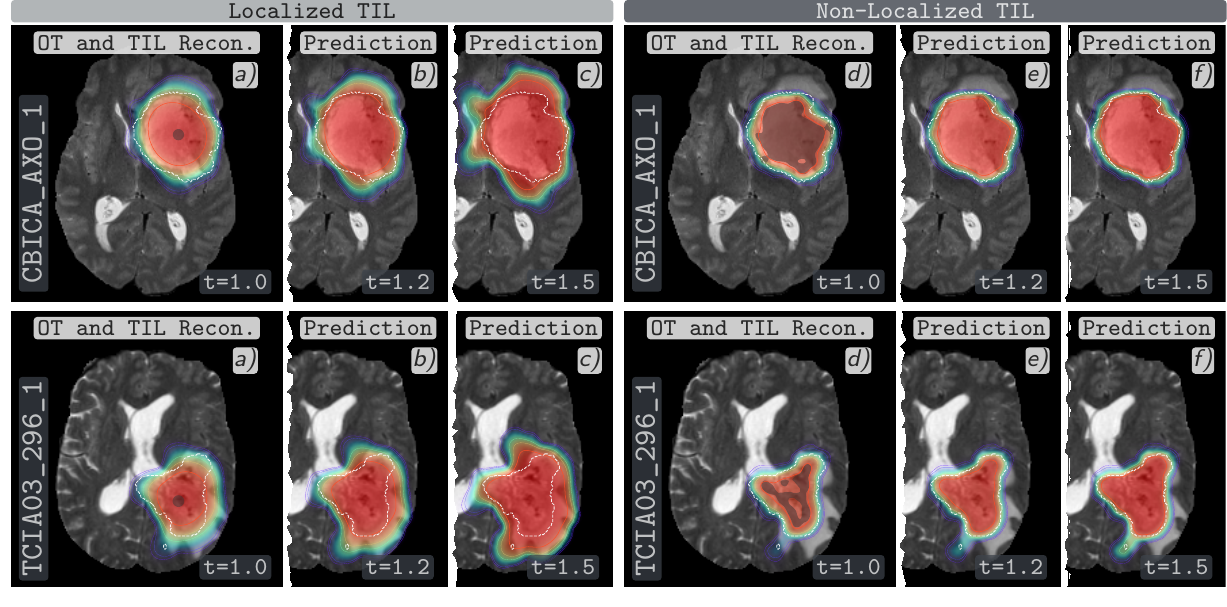}
\caption{
\ipoint{Comparison of localized (sparse) and non-localized TIL reconstruction and tumor prediction}. Each row corresponds to a different BRATS subject. The first three columns are localized TIL; the last three are non-localized TIL. The observed tumor (OT) segmentation that drives the inversion  is the white (dashed) contour line. Subfigures (a) and (d) show the observed tumor (OT) reconstruction at the scan time (t=1) as the solution of a tumor growth problem colored from red (100\% tumor saturation) to violet (5\% saturation). The TIL is in gray and contained in the support of OT. We see that the predictions of the observed tumor concentration at two future times (t=1.2 and t=1.5) differ a lot when comparing the localized (c--d) and non-localized (e--f) TIL reconstructions. The non-localized TIL underpredicts infiltration.
\label{fig:l1vsl2-pred}\vspace{-5.5pt}}
\vspace{-8pt}
\end{figure}

In this  paper, we focus on the problem of calibrating a tumor-growth model from  a single scan. Indeed, for the majority  of GBM cases, preresection clinical decisions are based only on a single scan.  Our reconstruction produces biomarkers that  can potentially offer additional diagnostic and prognostic value, as they might relate to the tumor’s aggressiveness and extent of infiltration beyond imaging-visible tumor boundaries. Given a multiparametric, tumor-bearing, pretreatment scan, we want to calibrate various parameters in a biophysical model. (This is the so-called \ipoint{inverse problem}.) We are particularly interested in the  \ipoint{tumor initiation location (TIL)}. Determining the TIL is mathematically required for robustly identifying tumor growth parameters (cf.\;\figref{fig:l1vsl2-pred}). Moreover, the tumor location could be related to distinct tumor mutations, so it could add complementary information to radiogenomics based analyses~\cite{roux2019mri,ren2012co,bilello-davatzikos-e16}. The  observed tumor location plays a key role in guiding surgical and radiation  treatment, and has prognostic value for patient survival~\cite{roux2019mri,li.sun.he-e19}. \emph{Our hypothesis is that TIL could play a similar role. Here, we are not doing an exhaustive test for this hypothesis. Instead, we focus on  an algorithm to  recover the TIL and perform a preliminary analysis of its correlation with survival.}  Although several methods have been proposed for estimating tumor parameters, there is surprisingly little work on robustly estimating the TIL.

\begin{figure}
\centering
\includegraphics[width=.495\textwidth]{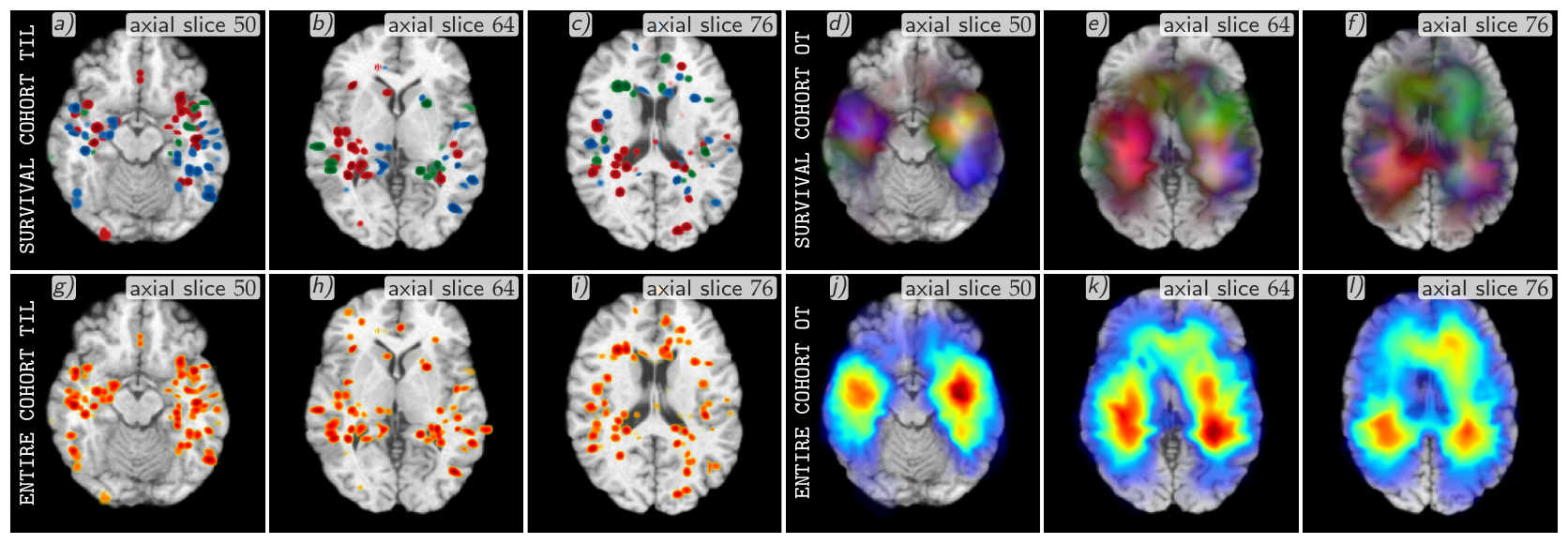}
\caption{
\ipoint{Cohort GBM statistical atlas for BratTS'18 data}. Top figures: Atlas showing relative frequency of tumor initiation location (TIL) (left, subfigures a--c), and observed tumor (OT) (right, d--f) for the Brats'18 survival data ($161$ cases; color coded by survival class: short survivors red; mid survivors green; long survivors blue). Bottom: Probabilities for TIL (g--i) and OT (j-l) across entire cohort of $206$ cases (TIL max probability is $0.007$; OT max probability is $0.1$). Applying our algorithm to larger datasets and more accurate models could help test a recent hypothesis that certain types of GBMs originate in the subventricular zone~\cite{Lee-18}.
\label{fig:gbm-atlas}\vspace{-3.5pt}}
\vspace{-8pt}
\end{figure}

\subsubsection{Contributions}
After an image segmentation preprocessing step, we solve an inverse problem for two scalar tumor parameters (\emph{migration rate} and \emph{proliferation rate}), and the \emph{TIL}, which is parameterized by a high-dimensional vector. In our recent work~\cite{Subramanian19aL1},  we presented theoretical results on the non-convexity and ill-posedeness of  the problem, introduced an overall regularization framework,  and verified the method on synthetic datasets. However, our method in~\cite{Subramanian19aL1} is not robust enough to be applied to generic clinical datasets.  Here, we expand~\cite{Subramanian19aL1} in several ways:
\begin{inparaitem}
    \item We stabilize and accelerate the solution of the inverse problem by introducing a \textit{(multilevel) coarse-to-fine} continuation scheme, and embed a two-step optimization scheme based on parameter decomposition. This is critical to prevent the method from producing infeasible or sub-optimal solutions.
    \item We enhance~\cite{Subramanian19aL1} to a \textit{fully automatic}, and \textit{robust} method for GBM biomarker extraction from actual clinical imaging data. We demonstrate robustness across 206 patients from the BraTS dataset~\cite{Bakas2018Brats}, featuring vastly heterogeneous GBM phenotypes along with multifocality. The extracted features are available in the supplementary material (SM). We correlate our results with a purely imaging based method to identify  multifocal tumor seeds (using connected components and center of mass).
    \item We create a \textit{GBM atlas of TIL} from the 206 case-cohort via atlas-registration (see~\figref{fig:gbm-atlas}). This helps to indentify high-risk tumor sites and correlations between the TIL and the severity and progression of the disease. We present preliminary results for patient survival predictions based on extracted GBM biomarkers.
\end{inparaitem}

\subsubsection{Related Work}
Mathematical modeling in cancer progression has a long tradition; in particular forward models for tumor progression are strongly represented in literature~\cite{Swanson:2008a,Lipkova2019a,HawkinsDaarud:2013a}. There is less work on patient-specific model calibration. Similar to our approach, most groups rely on reaction-diffusion models and calibrate parameters using MRI scans~\cite{Clatz:2005a,Hogea:2008b,Konukoglu:2010b,Swanson:2008a,Rekik:2013a}. Despite its simplicity, this framework has been shown to be effective in medical MR image analysis, segmentation, and tumor characterization~\cite{Swanson:2008a,Jackson:2015a,Lipkova2019a,Gooya:2012a,macyszyn-e16}.

Several parameter calibration studies assume access to subject-specific longitudinal imaging data (from two or more time points), which are rarely available for GBMs. In contrast, our approach requires one single pretreatment scan only. Besides our work in~\cite{Subramanian19aL1}, approaches for single-snapshot data have been presented in~\cite{Lipkova2019a,Scheufele:2018,Gooya:2012a} (and the references therein for older works). The groups in \cite{Scheufele:2018,Gooya:2012a}  integrate image registration with tumor inversion methods; Scheufele et al.\;\cite{Scheufele:2018}  use an $\ell_2$  penalty that results in non-localized TIL (see~\figref{fig:gbm-atlas}) that lead to erroneous predictions of infiltration and proliferation; Gooya et al.\;\cite{Gooya:2012a}  manually select tumor seeds, which is arbitrary, error prone, and not-reproducible. Lipkova et al.\;\cite{Lipkova2019a} employ a Bayesian framework for parameter calibration, but restrict tumor initiation to only one distinct location, which limits practical applicability to multifocal pathologies. For more discussion, see~\cite{Subramanian19aL1}.

Tumor location carries important information for low and high-grade gliomas~\cite{parisot-paragios-e16,bilello-davatzikos-e16}. For GBMs, Le et  al.\;\cite{li.sun.he-e19} analyzed the correlation between tumor location and clinical properties of glioblastomas in frontal and temporal lobes for 406 patients; Ren et al.\;\cite{ren2012co} analyzed 528 patients with gliomas with scans and histological data and analyzed  incidences of 1p/19q co-deletion and IDH1/2 mutation in gliomas by regions and hemispheres in the brain. Roux et al.\;\cite{roux2019mri} found longer survival in those patients with tumors in the left temporal lobe regardless of whether the patient received resection or biopsy; moreover, they found distinct spatial distributions of tumor determined by molecular biomarkers like MGMT.  Binder et al.\;\cite{binder-davatzikos-e18} correlated the shape of tumor with EGFR mutations, and Bilello et al.\;\cite{bilello-davatzikos-e16} correlated such mutations  with the observed tumor location. Further evidence suggesting that the location of tumor initiation relates to distinct tumor mutations and might correlate with patient survival can be found in~\cite{Jungk-19, Lee-17, Lee-18, Adeberg-14, Chaichana-08,macyszyn-e16}.

\subsubsection{Outline}
In \secref{sec:methods}, we introduce the model and the new calibration algorithm. In~\secref{sec:results}, we present results from the BraTS dataset and analyze correlations to patient survival. More technical details and  results can be found  in the SM.

\vspace{-8pt}
\section{Methods}
\label{sec:methods}

\vspace{-3pt}
\subsection{Tumor growth model}\label{sec:tumor-model}

\vspace{-5pt}
\subsubsection{Model and Materials}
Our tumor-growth model is a standard single-species  nonlinear reaction-diffusion partial differential equation (PDE)~\cite{Murray:1989a}
\vspace{-1pt}
\begin{subequations}\label{eq:tu-state}
\begin{align}
& &\p_t c - \kappa \D{D}c - \rho \D{R} c &=  0 & &\mbox{in}~ \Omega_{\D{B}} \times (0,1],
\label{eq:fwd} \\
& & c(0)  &= c_0.
\label{eq:fwd_initial}
\end{align}
\end{subequations}
Here $c=c(\vect{x},t) \in [0,1]$ denotes the tumor concentration
evolving over time  $t\in[0,1]$ in the three-dimensional brain domain $\Omega_{\D{B}} \subset \Omega$, embedded into the simulation domain,  $\Omega = [0, 2\pi]^3$. Tumor invasion and proliferation are modeled by the \iemph{diffusion} operator $\D{D}c = \idiv \mat{\kappa_{\vect{m}}}\igrad c$, and a logistic \iemph{reaction} term $\D{R}c = \mat{\rho_{\vect{m}}}c(1-c)$, respectively, with spatially variable coefficients  $\mat{\rho_{\vect{m}}}$ and $\mat{\kappa_{\vect{m}}}$ defined based on the underlying material properties $\vect{m}$. The latter indicate a vector of voxelwise probabilities $m_i$ for  \iemph{white matter} $(W)$, \iemph{gray matter} $(G)$, and \iemph{cerebrospinal fluid with ventricles} $(F)$, which are inferred from a deep convolutional neural network~\cite{brats18} using mpMRI. Tumor growth is assumed along the white matter fiber tracts only. We denote characteristic cell proliferation and cell migration in white matter by $\rho_w$ and $\kappa_w$, respectively. Tumor cells do not grow or invade into the cerebrospinal fluid-filled ventricles. Details and examples of tumor growth for different values for $\rho_w$ and $\kappa_w$ are given in the SM.
Although the time horizon is not known, in~\cite{Subramanian19aL1} we show that, after non-dimensionalization, only $\kappa$ and $\rho$ are independent variables and we can set the time horizon to one by simple rescaling. We approximate no-flux boundary conditions $\partial \Omega_{\D{B}}$ via a penalty approach~\cite{Gholami:2016a}, and use periodic boundary conditions on $\partial \Omega$. Finally, $c(0)=c(\vect{x},0)$ is the initial concentration of the tumor $c_0$. Our hypothesis is that it is highly localized in space.

Following~\cite{Subramanian19aL1}, we introduce a parametrization of $c_0$ as follows
\vspace{-4pt}
\begin{equation}\label{eq:parametrization}
c_0 = \mat{\Phi}^T(\vect{x})\vect{p} = \sum_{i=1}^{n_p} \Phi_i(\vect{x})p_i,~~ \vect{p} \in \ns{R}^{n_p},
\vspace{-2pt}
\end{equation}
where $\mat{\Phi}(\vect{x}) = \{\Phi_i(\vect{x})\}_{i=1}^{n_p}$ is a set of Gaussian basis functions with $\Phi_i(\vect{x}) = \Phi_i(\vect{x} - \vect{x}_i, \sigma)$, centered at location $\vect{x}_i$.
The standard deviation $\sigma$ controls the width of the Gaussian basis functions. Here, we make the assumption that the spatial distribution of the initial tumor is given by $\Phi_i$ for some $i$, and sufficiently small $\sigma$.%
\footnote{
The continuum hypothesis of our model breaks down at the microscopic scale of tumor initiation. Thus, we use our PDE model after the tumor has grown sufficiently large while remaining localized, and assume the initial tumor concentration can be approximated sufficiently well by a Gaussian distribution. To allow for multifocality and more complex shapes, we allow for a small number of Gaussians to be activated at time $t=0$.}
We invert for
\vspace{-4pt}
\begin{equation}\label{eq:inv-parameters}
\vect{g} = (\rho_w, \kappa_w, \vect{p}),
\vspace{-2pt}
\end{equation}
i.e., the tumor \iemph{proliferation} (a scalar) and \iemph{infiltration} (a scalar) rate in white matter, and the \iemph{TIL} (a vector) parametrized by $\vect{p}$.

\subsubsection{Calibration and Extraction of Biomarkers}
We calibrate biophysical parameters $\vect{g}$ so that the predicted tumor evolution of the model in~\eqref{eq:tu-state} matches the patient segmentation. As input, the inverse solve takes the mpMRI segmentation~\cite{brats18} and derives probability/cell concentration maps for the healthy tissue anatomy labels and tumor sub-regions, i.e., we consider the tumor core $m_T$ (necrotic core, enhancing tumor, and non-enhancing tumor), and the  peritumoral edema $m_E$; detailed in~\secref{sec:data-processing}.
Then we solve the PDE-constrained nonlinear least squares problem
\vspace{-5pt}
\begin{subequations}\label{eq:inv-tumor-abstract}
\begin{flalign}
&\min_{\vect{g}} \D{J}(\vect{g}) \defeq \frac{1}{2}\|\D{O}c(1) - m_{T} \|_{L_2(\Omega)}^2 + \beta \D{S}(\vect{g}) \\
&\mbox{\text{subject to}}~\D{F}(\vect{g})~\mbox{given by}~\eqref{eq:tu-state},~\mbox{and}~\D{C}(\vect{g}),
\end{flalign}%
\end{subequations}
minimizing the discrepancy between model prediction $c(1)$ and target data $m_T$ in a least squares sense, balanced with a regularization term  $\D{S}(\vect{g})$ (with weight $\beta$) to alleviate ill-conditioning. $\D{C}(\vect{g})$ is a set of additional constraints, allowing us to \iemph{simultaneously invert} for the set of parameters $\vect{g}$, and promote \iemph{sparse localization} of the TIL. Data discrepancy is only measured at certain \iemph{observation points}, defined by an \ipoint{observation operator} $\D{O}$, accounting for the fact that the true extent of tumor infiltration is not directly  observable. This operator defines a region with clearly observable tumor margin (while the true tumor may extend beyond this margin); details are given in~\secref{sec:continuation_scheme}.  We introduce a robust algorithm for the solution of the inverse problem by employing a multilevel continuation scheme based on spatial grid-coarsening. This reduces the risk of early termination in (bad) local minima, and improves the practicality of the method for real clinical data, featuring single- and multifocal pathologies with great heterogeneity.

 Reconstructing these parameters is a hard problem.  It is well known that inversion for all parameters of a reaction-diffusion problem is ill-posed, since the forward model $\D{F}$ is non-unique in the sense that different parameter configurations produce the same solution~\cite{Konukoglu:2010b,Subramanian19aL1}. Thus, additional constraints need to be imposed to alleviate the ill-posedness. We enforce $\max_{\vect{x}} c_0(\vect{x}) = 1$, corresponding to the modeling assumption that at $t=0$ the tumor volume concentration reaches $100\%$ of a voxel for the first time. Additionally, we need to ensure the tumor initial condition to be small and localized. We impose an $\ell_0$ constraint $\|\vect{p}\|_{L_0(\ns{R}^{n_p})} \leq \nu$, restricting the initial tumor to consist of a small number $\nu$ of initiations (i.e., active Gaussians) only.
For numerical optimization, we introduce additional bound constraints on the inversion variables to safeguard against infeasible solutions and bad local minima.
We complete the formulation by defining the regularization term $\D{S}(\vect{g})$ as $\frac{1}{2}\|\mat{\Phi}^T\vect{p}\|_{L_2(\Omega)}^2$, a standard Tikhonov regularization model.

Hence, our \ipoint{inverse problem} reads:
\begin{subequations}\label{eq:inv-tumor-formulation}
    \begin{equation}
    \min_{(\rho_w, \kappa_w, \vect{p})}
    \frac{1}{2}\|\D{O}c(1) - m_{T} \|_{L_2(\Omega)}^2 + \frac{\beta}{2}\|\mat{\Phi}^T\vect{p}\|_{L_2(\Omega)}^2  \label{eq:objective}
    \end{equation}
    \begin{numcases}{\mbox{s.t.}~}
       \D{F}(\rho_w, \kappa_w, \vect{p}) &\mbox{given by}~\eqref{eq:tu-state}, \label{eq:state_constraint} \\
       \max \mat{\Phi}^T\vect{p} = 1 &\mbox{in}~ $\Omega$, \label{eq:max_constraint} \\
       \|\vect{p}\|_{L_0(\ns{R}^{n_p})} \leq \nu &\mbox{in}~ $\ns{R}^{n_{\vect{p}}}$, \label{eq:sparsity_constraint} \\
       0 \leq \vect{p}  &\mbox{in}~ $\ns{R}^{n_{\vect{p}}}$, \label{eq:bound_p} \\
       \kappa_{min} \leq \kappa_w \leq \kappa_{max} &\mbox{in}~ $\ns{R}$, \label{eq:bound_kappa} \\
       \rho_{min} \leq \rho_w \leq \rho_{max}  &\mbox{in}~ $\ns{R}$. \label{eq:bound_rho}
    \end{numcases}
\end{subequations}
Next, we discuss the numerical solution of~\eqref{eq:inv-tumor-formulation}.

\vspace{-10pt}
\subsection{Numerical Scheme}
\label{sec:numerical_scheme}
\vspace{-3pt}
In \secref{sec:inverse_problem}, we summarize the scheme from~\cite{Subramanian19aL1} for solving~\eqref{eq:inv-tumor-formulation}.
Then, in~\secref{sec:continuation_scheme} we discuss the novel algorithmic modifications to enhance robustness in terms of avoiding stagnation, divergence of the optimizer, arbitrary parameter selection, and challenges with multifocal tumors.
\vspace{-3pt}
\subsubsection{Biophysical Parameter Inversion}
\label{sec:inverse_problem}

\smallskip
\paragraph{Selection of Gaussian Basis Functions}
The set $\{\Phi_i(\vect{x})\}_{i=1}^{n_p}$ of Gaussian basis functions is determined automatically; that is, the number $n_p$ and spatial distribution (i.e., centers $\{\vect{x}_i\}_{i=1}^{n_p}$) are defined based on an input tumor concentration map $m_{\Phi}$, e.g., tumor core input data $m_{\Phi} = m_T$, or the tumor initial condition $m_{\Phi} = c(0)$.
We refer to the SM for details.

\smallskip
\paragraph{Enforcing the Sparsity Constraint}
Solving the $L_2$-regularized problem~\eqref{eq:objective}-\eqref{eq:state_constraint} typically results in tumor initial conditions which closely resemble the input data $m_T$ up to scaling. As a result, the parameters $\rho_w$ and $\kappa_w$ are underestimated~\cite{Subramanian19aL1}.
The sparsity constraint~\eqref{eq:sparsity_constraint} restricts TIL to a few, small, and localized positions, i.e., the solution $\vect{p}$ is constrained to have at most $\nu$ non-zero entries. The corresponding active Gaussians form the support of $\vect{p}$ or $\mbox{supp}(\vect{p})$, denoting the set of non-zero indices of $\vect{p}$.
We employ an iterative recovery strategy based on compressive sampling~\cite{Nedell09a}, and determine important components in $\vect{p}$ based on the gradient of the objective~\eqref{eq:objective} (i.e., large gradient corresponds to large data-misfit).
A solution to~\eqref{eq:inv-tumor-formulation} is found iteratively by alternating between \bipa \item identifying a set of components $\D{V}$, $|\D{V}| \leq 3\nu$ using the gradient, and \item solving an $L_2$-regularized problem~\eqref{eq:objective}-\eqref{eq:state_constraint}, \eqref{eq:bound_p}-\eqref{eq:bound_kappa} for $\vect{g}$ on these components. \eipa That is, we iterate:
\begin{enumerate}[1.]
    \item \iemph{Identify Significant Gaussians}, i.e., Gaussians corresponding to the $2\nu$ largest components of $\igrad\D{J}$.
    \item \iemph{Merge Support} of current solution with new Gaussians from step 1; this forms $\D{V}$.
    \item \iemph{Find Solution} $\vect{g}$ of~\eqref{eq:objective}-\eqref{eq:state_constraint}, \eqref{eq:bound_p}-\eqref{eq:bound_kappa} within the subspace defined by $\D{V}$.
    \item \iemph{Restrict Solution} to $\nu$ largest components in $\vect{p}$, and set $p_i=0$ for remaining components.
\end{enumerate}
The details can be found in~\cite{Subramanian19aL1}.

\smallskip
\paragraph{Enforcing the Max Constraint}
We implement the max constraint $\max \mat{\Phi}^T\vect{p} = 1$ in $\Omega$ by following a two-step strategy, keeping either $\rho_w$ or $\vect{p}$ fixed:
\begin{enumerate}[I.]
    \item \iemph{$\ell_0$-phase:} Given an estimate for $\rho_w$, we solve steps 1--4 from above for parameters $(\kappa_w, \vect{p}) \subset \vect{g}$, and keep $\rho_w$ constant during this phase.
    \item \iemph{$\rho/\kappa$-phase:} Given an estimate for $(\kappa_w, \vect{p})$, we rescale $\vect{p}$ such that $\max \mat{\Phi}^T\vect{p} = 1$ is fulfilled, and solve~\eqref{eq:inv-tumor-formulation} for parameters $(\kappa_w, \rho_w) \subset \vect{g}$, keeping the (rescaled) vector $\vect{p}$ fixed.
\end{enumerate}

\paragraph{Numerical Optimization of the $L_2$-Regularized Problem}
We solve~\eqref{eq:objective}-\eqref{eq:state_constraint}, \eqref{eq:bound_p}-\eqref{eq:bound_kappa} in an optimize-then-discretize fashion using an adjoint-based approach to compute gradients. The resulting optimization problem is solved using a bound constrained, limited memory BFGS quasi-Newton solver globalized with Armijo line-search.
Details on the numerical solution of the forward problem~\eqref{eq:tu-state} are given in the SM.

Next, we describe how~\eqref{eq:inv-tumor-formulation} is solved using a \ipoint{coarse-to-fine multigrid approach with parameter decomposition}, to improve robustness and practicality of the scheme.

\smallskip
\subsubsection{A Robust Continuation Scheme with Parameter Decomposition}
\label{sec:continuation_scheme}
We stabilize and speed up the solution of~\eqref{eq:inv-tumor-formulation} by a \iemph{multilevel coarse-to-fine continuation scheme}. We consider image resolutions $n^{\ell}\in\{64^3,128^3,256^3\}$, $\ell=1,2,3$,
and solve a series of three optimization problems, where the coarse-level solution serves as an initial guess, increasing the chance of keeping the Newton method within the basin of attraction of the global minimum.
Besides coarsening the spatial resolution, we also coarsen the bandwidth $\sigma^{\ell} \rightarrow 2\sigma^{\ell} = \sigma^{\ell-1}$ of the Gaussian basis functions across levels.
On levels $128^3$ and $256^3$, we select the Gaussians basis functions based on the coarse level solution for the tumor initial condition, i.e., we set $m_{\Phi} \defeq c(0)^{\ell-1}$, while on the coarsest level $64^3$, we use the tumor core input data $m_{\Phi} \defeq m_T$ for Gaussian selection. This strategy allows to robustly refine to a very small target bandwidth of $\sigma = \nf{2\pi}{256}$ on the original image resolution $n^3=256^3$, to enable very localized (point-like) tumor initial conditions.
Furthermore, high resolution is required to reliably invert for the diffusion coefficient $\kappa_w$~\cite{Subramanian19aL1}.

\smallskip
\paragraph{Injection of Coarse Level Solution $\vect{g}^{\ell-1}$}
Parameters $c(0)$ and $\rho_w$ are correlated: a non-localized initial condition results in underestimation of $\rho_w$ (cf. SM). Changing the bandwidth $\sigma$ across levels changes the scale of $\rho_w$. To compensate this, we modify the max constraint~\eqref{eq:max_constraint} to  $\max \mat{\Phi}^T\vect{p} = \delta^{\ell}$ with $\delta^{\ell}\in\{\nf{1}{4},\nf{1}{2},1\}$ for $\ell=1,2,3$, such that the integral $\int_{\Omega} \mat{\Phi} \dx = \mbox{constant}$, across levels. The estimation of $\rho_w^{\ell-1}$ and $\kappa_w^{\ell-1}$ on the coarse level is then used as initial guess on the fine level. Similarly, $\vect{p}^{\ell-1}$ is injected as initial guess on the fine level.

\smallskip
\paragraph{Enabling Multifocality}
The  ``sparsity level'' $\nu$ of the initial tumor $c(0)$ is an important parameter. If $\nu$ is too large the TIL is not localized, which is not physical and leads to wrong growth parameters. If $\nu$ is too small we cannot reconstruct complex tumor shapes, for example multifocal tumors. To facilitate an \iemph{a priori} choice for $\nu$, we perform a connected component analysis (see SM for details) on the tumor input data $m_T$, and define a target sparsity $\nu_{c_j}$ per component $j \in \{1,\ldots,n_c\}$ as follows:%
\footnote{
$n_c$ is the number of components with relative mass above $0.1\%$, as obtained from connected component analysis (see SM for details).}%
We set the overall sparsity of $c(0)$ to $\nu=5n_c$, and sub-divide it into a \iemph{per-component sparsity} by attributing $3$ DOFs to each component, and distributing the remaining $2n_c$ DOFs according to the relative mass.%
\footnote{
To account for potentially erroneous segmentations used to derive the target data $m_T$, we additionally attribute one degree of freedom to components with relative mass below $0.1\%$.}
Steps 1--4 in the \iemph{$\ell_0$-phase} are then performed as per-component. This strategy allows for robustly dealing with multifocal tumors of vastly heterogeneous size and shape.

\smallskip
\paragraph{Observation Operator}
Using the correct observation operator is critical, and its choice strongly affects the inversion. The  extent of tumor infiltration cannot be observed from MRI. Further, we are lacking a model for the formation of peritumoral edema. While edema usually presents a higher concentration of malignant cells, it cannot be considered cancer tissue. For our model to capture formation of edema, we would have to define a value for the tumor concentration within this region, that we want the model to match. Such a value is not known.  Instead, we employ the concept of partial observations, using an observation operator based on tumor sub-regions as $\D{O} \defeq m_T + \lambda (1-m_T-m_E)$. In other words, we have full observation within the patient's tumor core $m_T$, but partial observations outside of the tumor core. $\lambda=0$, e.g., reflects lack of observations everywhere but for the tumor core and, thus, does not penalize tumor growth outside of this region (predicting a tumor concentration of one everywhere, would, e.g., give zero data-misfit for this choice). $\lambda=1$, on the other hand reflects lack of observations only in the region of peritumoral edema. We experimentally assess the performance for different values of $\lambda$ in the SM. We found $\lambda=1$, (i.e., $\D{O} \defeq 1-m_E$) to yield best results (in terms of matching the observed tumor) across the BraTS dataset.

\smallskip
\paragraph{Summary of Inversion Scheme}
We summarize the overall scheme in~\algref{alg:pseudocode}. The grid-continuation strategy provides good initial guesses for the inversion parameters $\rho_w$, $\kappa_w$, and $\vect{p}$. On the coarsest level, Gaussians are selected based on the input data $m_T$, while on subsequent levels we use the coarse level initial condition. If a coarse solution is available, we invert for parameters $\rho_w$ and $\kappa_w$ (line~\ref{alg:l:pre-rho-kappa}), before seeking for a sparse initial condition (lines~\ref{alg:l:l0-phase-start}--\ref{alg:l:l0-phase-end}). This stabilizes the inversion for the location of tumor initiation, which is largely affected by $\rho_w$ and $\kappa_w$.
\begin{algorithm}[tb]
  \centering\scriptsize
  \begin{algorithmic}[1]
    \Statex{}
    \Require{~\, Initial guess $\vect{g}^0$, and label maps $\vect{m},\, m_T,\, m_E$}
    \Ensure{Solution $\vect{g} = (\rho_w,\;\kappa_w,\;\vect{p}),~ c(0)$}
    \vspace{1mm}\hrule\vspace{1mm}
    \State{down-sample $\{\vect{m}\}_{\ell=1}^3,\, \{m_T\}_{\ell=1}^3, \{m_E\}_{\ell=1}^3$}
    \State{$m_{\Phi} ~\leftarrow~ m_T^{\ell=1}$}
    \For{$\ell = 1, 2, 3$}
        \State{select Gaussians based on $m_{\Phi}$}
        \State{$\D{O}\defeq 1-m_E^{\ell},~ \sigma^{\ell} = \nf{2\pi}{n^{\ell}},~ \D{V} ~\leftarrow~ \{\}$}
        \State{$\vect{\nu}_c,\; n_c,\; \vect{c} ~\leftarrow~ $\funccall{concomp}($m_T^{\ell}$)} \label{alg:l:concomp}
        \If {$\ell > 1$}
            \State{$\vect{g}^{\ell} \leftarrow \vect{g}^{\ell-1} (injection),~ \D{V} ~\leftarrow~ \mbox{supp}(\vect{p}^{\ell-1})$}
            \State{rescale $\vect{p}$ s.t. $\max \mat{\Phi}^T\vect{p} =  \nf{n^{\ell}}{256}$}
            \State{$\rho_w,\; \kappa_w ~\leftarrow~ $\funccall{$\rho/\kappa$-phase}}
            \Comment[.25\linewidth]{$\vect{p}\;\;=$ fixed} \label{alg:l:pre-rho-kappa}
        \EndIf
        \For{$k = 1, \ldots, \iota$} \Comment[.25\linewidth]{\funccall{$\ell_0$-phase}}\label{alg:l:l0-phase-start}
            \State{$\D{V} ~\leftarrow~ \mbox{supp}(\vect{p}) \cup \mbox{supp}(\igrad\D{J}(\vect{g})|^{2\vect{\nu}})$} \label{alg:l:grad-thresh}
            \State{solve~\eqref{eq:inv-tumor-formulation} in $\D{V}$ for $(\kappa_w,\vect{p})$}
            \Comment[.25\linewidth]{$\rho_w=$ fixed}
            \State{$\vect{p} ~\leftarrow~ \vect{p}|^{\vect{\nu}} (restriction),~\D{V} ~\leftarrow~ \mbox{supp}(\vect{p})$} \label{alg:l:l0-phase-end}
        \EndFor
        \State{solve~\eqref{eq:inv-tumor-formulation} in $\D{V}$ for $(\kappa_w,\vect{p})$}
        \Comment[.25\linewidth]{$\rho_w=$ fixed} \label{alg:l:l2-post}
        \State{rescale $\vect{p}$ s.t. $\max \mat{\Phi}^T\vect{p} =  \nf{n^{\ell}}{256}$}
        \State{$\rho_w,\; \kappa_w ~\leftarrow~ $\funccall{$\rho/\kappa$-phase}}
        \Comment[.25\linewidth]{$\vect{p}\;\;=$ fixed} \label{alg:l:kappa-rho-post}
        \State{up-sample $c(0)^{\ell}$ and set $m_{\Phi} ~\leftarrow~ c(0)^{\ell}$}
     \EndFor
  \end{algorithmic}
  \caption{\label{alg:pseudocode}\small \textit{Outline of inversion scheme using grid-continuation and parameter decomposition in pseudocode. During the \funccall{$\rho/\kappa$-phase}, the vector $\vect{p}$ is fixed, and estimates for $\rho_w$ and $\kappa_w$ are computed. In the \funccall{$\ell_0$-phase}, $\rho_w$ is held constant to solve for a sparse $\vect{p}$, and $\kappa_w$. \funccall{concomp} performs a connected component analysis on $m_T$ (see SM for details), returning the number of components $n_{c}$, and the sparsity per component $\vect{\nu}_{c}$.} \vspace{-14pt}}
\end{algorithm}
On every level, connected component analysis is performed on the data $m_T$ (line~\ref{alg:l:concomp}), defining the sparsity level $\vect{\nu} = \{\nu_{c_j}\}_{j=1}^{n_c}$ per component. Gradient and solution thresholding (lines~\ref{alg:l:grad-thresh} and~\ref{alg:l:l0-phase-end}) is then performed \iemph{per component}, i.e., the largest $\nu_{c_j}$ entries from $\igrad\D{J}(\vect{g})$ and $\vect{p}$ are selected, respectively.

\smallskip
\paragraph{Remarks on Robustness and Convergence}
The discussed extensions to the original inversion scheme~\cite{Subramanian19aL1} are inevitable, to cope with complicated multifocal tumors and great heterogeneity when targeting real clinical data. In fact, if omitted, the inversion diverges or terminates in bad local minima.
In particular, a good initial guess for $\vect{g}$ is critical to keep the optimizer within a basin of attraction of the solution, and helps to cope with the severe non-convexity of the problem.
Further, we found that (incorrect) values for $\rho_w$, and $\kappa_w$ largely affect the inversion for a sparse initial condition, and vice versa and, thus, the parameter decomposition scheme has to be chosen carefully.
Together, two concepts deliver the required robustness: \bipa \item the coarse level initial guess, and \item the choice of parameter decomposition. \eipa For the  latter, it is important to solve for $\rho_w$ and $\kappa_w$ \iemph{before} and \iemph{after} inversion for $\vect{p}$, and to only keep $\rho_w$ and $\vect{p}$ alternately fixed, but always invert for  $\kappa_w$.

\smallskip
\subsubsection{Atlas Registration}\label{sec:methods:atlas-reg}
%
To analyze our cohort of 206 patients, we co-registered them to a template using our  large-deformation diffeomorphic image registration software \texttt{CLAIRE}~\cite{Mang2018b:CLAIRE}. Patient images are first affinely registered to the atlas space, followed by large-deformation, diffeomorphic registration of tissue label maps $m_W$, $m_G$, and $m_F$ using \texttt{CLAIRE}. Since we do not have a mpMRI of the healthy brain of the patient, we approximate it using the tumor-bearing scan in which edema and tumor regions in the current scan are replaced by white matter. More sophisticated methods that account for mass-effect exist, but here we opted for a simpler solution.
The reconstructed locations of tumor initiation are then mapped and superimposed in a common atlas space.

\vspace{-8pt}
\subsection{Workflow Summary}
\label{sec:implementation_details}
Before evaluating our methodology, let us summarize some remaining implementation details.

\subsubsection{Imaging Data Processing}\label{sec:data-processing}
We consider clinical 3D MRI data of 206 GBM patients from the BraTS'18 training data set~\cite{Bakas2018Brats}. The dataset provides expert segmentation of different GBM phenotypes into necrotic tumor core (NEC), enhancing tumor (EN), and peritumoral edema (ED). In a pre-processing step, we infer binary segmentations of healthy tissue labels, i.e., white matter (WM), gray matter (GM), cerebrospinal fluid (CSF), and ventricles (VE) (or, if not available segmentations of tumor regions, respectively) from mpMRI data, using a deep convolutional neural network~\cite{brats18}.
The observations used for calibration are the tumor core (TC), composed of the labels (NEC) and (EN).
Solver settings and parameter choices are given in the SM.

\vspace{-8pt}
\section{Results}
\label{sec:results}
\begin{figure*}[h!]
\centering
\includegraphics[width=\textwidth]{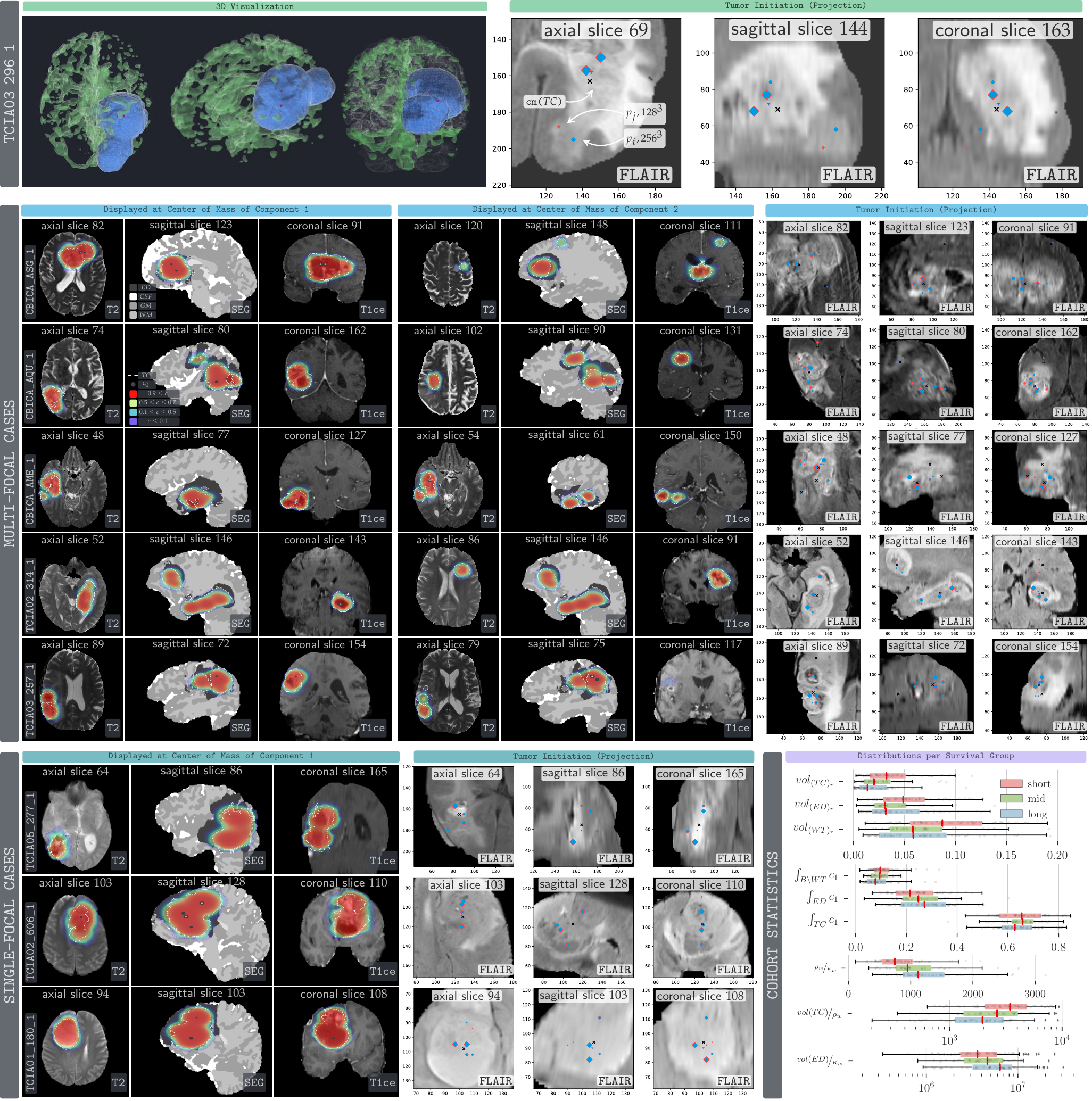}
\caption{
\ipoint{BraTS'18 Results}. \crule[pastelgreen!80]{0.2cm}{0.2cm} Left: Shown are axial, coronal, and sagittal 3D views of the predicted tumor (blue), its estimated origin (red), and $CSF$ with ventricles (green). Right: Projection of estimated TIL $\vect{p}$ onto axial, coronal, and sagittal plane (red corresponds to a $128^3$ image resolution; blue corresponds to a  $256^3$ resolution). The size of the markers indicates the magnitude of $\vect{p}$. The center of mass of tumor core input data components are indicated with a black cross.
\crule[babyblue]{0.2cm}{0.2cm} \ipoint{Multifocal Cases}. The slices are selected
at coordinates of the center of mass of the largest (left) and second largest (middle) $TC$ component.
The biophysical  model-predicted tumor $c(1)$  is given in contours from red to violet; the reconstructed TIL, $c(0)$, is shown in dark gray and is always in the support of $c(1)$; the $TC$ input data is indicated as a dashed white line.
The red $c(1)$ contour line corresponds to $0.9 \leq c(1)$, and matches well with the input $TC$ data. The right-hand side shows TIL projections as explained above.
\crule[colorC]{0.2cm}{0.2cm} \ipoint{Single-focal Cases}.
\crule[c1color1!60]{0.2cm}{0.2cm} \ipoint{Cohort Statistics}. Depicted are median and variance of relative size, and reconstruction measures $\int_{X} c_1 \defeq \nf{\int_{X} c(1)\dx}{\int_{\Omega}c(1)\dx}$, with $X \in \{TC, ED, B\setminus WT\}$ of tumor subregions, subdivided into three survival classes (long, mid, short). We also report the distribution of $\nf{\rho_w}{\kappa_w}$, $\nf{vol(TC)}{\rho_w}$, and $\nf{vol(ED)}{\kappa_w}$ for the three classes. Boxes range from lower to upper quartile.
\label{fig:sim-img-brats}\vspace{-16pt}}
\end{figure*}

We examine robustness of our methodology against 206 GBM cases from the BraTS'18 training cohort, featuring great inter-subject variability and vastly heterogeneous GBM phenotypes. We report distributions of calibrated model parameters and other image based features. For 161 cases, we  also have  patient age, treatment status, and patient overall survival. We investigate correlations between image-based and biophysical model features (such as tumor origin, growth rates, tumor size, and others) with survival. In summary, we report the following:

\begin{itemize}\zapspace
\item \emph{Inversion~(\secref{sec:results:inversion}).}\\
{\bf \tabref{tab:brats-sim-results}:} proliferation and migration rates; reconstruction errors; statistics for the BraTS'18 dataset.
{\bf \figref{fig:sim-img-brats}:} reconstruction of the OT and TIL; comparison with center of mass of tumor components; variability of several global parameters across BraTS'18 subjects.
\item \emph{Atlas construction~(\secref{sec:results:gbm-atlas}).}\\
{\bf \figref{fig:gbm-atlas-3d}:} localized TILs for all BraTS'18 subjects in atlas space (TIL probability atlas).

\item \emph{Survival analysis~(\secref{sec:results:prediction}).}\\
{\bf \tabref{tab:survival-prediction-results}:} results using hand-crafted features.
{\bf \tabref{tab:feature-importance}:} feature importance.
{\bf \figref{fig:func-atlas-analysis-top16-labels}:} regional correlations using TIL and OT probability atlases.
\end{itemize}
In the supplementary material (SM) we report the following: examples of sensitivity of solution to tumor parameters; sensitivity to the observation operator (\secref{sec:continuation_scheme}); examples of TIL localization; Atlas TILs grouped per survival class; detailed regional analysis survival results; and all reconstruction results for the 206 BraTS'18 subjects.

Before discussing the results in detail, let us introduce some notation. To compare the prediction of the biophysical model ($c(1)$)  to the observed tumor segmentation ($m_T$), we compute  $\mu=\nf{\|{c}(1)-m_T\|_{L^2(m_T)}^2}{\|m_T\|_{L^2(m_T)}^2}$.%
\footnote{We prefer this over a DICE coefficient, which would require arbitrary thresholding of $c(1)$.}
With $\int_{X} c_1$, we denote integrals over  subregions $X \in \{TC, ED, $ $  B\setminus WT\}$. These are used for feature extraction. $D_{cm}$ is the distance between the center of mass (CM) of $c(0)$ (TIL), and the CM of the tumor input data $m_T$ (in voxels);%
\footnote{The distance is computed per (tumor) component (see~\secref{sec:numerical_scheme}). We only report the distance for the largest component here. The distance for all components can be found in the supplementary material.}
$D_{\vect{p}}$  is the distance between the two most weighted locations $\vect{x}_{{p_i}_1}$ and $\vect{x}_{{p_i}_2}$ of TIL. $D_{cm}$ and $D_{\vect{p}}$ are used to quantitatively assess variability of estimated TIL and, in particular, the deviation from the CM of the input data.

\vspace{-10pt}
\subsection{TIL computation for BraTS'18 High Grade Gliomas}\label{sec:results:inversion}
\vspace{-2pt}
\paragraph{Results}
We report representative results in~\tabref{tab:brats-sim-results}. We qualitatively examine \bipa \item the sparse localization and spatial distribution of TIL, as well as \item the reconstruction ability and robustness when given complicated shapes, and large, multifocal tumors \eipa by visual inspection of various 2D slices given in~\figref{fig:sim-img-brats}.
We report statistics on the data mismatch and distance of TILs in~\tabref{tab:brats-sim-results} from 206 cases. \figref{fig:sim-img-brats} also shows distributions of the size of tumor regions, the reconstruction measures $\int_{X}c_1$, and the fractions $\nf{\rho_w}{\kappa_w}$, $\nf{vol(TC)}{\rho_w}$, and $\nf{vol(ED)}{\kappa_w}$, respectively, for each survival class from 161 patients (with survival data). Qualitative and quantitative results for the entire cohort are given in the supplementary.

\begin{table}
    \caption{
    \ipoint{Top: Quantitative results.}
    We report the estimated tumor proliferation and infiltration rates $\rho_w$, and $\kappa_w$, the relative $\ell_2$-data misfit at observation points $\mu$, the distance $\distCM$ (in voxels) between the center of mass (CM) of tumor initiation sites $c(0)$ and the CM of $TC$ (of the largest component), and the distance $\distPi$ (in voxels) between the two most important tumor initiation sites. Mean and standard deviation are given for the entire  BraTS'18 cohort of 206 cases.
    \ipoint{Bottom:} Median and variance of the misfit $\mu$, and voxel distances $\distCM$, and $\distPi$. Boxes range from lower to upper quartile.
    \label{tab:brats-sim-results}}
\centering\scriptsize\setlength\tabcolsep{2pt}
\begin{tabular}{l|LLlLll}
    \toprule
    BraTS ID                     & $\rho_w$        & $\kappa_w$    & $\mu$  & $\distCM$ & $\distPi$  \\
    \midrule
    {\scriptsize\texttt{CBICA\_ASG~~(2)}} & 11.1 & \num{1.526e-02} & \num{1.734e-01}  & $2.8$ & $19.6$   \\
    {\scriptsize\texttt{CBICA\_AOZ~~(3)}} & 9.49 & \num{9.96e-03}  & \num{1.915e-01}  & $3.9$ & $19.6$   \\
    {\scriptsize\texttt{CBICA\_AQU~~(2)}} & 9.66 & \num{1.138e-02} & \num{1.882e-01}  & $1.3$ & $25.3$   \\
    {\scriptsize\texttt{CBICA\_AME~~(3)}} & 10.1 & \num{1.040e-02} & \num{1.930e-01}  & $3.6$ & $19.6$   \\
    {\scriptsize\texttt{TCIA02\_314~(1)}} &  9.7 & \num{1.069e-02} & \num{2.617e-01}  & $4.1$ & $49.3$   \\
    {\scriptsize\texttt{TCIA03\_257~(2)}} &  9.2 & \num{1.113e-02} & \num{2.190e-01}  & $4.2$ & $17.8$   \\
    {\scriptsize\texttt{TCIA05\_277~(1)}} & 10.9 & \num{2.485e-02} & \num{1.636e-01}  & $8.1$ & $51.2$   \\
    {\scriptsize\texttt{TCIA03\_296~(1)}} & 11.6 & \num{2.764e-02} & \num{1.800e-01}  & $7.1$ & $19.6$   \\
    {\scriptsize\texttt{TCIA02\_606~(1)}} & 11.6 & \num{2.683e-02} & \num{1.741e-01}  & $5.4$ & $40.0$   \\
    {\scriptsize\texttt{TCIA01\_180~(1)}} & 12.7 & \num{1.392e-02} & \num{1.371e-01}  & $6.7$ & $17.8$   \\
    {\scriptsize\texttt{CBICA\_BHM~~(1)}} & 9.17 & \num{9.308e-03} & \num{1.666e-01}  & $5.5$ & $16.0$   \\
    \midrule
    \ipoint{Mean}          & \num{1.032728014423077E1} & \num{1.37422319375E-2}  & \num{1.9009879013798198E-1} & \num{3.9917188673551105E0} & \num{1.72965086819806E1} \\
    \ipoint{Std.\;Deviation} & \num{2.654222125049975E0} & \num{9.7547626366E-3} & \num{5.520869343859957E-2} & \num{2.9827639391904976E0} & \num{8.72243412295542E0} \\
    \bottomrule
\end{tabular}
\includegraphics[width=.215\textwidth]{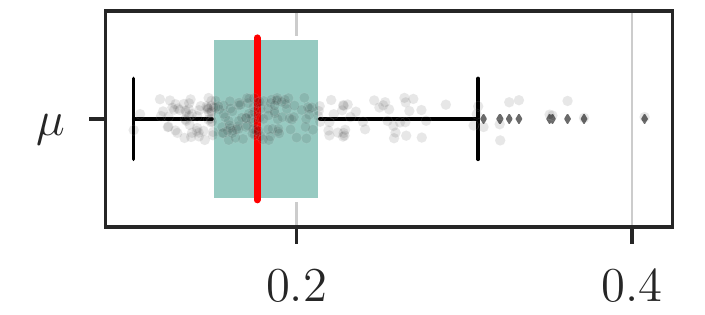}
\includegraphics[width=.24\textwidth]{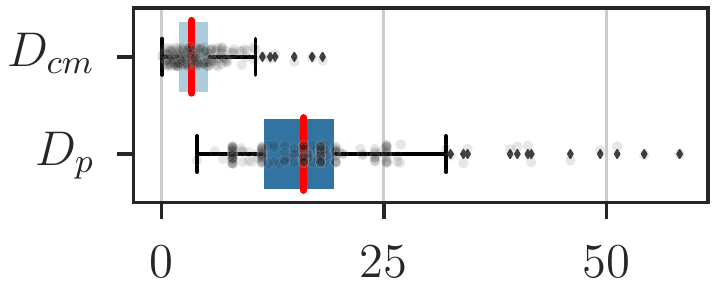}
\vspace{-20pt}
\end{table}

\paragraph{Discussion}
Our scheme runs on all scans without any failures and was able to reconstruct well-localized TILs. The model-predicted grown tumor (from estimated growth and infiltration rate) originating from those initial sites, is in good agreement with the observed $TC$ data, as can be seen from \figref{fig:sim-img-brats}. This is quantitatively supported by a small (mean) data-misfit $19\%$ error for $\mu$  in~\tabref{tab:brats-sim-results}, with small variations of $\pm 5\%$ across the entire data set.
The measure $\int_{ED}c_1$ from~\figref{fig:sim-img-brats} furthermore indicates fair extent of reconstruction of peritumoral edema, while the amount of spurious (false) tumor growth is small with a mean of around $0.1$ for $\int_{B\setminus WT}c_1$ across all cases.

Regarding the reconstructed TILs, one question that arises is how to evaluate their correctness. Of course, we don't have a ground truth. In~\cite{Subramanian19aL1} we performed tests with synthetic ground truth. Here the unknown  biophysical modeling errors  and the unknown healthy anatomy  of the subjects prevent us from conducting such analysis for the BraTS'18 set. So as a way to estimate the validity of our estimates we compare the reconstructed $\vect{p}$ with the center of mass computed using a connected-component analysis of the segmented tumor (see SM).
We  also check whether our solutions are sparse and localized. Indeed, we observe \bipa \item our estimated $c(0)$ are  sparse, and localized, and that \item $\vect{p}$ is spread out and often quite different from the center of mass of the segmented tumor. \eipa This can be seen from the visualization of projected $p_i$ activation (blue dots) in axial, coronal and sagittal views of \figref{fig:sim-img-brats}. On average the center of mass of tumor initiations (per component) deviates from the center of the $TC$ input data by only around 5 voxels. The distance between the TIL activation sites, however, averages at $17$ voxels with large outliers (cf.\;\tabref{tab:brats-sim-results} and \figref{fig:sim-img-brats}). The former attests plausibility of our solution, while the latter clearly indicates a surplus value of identifying distinct sites of tumor initiation. Lastly, the solution vector $\vect{p}$ converges across levels of our coarse-to-fine scheme (cf.\;projections of red ($128^3$ and blue ($256^3$) $p_i$ activation in~\figref{fig:sim-img-brats}).

\subsection{Construction of GBM Atlas}\label{sec:results:gbm-atlas}
\vspace{-2pt}
We construct a cohort GBM atlas  by mapping the reconstructed TILs, $c(0)$,  and  the predicted grown tumors, $c(1)$, of 161 individuals back into a common atlas space. In a second step, we analyze the distribution of TILs in particular structural and functional regions of the brain. We investigate correlations between probability of tumor occurrence in certain regions with indication towards patient survival. The probabilities are computed per voxel and per region. For a voxel $x$, we define the probability as $\sum_i c_i(x) / \sum_{ix} c_i(x)$, where $i$ indicates the case. Similar frequencies are computed for regions by summing all the voxels of a region.

\begin{figure}[tb]
\centering
\includegraphics[width=.49\textwidth]{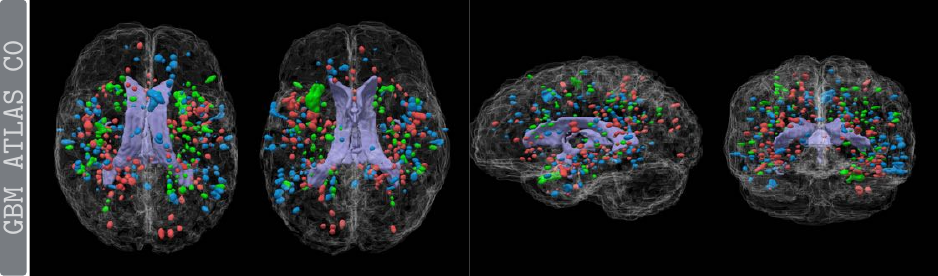}\vspace{1pt}
\includegraphics[width=.49\textwidth]{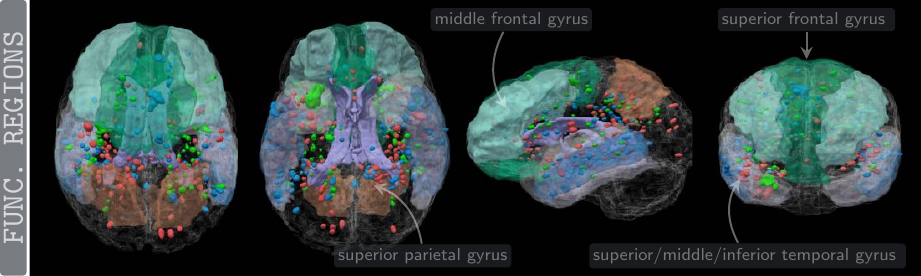}
\caption{
\ipoint{Cohort GBM atlas for BratTS'18 survival data}. \iemph{Top:} Tumor Initiation Locations (TILs) of $161$ patients superimposed and displayed in the atlas space (short survivors in red; mid survivors in green; long survivors in blue).
\iemph{Bottom:} Frequently affected functional regions.
\label{fig:gbm-atlas-3d}\vspace{-12pt}}
\end{figure}

To compute the tumor occurrence frequencies,  we register the scans to a healthy template using deformable registration (see~\secref{sec:methods:atlas-reg}). The template image is segmented into structural and  functional regions of the brain. We map the TILs,  $c(1)$, and segmentation labels to the atlas space and compute tumor origination frequencies for each region and survival class (short, mid, long).

\begin{figure}[htb]
    \vspace{-13pt}
    \centering
    \includegraphics[width=0.49\textwidth, trim=5 5 0 10,clip]{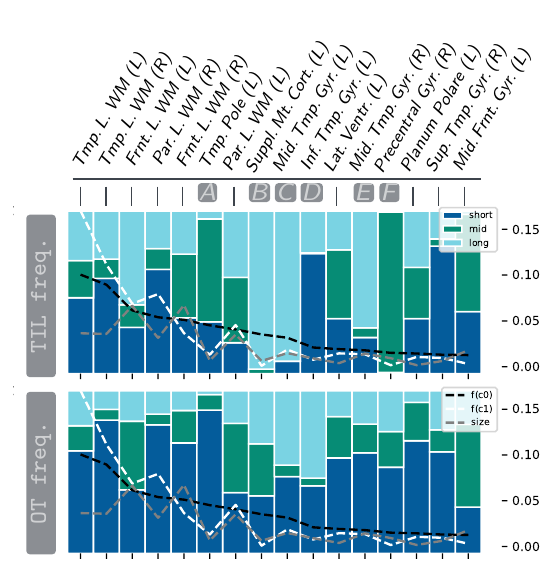}
    \caption{\ipoint{Correlation of patient survival with TIL in certain functional regions}
     based on the probabilistic GBM atlas $a$ (cf.\;\figref{fig:gbm-atlas}), we compute TIL and OT frequencies $\nf{\int_{\D{L}} a_X(\vect{x}) \d{\vect{x}}}{\int_{\Omega}a(\vect{x}) \d{\vect{x}}}$ per functional region $\D{L}$; $a_X$ for $X\in\{S,M,L\}$ denotes the probabilistic atlas of short, mid, and long survivors.
     We report  the ratio of (TIL/OT) frequencies for short (blue), mid (green), and long (cyan) survivors per label (imbalances indicate higher likelihood for survival class in this region), along with the rel.\,size and overall percentage (summed across short, mid, long) of TIL ($f(c_0)$) and OT ($f(c_1)$).
     We display the top 16 most frequently affected regions of 145 total (see SM for detailed results).
    \label{fig:func-atlas-analysis-top16-labels}
    \vspace{-8pt}
    }
\end{figure}

\paragraph{Results}
\figref{fig:gbm-atlas} and \figref{fig:gbm-atlas-3d} depict the cohort GBM probability atlas. In~\figref{fig:func-atlas-analysis-top16-labels}, we report tumor occurrence probabilities for the most affected brain regions.

\paragraph{Discussion}
\figref{fig:gbm-atlas} indicates regions commonly affected by tumors, which lead to overall survival of less than 10 months (\textit{short} survivors; red), between 10 and 15 months (\textit{mid} survivors; green), or more than 15 months (\textit{long} survivors; blue); (see SM for detailed results). We observe an indication for correlation of short patient survival and tumor origin close to the ventricles (especially lower left ventricle), while tumors originating in white matter closer to the skull correlate with a longer life expectancy. This is in accordance with typically more severe brain damage caused by the tumor mass-effect for tumors originating close to the ventricles ~\cite{Jungk-19,Lee-17,Lee-18} and difficulties with resecting such tumors.

In~\figref{fig:func-atlas-analysis-top16-labels}, we see the tumor occurence probabilities grouped per survival class. Observe that these probabilites  differ between the TIL and OT atlases. The  OT regional  probabilities are not as discriminative as the TIL and are biased due to large tumors from short-term survivors. These large tumors ``touch'' several regions. Using  TIL frequencies allows for more discriminative regional probabilities. For example, see \;\figref{fig:func-atlas-analysis-top16-labels} where TIL shows high frequency for mid (A,F), long (B,D,E), and short (D)---while OT frequencies are uninformative. Of course, due to the quite limited data in this study the results are not definite and suffer from statistical noise. They do suggest, however, issues with using the OT data and the potential benefits of using localized TIL information.

\begin{table}[bt]
    \caption{\ipoint{Survival classification} for (i) image-based, and (ii) biophysically augmented features. The total number of samples is 161 (64 short, 41 mid, and 56 long survivors); scores are reported as weighted mean. We use an 80/20 train/test split, and also report the average accuracy for a 5-fold CV.
    \label{tab:survival-prediction-results}
    }
{\scriptsize\setlength\tabcolsep{3pt}
\begin{tabular}{l|c|ccccc}
\toprule
          & \textbf{\textit{Training Data}}  & \multicolumn{4}{c}{\textbf{\textit{Test Data}}}           \\
 \textit{features} & \textit{CV avg.-accuracy} & \textit{accuracy} & \textit{precision} & \textit{recall} & \textit{f1-score} \\
 \midrule
 Age (A)          & $0.43\pm 0.11$ & $0.33$ & $0.33$ & $0.34$ & $0.33$ \\
 A+Phys.        & $0.46\pm 0.15$ & $0.46$ & $0.45$ & $0.46$ & $0.45$ \\
 A+Img.         & $0.46\pm 0.18$ & $0.39$ & $0.37$ & $0.39$ & $0.37$ \\
 A+Img.+Phys.   & $0.51\pm 0.12$ & $0.54$ & $0.53$ & $0.54$ & $0.53$ \\
 \bottomrule
\end{tabular}
}
\vspace{-11pt}
\end{table}

\vspace{-10pt}
\subsection{Survival Prediction}\label{sec:results:prediction}
\vspace{-2pt}
We conduct a preliminary study for survival prediction for the BraTS'18 survival data.
We extract image and physics-based features from 161 individuals\footnote{64 short survivors, 41 mid survivors, and 56 long survivors.} with known overall survival
to build a classifier labeling a new subject as either \textit{short}, \textit{mid}, or \textit{long} survivor. We employ XGboost~\cite{Chen:2016}, an advanced gradient boosted random forest algorithm for multiclass classification. We consider an 80/20 train/test split, and train the model using $100$ rounds, at a learning rate of $\eta=0.01$ with a maximal tree depth of $5$.\footnote{The remaining parameters are set to the default values form XGboost's Python API.}
The model is trained using the following two sets of features:
\begin{enumerate}[A.]
\item \ipoint{Image-based features:}
\bipa
    \item volume of $TC$, $ED$, and $NE$ relative to brain;
    \item volume fraction $\nf{vol(EN)}{vol(TC)}$ of enhancing tumor relative to tumor core;
    \item patient age, and treatment status.
\eipa
\item \ipoint{Physics-based features (tumor inversion):}
\bipa
    \item ratio $\nicefrac{\rho_w}{\kappa_w}$;
    \item ratios $\nf{vol(TC)}{\rho_w}$, and $\nf{vol(ED)}{\kappa_w}$ (relates size of tumor core to cell proliferation, and size of edema to cell infiltration).
\eipa
\end{enumerate}

\paragraph{Results} A classification report on the test set considering both set of features is given in~\tabref{tab:survival-prediction-results}. Additionally, we report the average accuracy with respect to a five-fold cross-validation on the training data. \tabref{tab:feature-importance} lists the determined feature importance by the respective classifier.

\paragraph{Discussion}
We observe an improvement of all classification scores when augmenting the set of  image-based features with additional biophysical features, acquired from our tumor inversion solver. Overall all classifiers generalize reasonably well as the average accuracy from cross-validation compares to the  accuracy on the left-out test set. In \tabref{tab:feature-importance} we report the feature importance;\footnote{For a single decision tree, importance is calculated by the amount that each attribute split point improves the performance measure, weighted by the number of observations the node is responsible for. Feature importances are averaged across all of the the decision trees in the model, see~\cite[p.\,367]{hastie2009elements}.} we observe that the biophysical features are quite important. In particular, the ratio of proliferation rate to infiltration rate is comparable to the patient's age importance.

We emphasize the preliminary nature of this study. The overall classification scores are fairly low, indicating the difficulty of ML methods to handle the large existing heterogeneity of the patient-specific information available, given the limited data. Similar conclusions have been drawn from the official BraTS survival prediction challenge~\cite{Bakas2018Brats}, where the winning teams report great variability in classification accuracy depending on the actual set of evaluation samples.

\begin{table}[tb]
    \caption{\ipoint{Feature importance}.
    \label{tab:feature-importance}
    }
{\scriptsize\setlength\tabcolsep{3pt}
\begin{tabular}{lccccccc}
\toprule
 \textit{features} & $\frac{\rho_w}{\kappa_w}$ & \textit{age} & $\frac{vol(TC)}{\rho_w}$ & $vol(TC)$ & $\frac{vol(ED)}{\kappa_w}$ & $vol(ED)$ & $\frac{vol(EN)}{vol(TC)}$ \\
 \midrule
 Age (A)        & --     & $1.00$ & --     &  --    & --     & --     & --     \\
 A+Phys.        & $0.28$ & $0.27$ & $0.23$ &  --    & $0.21$ & --     & --     \\
 A+Img.         & --     & $0.28$ & --     & $0.28$ & --     & $0.21$ & $0.20$ \\
 A+Img.+Phys.   & $0.18$ & $0.17$ & $0.15$ & $0.12$ & $0.12$ & $0.11$ & $0.11$ \\
 \bottomrule
\end{tabular}
}
\vspace{-11pt}
\end{table}

\vspace{-8pt}
\section{Conclusion}
\label{sec:conclusion}
We presented a methodology for reconstructing tumor growth parameters and sparse  tumor initiation locations (TILs) from clinical glioblastoma MR imaging data by means of tumor-growth inversion methods. We applied it on 206 subjects from  the BraTS'18 training set.  We simultaneously estimate \bipa \item the specific tumor cell proliferation rate, \item the tumor cell migration rate, and \item the original, sparse tumor initial locations. \eipa These quantities, combined with image features, were used  in a shallow classifier for survival prediction. Also, the TILs and the observed tumor margins where used  to construct two different probabilistic atlases of tumor occurrence. We presented preliminary results, relating patient survival to the frequency of tumor occurrence in a specific functional region of the brain.

Some key findings of our study are the following: \textbullet{} Our method can be reliably applied to clinical data in a  fully automatic fashion. \textbullet{} TIL localization results in dramatically different infiltration prediction and tumor growth parameters. \textbullet{} The localized biophysical inversion TIL correlates with an entirely image-based TIL computed with  connected-component analysis of the tumor segmentation. However, the image-based are quite often different from the TIL and cannot reliably be used to estimate tumor parameters. \textbullet{} The probabilistic tumor occurrence atlas using the TIL differs from the atlas reconstructed using the observed tumor margins. In particular, the latter seems to be biased by the tumor size of particular subjects. \textbullet{} Preliminary results in classification of survival are encouraging but inconclusive due to  the small sample size.  \textbullet{}  We observed that the area around the lower left ventricle (left temporal lobe)  shows a clustering of aggressive tumors, while tumors resulting in longer survival are generally found closer to the skull. 

Here are some of  the \iemph{limitations} of our work. We use a phenomenological, single-species model, that cannot distinguish between different glioma phenotypes observed from medical MRI such as necrotic core, enhancing and non-enhancing tumor, and peritumoral edema. We do not  account for mass effect; we have derived more complex models that do so~\cite{subramanian-gholami-biros19}, and their integration with the sparse TIL reconstruction is ongoing.  Our dataset is quite small and voxelwise/regional information is questionable.  We have not included comparisons with postoperative recurrence, which is  another way to validate the model. (Such postoperative datasets are not available for the BraTS'18 images.) The software is not yet open source but our plan is to make it available through the \href{https://www.med.upenn.edu/cbica/captk/}{CaPTk} interface in the  near future.

Our method lays the basis for reliable biophysical model calibration from a single preoperative scan.  In particular the location of tumor initiation could be used to differentiate tumor mutations, and could (among other radiographic features and biomarkers) contribute towards important clinical diagnostic and prognostic outcomes such as prediction of patient survival times, and identification of brain regions that are of high risk to develop tumors. However, testing  these hypotheses conclusively requires further studies with larger datasets and richer biophysical models. Our next steps include the combination of machine learning methods with more complex multispecies tumor models.


%



\section*{Acknowledgment}
The authors would like to thank Naveen Himthani and Andreas Mang for their valuable contributions and the fruitful discussions.

\ifCLASSOPTIONcaptionsoff
  \newpage
\fi




\vspace{-8pt}
\bibliographystyle{bib/IEEEtran}
\interlinepenalty=10000

\bibliography{bib/IEEEabrv,bib/literature.bib}
\end{document}


%
\title{Supplementary Materials for \\
 {\large\textbf{Automatic MRI-Driven Model-Calibration for Advanced Brain Tumor Progression Analysis}}}
%
%
%

\author{Klaudius Scheufele, Shashank Subramanian, George Biros, \IEEEmembership{Member~IEEE}
\thanks{K. Scheufele, Oden Institute, University of Texas at Austin, Texas, USA e-mail: kscheufele@{austin}.utexas.edu.}
\thanks{S. Subramanian, Oden Institute, University of Texas at Austin, Texas, USA e-mail: shashanksubramanian@{utmail}.utexas.edu.}
\thanks{G. Biros, Oden Institute, University of Texas at Austin, Texas, USA e-mail: biros@{oden}.utexas.edu.}
\thanks{Manuscript received April 19, 2005; revised August 26, 2015.}}

\markboth{IEEE Transactions on Medical Imaging}
{Scheufele \MakeLowercase{\textit{et al.}}: Automatic MRI-Driven Model-Calibration for Advanced Brain Tumor Progression Analysis}
%

\maketitle


%

%
\IEEEpeerreviewmaketitle

\textbf{The PDF file includes:}\\
\begin{itemize}
    \item \textbf{Methods~(\secref{sec:sm:methods})} \\
    ({\bf\secref{sec:tumor-model}}) \ipoint{Tumor Growth Model}.
    \begin{itemize}[~]
        \item Details on tumor progression model and materials.
    \end{itemize}
    ({\bf\secref{sec:numerical_scheme}}) \ipoint{Numerical Scheme.}
    \begin{itemize}[~]
    \item Details on numerical solution of the forward problem, and biophysical parameter inversion, including parametrization of the TIL.
    \end{itemize}
    ({\bf\secref{sec:implementation_details}}) \ipoint{Implementation Details.}
    \begin{itemize}[~]
    \item Details on the spatial and temporal discretization method, justification of important parameter choices, and limitations.
    \end{itemize}
    \vspace{3pt}
    \item \textbf{Numerical Results~(\secref{sec:sm:results})} \\
    ({\bf\secref{sec:fwd-model-sensitivity}}) \ipoint{Parameter Sensitivity Forward Model}.
    \begin{itemize}[~]
      \item {\bf\figref{fig:forward-vary-tho-k}:} Simulated tumor concentration for varying  cell proliferation and cell migration rates.
    \end{itemize}
    ({\bf\secref{sec:sparse-TIL}}) \ipoint{Enforcing Localized TIL}.
    \begin{itemize}[~]
    \item {\bf\figref{fig:l1l2}:} Localized TIL versus non-localized TIL, and effects on estimated cell proliferation and cell migration rate.
    \end{itemize}

    ({\bf\secref{sec:sm:obs-op}}) \ipoint{Sensitivity to Observation Operator}.
    \begin{itemize}[~]
    \item {\bf\figref{fig:obs-operator}:} Numerical results (qualitative and quantitative) demonstrating sensitivity of the inversion scheme to the employed observation operator (2+9 cases).
    \item {\bf \tabref{tab:obs-operator}:} Qualitative results showing sensitivity of the parameter estimation (tumor inversion) with respect to the observation operator (14 cases).
    \item {\bf \figref{fig:obs-operator}}  Quantitative results showing sensitivity of the tumor inversion with respect to the observation operator (14 cases).
    \end{itemize}

    ({\bf\secref{sec:sm:inversion}}) \ipoint{Inversion and Cohort GBM Statistical Atlas}.
    \begin{itemize}[~]
        \item Performance measures for biophysical tumor inversion, and details on construction of TIL and OT GBM atlas of BraTS'18 cohort. Detail on computation of TIL frequencies in specific (functional) brain regions.
        \vspace{5pt}
        \medskip
        \medskip
        \medskip
        \item \emph{1) Parameter Estimation and TIL Reconstruction:}
        \item {\bf\tabref{tab:brats-sim-results-ALL}:} Quantitative simulation results (estimated model parameters, misfit, TIL) for the entire BraTS'18 cohort.
        \item {\bf \figref{fig:brats18-sim-1}--\figref{fig:brats18-sim-6}:} Qualitative results of tumor growth inversion and sparse TIL reconstruction for the entire BraTS'18 cohort.
        \vspace{3pt}
        \item \emph{2) Cohort GBM Statistical Atlas for BraTS'18 Data:}
        \item {\bf \figref{fig:brats18-gbm-atlas-c0-4x9}:} Tumor initiation location (TIL) GBM atlas of BraTS'18 survival data cohort, visualized as 2D cuts in axial plane. Probabilities for short, mid, and long survivors are outlined separately.
        \item {\bf\figref{fig:brats18-gbm-atlas-c1-4x9} and \figref{fig:brats18-gbm-atlas-c1-sml-1}--\figref{fig:brats18-gbm-atlas-c1-sml-2}:} Observed Tumor (OT) GBM atlas of BraTS'18 survival data cohort, visualized as 2D cuts in axial plane. Probabilities for short, mid, and long survivors are outlined separately.
        \item {\bf\figref{fig:gbm-atlas-3d_v2}:} 3D rendering of tumor initiation location (TIL) and observed tumor (OT) GBM atlas for BratTS'18 survival data, subdivided into short, mid, and long survivors.
        \vspace{3pt}
        \item \emph{3) Correlation of Patient Survival and TIL:}
        \item {\bf \tabref{tab:func-atlas-freq}:} Correlation of patient survival with TIL in certain functional regions (study based on the SRI atlas brain~\cite{Rohlfing:2010sri} segmentation of functional brain regions).
        \item {\bf\figref{fig:gbm-atlas-3d}:} 3D rendering of tumor initiation location (TIL) cohort GBM atlas for BratTS'18 survival data with frequently affected brain regions highlighted.
        \item {\bf\tabref{tab:func-atlas-freq-muse} and \figref{fig:func-atlas-analysis-all-labels}:} Correlation of patient survival with TIL in certain functional regions (study based on expert segmentation of atlas template provided by the University of Pennsylvania).
    \end{itemize}
\end{itemize}

\medskip
\noindent
\textbf{Other supplementary materials for this manuscript includes the folowing:}\\
{\small( \textit{available online along with this PDF)} \\
\begin{itemize}
    \item ({\bf\secref{sec:sm:features}}) \ipoint{BraTS'18 Feature Extraction}.
    \begin{itemize}[~]
        \item Handcraftet image-based and biophysics-based \href{https://docs.google.com/spreadsheets/d/e/2PACX-1vTa96OpAYMAjJOmIwKqV_yupCPbB4YWNwsIdYcrZIhzNptJQLVl0OFwUxt_hEOypFX9nXBBmy9o4fNy/pubhtml?gid=1922701037&single=true}{features} of BraTS'18 cohort provided in \verb!.xls! data file (Microsoft Excel format), and \href{https://docs.google.com/spreadsheets/d/e/2PACX-1vTa96OpAYMAjJOmIwKqV_yupCPbB4YWNwsIdYcrZIhzNptJQLVl0OFwUxt_hEOypFX9nXBBmy9o4fNy/pub?gid=1922701037&single=true&output=csv}{.csv} data file. For the description of the reported features, refer to {\bf \tabref{tab:feature-descr}}.
    \end{itemize}
\end{itemize}

\newpage

\section{Methods}\label{sec:sm:methods}

\subsection{Tumor growth model}\label{sec:tumor-model}

\medskip
\subsubsection{Model and Materials}
 Malignant cell invasion and proliferation are modeled by the \iemph{diffusion} operator $\D{D}(c) = \idiv \mat{\kappa_{\vect{m}}}\igrad c$, and a logistic \iemph{reaction} term $\D{R}(c) = \mat{\rho_{\vect{m}}}c(1-c)$, respectively. Material properties are derived using voxelwise probabilities $m_i$ for the \iemph{white matter} $(W)$, \iemph{gray matter} $(G)$, and \iemph{cerebrospinal fluid with ventricles} $(F)$ (see SM for details). We use $\vect{m}$ to indicate the voxelwise probability vector.
\begin{equation}
\vect{m}(\vect{x},t) =
\left( m_i(\vect{x},t) \right)_{i = W, \,G, \,F} \in\ns{R}^3, \;
\label{eq:vec_m}
\end{equation}
Probability maps are inferred from a deep convolutional neural network~\cite{brats18}.
Using $\vect{m}$ the diffusivity is defined as $\mat{\kappa}_{\vect{m}}(\vect{x})$ as $\kappa_w  m_{W}(\vect{x}) + \kappa_g m_{G}(\vect{x})$ with characteristic cell migration rates $\kappa_w$ and $\kappa_g$ in white matter and gray matter, respectively. Similarly, $\mat{\rho}_{\vect{m}}(\vect{x})$ is defined as $\rho_w  m_{W}(\vect{x}) + \rho_g m_{G}(\vect{x})$ with characteristic cell proliferation rates $\rho_w$ and $\rho_g$ in white matter and gray matter, respectively. Tumor cells do not grow or invade into the cerebrospinal fluid-filled ventricles $(F)$. Exemplary simulations of tumor cell density using different proliferation and migration rates are given in the supplementary material.

\subsection{Numerical Scheme}
\label{sec:numerical_scheme}

\medskip
\subsubsection{Numerical Solution of the Forward Problem}
We employ a pseudo-spectral Fourier approach on a regular grid for spatial discretization and apply periodic boundary conditions on $\partial\Omega$. All spatial differential operators are computed via a 3D fast Fourier Transform. For numerical solution of the parabolic PDEs in the forward equation (1) (and similarly adjoint eq.), we use an unconditionally stable second-order Strang operator-splitting method. For the diffusion part, we use an implicit Crank-Nicolson method, and solve the linear system with a preconditioned CG method.

For further details including derivation of optimality conditions, discretization  of PDE solvers, and numerical optimization, we refer the reader to our previous works~\cite{Subramanian19aL1,Gholami:2016a,Scheufele:2018,Scheufele:2019b}.

\medskip
\subsubsection{Biophysical Parameter Inversion}
\label{sec:inverse_problem}

\medskip
\paragraph{Selection of Gaussian Basis Functions}
The set $\{\Phi_i(\vect{x})\}_{i=1}^{n_p}$ of Gaussian basis functions is determined automatically; that is, the number $n_p$ and spatial distribution (i.e., centers $\{\vect{x}_i\}_{i=1}^{n_p}$) are defined based on an input tumor concentration map $m_{\Phi}$. Let us assume, for now, that $m_{\Phi} = m_T$ is the tumor core input data. From an equally spaced candidate set of Gaussians,\footnote{Candidate Gaussians are centered around a subset of regular grid points, equally spaced with $\delta = 2\sigma$. The spacing $\delta$ is chosen empirically to guarantee good conditioning of $\int_{\Omega}\mat{\Phi}\mat{\Phi}^T\dx$.
} we select all those functions $\Phi_i$ for which the tumor volume fraction (determined by $m_{\Phi}$) within the $\sigma$-ball $\D{B}^{\sigma}(\vect{x}_i)$ around $\vect{x}_i$ exceeds a certain threshold $\tau_V$, i.e., $\int_{\D{B}^{\sigma}(\vect{x}_i)} m_{\Phi} \dx > \tau_V \int_{\D{B}^{\sigma}(\vect{x}_i)} \dx$. This forms the initial support $\D{G}$ of $\vect{p}$. To prevent spurious infiltration of cancerous cells into CSF, the selected basis functions are set to zero in regions overlapping the CSF. See~\cite{Subramanian19aL1,Scheufele:2019b} for details.

\subsection{Implementation Details}
\label{sec:implementation_details}

\medskip
\subsubsection{Discretization}
We discretize the space-time domain $\Omega \times [0,1], \Omega= [0,2\pi)^3$, using regular grids. In space, we consider $n = \prod_{j=1}^d n_j, n_j \in \ns{N}$ regular grid points, with $\vect{x}_{\vect{i}} = \nicefrac{2\pi \vect{i}}{n}$, $\vect{i} = (i_1,\ldots, i_d) \in \ns{R}^d$ and $0 \leq i_j \leq n_j-1$, $j=1,\ldots, d$. For all spatial operations, we use a spectral projection scheme~\cite{Gholami:2017a}.
We assume all functions/images in our formulation to be periodic, sufficiently smooth, and continuously differentiable in space. We apply appropriate filters and smoother, and, if necessary, periodically extend (zero-padding) or mollify our discrete data to meet these requirements. All spatial derivatives and differential operators are then applied in the spectral domain using Fourier transforms.

In time, we use a nodal discretization, resulting in $n_t+1$ discretization points, and employ a unconditionally stable, second order Strang-splitting method~\cite{Strang:1986,Hogea:2008b} for the parabolic forward and adjoint equations. We take advantage of the splitting approach by applying an implicit Crank-Nicolson method for the diffusion part and plugging in the analytical solution for the reaction part of the PDE.
For stability reasons, and to limit aliasing effects from FFT computations, we use $n_t=40$ time steps of size $\Delta t=0.025$ for the time integration in forward and adjoint operator.

\medskip
\subsubsection{Truncation of Gaussian Basis Functions}
Since Gaussians have global support, the inversion for the infiltration rate $\kappa_w$ may be affected by the reaction coefficient $\rho_w$. That is, due to the global support, there is small positive tumor concentration everywhere in the brain. Thus by increasing the reaction parameter $\rho_w$, an effect similar to diffusion can be emulated to a certain extend.
To decouple the effects of cell proliferation and cell migration, and allow for reliable estimation of $\kappa_w$, we truncate the Gaussian basis functions after a radius of $5\sigma$.

\medskip
\subsubsection{Computation of Connected Components}
To define the target sparsity level $\nu$ of the reconstructed TIL, and also for comparison of TIL and the center of mass of tumor core input data, we perform connected component analysis on the tumor core ($m_{TC}$) input data. We use scipy's \verb!scipy.ndimage.measuremnts.label! algorithm, which implements the standard two-pass binary connected-component labeling algorithm, generalized to ND. For connectivity, we consider eight (axis aligned) neighbors.

\medskip
\subsubsection{Parameter Choices}\label{sec:param-choices}
\label{sec:parameter_choices}
We justify some of the remaining parameter choices of our method.
We set the regularization parameter for $L_2$-inversion in the restricted subspace to $\beta=1$. We found this value via an L-curve test, and manual inspection of the optimizer convergence, and solution.
We terminate the $L_2$-inversion in the restricted subspace after a gradient reduction of $4$ orders of magnitude, or after $50$ quasi-Newton iterations.\footnote{This is to prevent excessive runtimes, but typically the optimizer converges in less than $50$ iterations.} To keep the optimizer within feasible bounds and prevent bad local minima, we define bound constraints $\kappa_{min}=\num{1E-4}$, $\kappa_{max}=1$, $\rho_{min}=0$, and $\rho_{max}=20$.\footnote{The lower bound for $\kappa_{w}$ is determined by the smallest recoverable diffusion coefficient on a resolution of $256^3$.} We terminate the compressive sampling algorithm after $\iota=4$ iterations on level $64^3$, and after $\iota=2$ iterations on finer levels. Due to the good initial guess, a sparse solution is typically found after one to two iterations.

\medskip
\subsubsection{Limitations}\label{sec:limitations}
Our method assumes knowledge of the healthy brain tissue structure, (i.e., a healthy patient scan before tumor emergence). In clinical practice such data is rarely available. As a crude approximation, we construct a ``healthy'' patient brain by replacing tumor core and peritumoral edema with white matter. In doing so, structural information underlying the tumor region is lost, and we furthermore have no means of reverting the tumor caused mass effect. A more advanced approach which couples tumor inversion methods and medical image registration to jointly approximate the healthy patient and estimate tumor parameters, is work in progress~\cite{Scheufele20Sibia2}.

\section{Results}\label{sec:sm:results}

\begin{figure*}[htb]
\includegraphics[width=\textwidth]{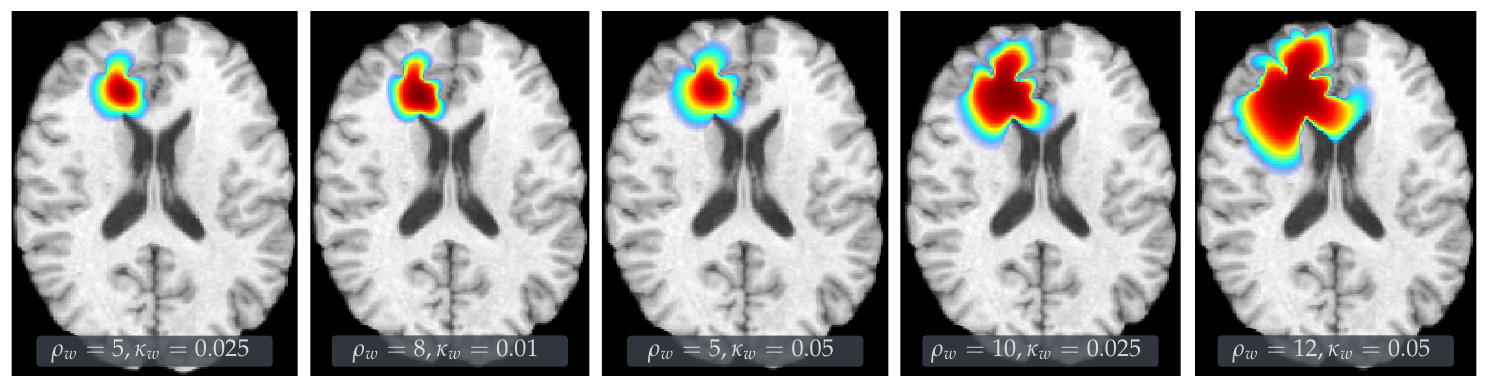}
\caption{\ipoint{Simulated tumor concentration for varying proliferation and migration rates.} We display axial slices of an exemplary T1 MRI scan, overlaid with tumor concentrations for varying proliferation and migration rates at time $T=1$, obtained from simulation using the single-species reaction-diffusion tumor progression model. The initial tumor is a single point source, and identical for all 5 displayed cases. The resulting tumor concentration, strongly varies with the assumed underlying cell proliferation and cell migration rates (in terms of size and shape of the tumor). Tumor is assumed to grow in white matter only.}
\label{fig:forward-vary-tho-k}
\end{figure*}

\begin{figure*}[htb]
\centering
\scriptsize\setlength\tabcolsep{1pt}
\begin{tabular}{p{0.56\textwidth} p{0.43\textwidth}}
  \vspace{-5pt}
  \hspace{-7pt}
  \includegraphics[width=0.57\textwidth]{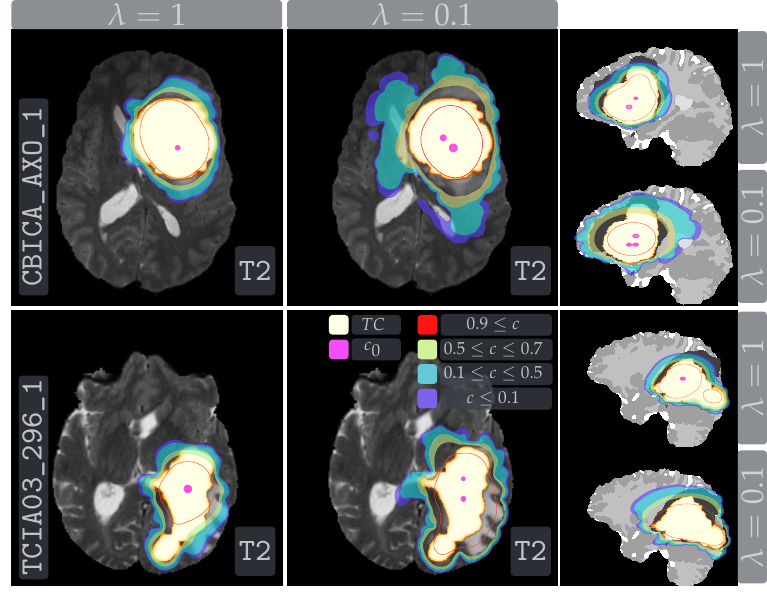}
  &
  \vspace{1pt}
  \begin{tabular}{cLllLLll}
      \toprule
      BraTS ID                     & $\lambda$ & $\rho_w$        & $\kappa_w$  & $\ellTwoObsMiss$ & $\ellTwoMiss$  & $\int_{ED}c_1$ & $\int_{B\setminus WT}c_1$  \\
      \midrule
      \cmidrule[0.05mm]{2-8}                                                                                                       %
      TCIA05 & $1$   & 10.9 & \num{2.485e-02}  & \num{2.776e-01}  & \num{5.631e-01}  & \num{3.347e-01} & \num{1.089e-01}    \\     
      277    & $0.1$ & 10.0 & \num{4.27e-02}   & \num{1.513e-01}  & \num{7.361e-01}  & \num{3.7076E-1} & \num{1.66829E-1}   \\     
      \cmidrule[0.05mm]{2-8}                                                                                                       %
      TCIA03 & $1$   & 11.6 & \CC \numb{2.764e-02}  & \num{2.972e-01}  & \CC\numb{5.061e-01}  & \num{2.457e-01} & \num{1.302e-01}   \\      
      296    & $0.1$ & 11.7 & \CC \numb{1.10e-01}   & \num{1.539e-01}  & \CC\numb{9.653e-01}  & \num{2.8490E-1} & \num{3.2224E-1}   \\      
      \cmidrule[0.05mm]{2-8}                                                                                                       %
      TCIA02 & $1$   & 11.6 & \num{2.683e-02}  & \num{3.226e-01}  & \num{5.059e-01}  & \num{2.276e-01} & \num{1.517e-01}   \\      
      606    & $0.1$ & 11.1 & \num{4.60e-02}   & \num{1.379e-01} & \num{6.919e-01}   & \num{2.4031E-1} & \num{2.6481E-1}   \\      
      \cmidrule[0.05mm]{2-8}                                                                                                       %
      CBICA  & $1$ &   \CC 13.1 & \CC\numb{3.010e-02} & \num{2.447e-01} & \num{3.944e-01} & \num{1.831e-01} & \num{1.131e-01}    \\
      AXO    & $0.1$ & \CC 9.9  & \CC\numb{2.590e-01} & \num{2.192e-01} & \num{7.242e-01} & \num{1.811e-01} & \num{4.180e-01}    \\
      \cmidrule[0.05mm]{2-8}
      CBICA & $1$   & 11.2 & \num{2.122e-02}   & \num{2.213e-01}  & \CC\numb{5.614e-01}  & \num{3.789e-01} & \CC\numb{5.683e-02}    \\     
      AAB   & $0.1$ & 10.6 & \num{3.73e-02}    & \num{1.338e-01}  & \CC\numb{8.138e-01}  & \num{1.3899E-1} & \CC\numb{2.8642E-1}    \\     
      \cmidrule[0.05mm]{2-8}                                                                                                       %
      CBICA & $1$   &  9.9 & \num{1.258e-02}   & \num{2.723e-01} & \CC\numb{6.796e-01}   & \num{4.570e-01} & \CC\numb{6.356e-02}   \\      
      AMH   & $0.1$ &  8.7 & \num{2.66e-02}    & \num{1.678e-01} & \CC\numb{9.542e-01}   & \num{5.3973E-1} & \CC\numb{1.4295E-1}   \\      
      \cmidrule[0.05mm]{2-8}                                                                                                       %
      TCIA01 & $1$   & \CC 14.3 & \num{1.483e-02}  & \num{2.117e-01} & \num{3.583e-01}   & \num{1.834e-01} & \CC\numb{6.403e-02}   \\      
      401    & $0.1$ & \CC 11.4 & \num{3.29e-02}   & \num{1.155e-01} & \num{5.905e-01}   & \num{3.2431E-1} & \CC\numb{1.1918E-1}   \\      
      \cmidrule[0.05mm]{2-8}                                                                                                       %
      CBICA & $1$   & \CC  9.11 & \CC \numb{7.722e-03} & \num{2.063e-01} & \num{3.038e-01}    & \num{1.417e-01} & \num{1.690e-01}  \\       
      ABB   & $0.1$ & \CC  6.55 & \CC \numb{1.55e-02}  & \num{2.014e-01} & \num{4.056e-01}   & \num{1.3899E-1} & \num{2.8642E-1}  \\       
      \cmidrule[0.05mm]{2-8}                                                                                                       %
      CBICA & $1$   &  9.7 & \CC \numb{6.070e-03}   & \num{2.514e-01} & \num{3.412e-01} &  \num{1.787e-01} & \num{1.007e-01}   \\       
      ATX   & $0.1$ &  8.7 & \CC \numb{1.29e-02}    & \num{1.259e-01} & \num{4.834e-01} &  \num{1.9927E-1} & \num{2.3846E-1}   \\       
      \bottomrule
  \end{tabular}
\end{tabular}
\caption{
\ipoint{Sensitivity to observation operator.} \ipoint{Left:} Qualitative results for different choices of the observation operator for $\lambda \in \{1, 0.1\}$. $\lambda=1$ reflects lack of observations only in peritumoral edema, while $\lambda=0.1$ also assumes lack of observations beyond the margin of edema.
Shown are axial, and saggital cuts of the 3D patient T2-MRI volume, overlaid with the $TC$ input data (yellow), the predicted tumor $c(1)$ from our calibrated model (contours from red to violet), and the tumor origin $c(0)$ (magenta). The red $c(1)$ contour line corresponds to $0.9 \leq c(1)$, and matches well with the input $TC$ data. \ipoint{Right:} Calibrated parameters $\rho_w$ and $\kappa_w$, and corresponding data mismatches $\ellTwoObsMiss$ (at observation points), and $\ellTwoMiss$ (everywhere) for variations $\lambda \in \{1,0.1\}$ of the observation operator. We also report the integral measures $\int_{ED}c_1 \defeq \nf{\int_{ED}c(1)\dx}{\int_{\Omega}c(1)}$, and $\int_{B\setminus WT}c_1 \defeq \nf{\int_{B\setminus WT}c(1)\dx}{\int_{\Omega}c(1)}$ indicating the approximation quality of edema, and erroneous tumor growth in healthy tissue, respectively.
\label{fig:obs-operator}}
\end{figure*}

\begin{figure}[htb]
\centering\scriptsize\setlength\tabcolsep{8.8pt}
\includegraphics[width=.49\textwidth]{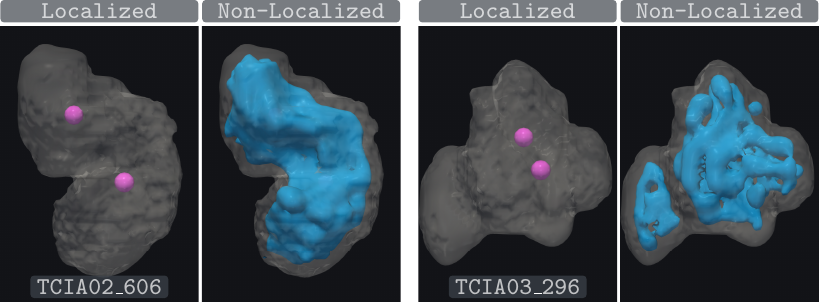}
\vspace{-4pt}

\begin{tabular}{lllll}
    \toprule
    BID & solver & $\rho_w$ & $\kappa_w$ & $\ellTwoObsMiss$ \\
    \midrule
    \multirow{2}{*}{TCIA02\_606} & localized     & $9.5$ & $\num{1.0E-2}$ & $\num{3.2E-1}$  \\
                & non-localized & $3.2$ & $\num{1.0E-4}$ & $\num{9.1E-2}$  \\
    \cmidrule{2-5}
    \multirow{2}{*}{TCIA03\_296} & localized     & $9.3$ & $\num{3.3E-2}$ & $\num{4.29E-1}$ \\
                & non-localized & $3.8$ & $\num{3.8E-4}$ & $\num{8.8E-2}$  \\
    \bottomrule
\end{tabular}
\caption{
\ipoint{Localized tumor initiation}. Shown are estimated tumor initial conditions.
The pink solution is obtained from the presented solver; the blue solution results if sparsity is not enforced~\cite{Gholami:2016a}.
The tumor core data is given for reference.
\label{fig:l1l2}}
\end{figure}
%
%
%

\subsection{Tumor Forward Model: Sensitivity to Proliferation and Migration Rate}\label{sec:fwd-model-sensitivity}
\figref{fig:forward-vary-tho-k} shows qualitative results for tumor progression simulations for varying values for cell proliferation rate and cell migration rate, using our single-species reaction-diffusion model. Displayed is the spatial tumor cell density at final time $T=1$. The solution of the reaction-diffusion forward model asymptotically approaches a traveling wave. The  shape of this traveling wave is determined by $\sqrt{\nf{\kappa_w}{\rho_w}}$, and its velocity by $2\sqrt{\kappa_w\rho_w}$ \cite{Harpold2007}.

\subsection{Enforcing Sparse and Localized Tumor Initiation}\label{sec:sparse-TIL}
Finding well localized tumor initiations is very important for various of reasons. First, the location of tumor initiation can be useful when integrated with more complex tumor progression models. Further, tumor initiation is suggested to correlate with patient survival~\cite{Adeberg-14, Chaichana-08, Weinberg-18}. Second, as we have shown in~\cite{Subramanian19aL1}, non-localized but spread out tumor initial conditions result in largely underestimated, and wrong tumor cell proliferation and migration rates. Moreover, omitting the sparsity constraint for the tumor initial condition, we have no control over the extent of this effect, and the estimated tumor initial condition tends to closely resemble the actual grown tumor target data. Restricting the initial tumor to a few, localized initial sites enables simultaneous and reliable estimation of the tumor emergence, proliferation, and migration rate.

In~\figref{fig:l1l2} we show exemplary results, comparing estimated tumor initial conditions using \bipa \item our presented methodology with sparse localization of tumor initiation, and \item the same methodology, if sparse localization is not enforced. \eipa
The table highlights dramatic underestimations of $\rho_w$, and $\kappa_w$, respectively.

\subsection{Sensitivity to the Observation Operator}\label{sec:sm:obs-op}
\label{sec:results:observation-operator}
As mentioned in the main text, the observation operator $\D{O} \defeq m_T + \lambda (1-m_T-m_E)$ is an important modeling assumption. To find an observation operator that works well with a large number of cases, we conduct numerical experiments for varying choices of $\lambda \in [0,1]$.
Exemplary results for $\lambda = 0.1$ and $\lambda = 1$ are given in~\tabref{fig:obs-operator}, along with qualitative results for two individuals. Further results are shown in ~\tabref{tab:obs-operator}, and~\figref{fig:obs-operator}.
We judge  quality by the ability to recover both, tumor core, and the region of peritumoral edema, while minimizing falsely predicted tumor growth.

\paragraph{Observations}
As expected, we observe sensitivity of the estimated values for $\rho_w$ and $\kappa_w$ to the choice of the observation operator. The scale of sensitivity, however, varies greatly across individuals: While $\kappa_w$ generally increases for smaller $\lambda$,
$\rho_w$ seems to be less sensitive. For both parameters, however, we also observe large deviations (highlighted in~\tabref{fig:obs-operator}).
Looking at the quality of the predicted grown tumor, we see a tendency of improved estimation of the peritumoral edema for lower values of $\lambda$ (cf.\;$\int_{ED}c_1$), but this comes at a price of overestimating tumor concentration in areas of healthy tissue. For some brains, the amount of predicted erroneous tumor concentration $\int_{B\setminus WT}c_1$ increases drastically for smaller $\lambda$, and alongside the data misfit $\ellTwoMiss$ with respect to the entire brain generally worsens. We found that overall $\lambda=1$, i.e., $\D{O} \defeq 1-m_E$, shows best outcomes, and nicely balances approximation of tumor core, and edema, while minimizing false predictions of tumor in healthy tissue.

\subsection{Calibration and Extraction of Biomarkers for the Brats'18 Cohort}\label{sec:sm:inversion}
Numerical results including estimated proliferation and migration rate for 206 individuals taken from the BraTS'18 training data set are given in~\tabref{tab:brats-sim-results-ALL}. Qualitative results are given in Figs.\;\ref{fig:brats18-sim-1}--\ref{fig:brats18-sim-6}.

We present axial, coronal, and saggital cuts of the three dimensional volume at the coordinates of the center of mass of the first component of the target data (tumor core). The tumor core target data is given in yellow, overlaid by (contour levels of) the predicted tumor cell density of our model (using estimated proliferation and migration rate) at time $T=1$. The red line corresponds to $0.9 < c(1)$ tumor cell density. The remaining contour lines correspond to 0.7, 0.5, and 0.1 (see main text for a description). The tumor initial condition is given in magenta. Projections of the tumor initial condition onto the axial, coronal, and saggital plane, are given in the right. Red markers indicate estimated tumor initial locations on the coarser level $n=128^3$, while blue makers correspond to the fine level $n=256^3$ solution.

For quantitative assessment, we consider the following performance measures:
\begin{subequations}\vspace{-6pt}\label{eq:performance-measures}
\begin{align}
    \ellTwoObsMiss &\defeq \nf{\|\D{O}c(1)-m_T\|_{L^2(\Omega)}^2}{\|m_T\|_{L^2(\Omega)}^2}, \label{eq:ell2-obs-misfit} \\
    \ellTwoMiss &\defeq \nf{\|c(1)-m_T\|_{L^2(\Omega)}^2}{\|m_T\|_{L^2(\Omega)}^2}, \label{eq:sell2-misfit} \\
    \scaledEllTwoDice &\defeq \nf{\|\overline{c(1)}-m_T\|_{L^2(m_T)}^2}{\|m_T\|_{L^2(m_T)}^2}, \label{eq:scaled-ell2-dice} \\
    \scriptstyle \int_{X} c_1 &\defeq \scriptstyle \nf{\int_{X} c(1)\dx}{\int_{\Omega}c(1)\dx}, ~~~X\;\in\;\{TC, ED, B\setminus WT\}, \label{eq:integral-measures} \\
    \scriptstyle D_{cm} &\defeq~\scriptstyle \mbox{dist}(\vect{x}_{cm}^{c_0},\vect{x}_{cm}^{TC}),~~ D_{\vect{p}} ~\defeq~\mbox{dist}(\vect{x}_{{p_i}_1},\vect{x}_{{p_i}_2}), \label{eq:dist-measures}
    \vspace{-6pt}
\end{align}
\end{subequations}
i.e.,
$\ellTwoObsMiss$ is the relative data misfit at observation points;
$\ellTwoMiss$ is the relative data misfit measured in the entire brain;
$\scaledEllTwoDice$ is a modified misfit term which quantifies the agreement (or overlap) of the predicted tumor, and the true data $m_T$;\footnote{The probability map $m_T$ is constructed from a binary segmentation, and thus has value one within the tumor core. To account for the difficulties of our reaction-diffusion growth model in matching a flat region of  saturated concentration, we re-scale the simulated $c(1)$ map between $[0,1]$, depicted by $\overline{c(1)}$.}
$\int_{X} c_1$, for sub-regions $X \in \{TC, ED, B\setminus WT\}$ are integral measures, reflecting the ability of our model to capture tumor core (TC) and edema (ED) sub-regions of the data ($\int_{TC} c_1$ and $\int_{ED} c_1$, respectively), and indicate erroneous tumor growth ($\int_{B\setminus WT} c_1$);
$D_{cm}$, is the distance  between the two centers of mass of $c(0)$, and the tumor input data $m_T$ (in voxels);\footnote{Distance is computed per component. We only report the distance for the largest component here. Distance for all components can be found in the supplementary material.}
and $D_{\vect{p}}$ is the distance between the two most important (two largest) locations $\vect{x}_{{p_i}_1}$ and $\vect{x}_{{p_i}_2}$ of tumor emergence. $D_{cm}$ and $D_{\vect{p}}$ are used to quantitatively assess variability of estimated locations of tumor emergence, and, in particular the deviation from the center of mass (CM) of the input data.

\subsubsection{Cohort GBM Atlas}\label{sec:sm:gbm-atlas}
We construct a GBM atlas of tumor initiation for the BratTS'18 survival data cohort.
\figref{fig:brats18-gbm-atlas-c0-4x9} shows axial slices $42$ through $112$ of the atlas with superimposed tumor initiations of 165 patients. Patients are divided into three survival classes: short (less than 10 month survival time) given in red, long (more than 15 months survival time) given in blue, and mid survivors (between 15 and 10 month survival time) given in green.

\figref{fig:brats18-gbm-atlas-c1-4x9} the spatial distribution of the average tumor cell density within the brain across the 165 patients, subdivided into short, mid and long survivors. The average tumor cell densities are blend together in a single RGB image with short survivors representing the red-channel, mid survivors the green-channel, and long survivors the blue-channel.
Figs.\;\ref{fig:brats18-gbm-atlas-c1-sml-1}--\ref{fig:brats18-gbm-atlas-c1-sml-2} also show the individual distributions for short, mid, and long survivors.

We also investigate patient overall survival in relation with tumor initiation in certain functional regions of the brain. The estimated site(s) of tumor initiation for each patient are mapped to a functional atlas, and accumulated per functional region and survival class.
For each functional region $\D{L}$ and each survival class $X$ in $S$ (short), $M$ (mid), and $L$ (long), we compute frequencies:
\begin{subequations}\label{eq:frequency-measures}
\begin{align}
q^{\;c_0} &\defeq  \nf{\int_{\D{L}} a^{TIL}_X(\vect{x}) \d{\vect{x}}}{\int_{\Omega}a^{TIL}(\vect{x}) \d{\vect{x}}}, \\ 
\hitFreqShortCMofIC &\defeq  \nf{\sum_{P\in X} [\vect{x}_{cm}^{c_0} \in \D{L}] }{\sum_{P \in S \cup M \cup L} 1}, \\
\hitFreqShortCMofTC &\defeq  \nf{\sum_{P\in X} [\vect{x}_{cm}^{TC} \in \D{L}] }{\sum_{P \in S \cup M \cup L} 1}, \\
q^{\;c_1} &\defeq  \nf{\int_{\D{L}} a^{OT}_X(\vect{x}) \d{\vect{x}}}{\int_{\Omega}a^{OT}(\vect{x}) \d{\vect{x}}}, 
\end{align}
\end{subequations}
for the initial localized tumor $c(0)$ (based on the TIL BraTS'18 cohort atlas $a^{TIL}$), for the center of mass $\vect{x}_{cm}^{c_0}$ of the initial tumor $c(0)$, for the center of mass $\vect{x}_{cm}^{TC}$ of the tumor core input data, and for the grown tumor $c(1)$ (instead of its origin $c(0)$).

For each of these frequencies, we then derive indications for a relation bewteen certain functional regions and patient overall survival.
Let $\vect{f}_{\D{L}} = (f_S, f_M, f_L)_{\D{L}}$ be the frequency of short, mid and long survivors for the functional region $\D{L}$, for either of the definititions in~\eqref{eq:frequency-measures}. Then, for each functional region $\D{L}$, we give
\begin{itemize}
\item a strong indication for \iemph{short}, \iemph{mid}, or \iemph{long} survivor, if $f_X > 0.8 (f_S + f_M = f_L)$, for $X\in\{S,M,L\}$
\item a weak indication for \iemph{short}, \iemph{mid}, or \iemph{long} survivor, if $f_X > f_M + f_L $,  for $X\in\{S,M,L\}$
\item a strong indication for \iemph{not} being a \iemph{short}, \iemph{mid}, or \iemph{long}, if $f_X < 0.2 (f_S + f_M = f_L)$,  for $X\in\{S,M,L\}$
\end{itemize}
and \iemph{uninformative} otherwise.

The full table for all functional regions is given in~\tabref{tab:func-atlas-freq}.  \figref{fig:gbm-atlas-3d} highlights the top 6 of those regions.

\subsection{Feature Extraction}\label{sec:sm:features}
For all 206 individuals, we extract a number of imaging-based and physics based features, which are made publicly available for further analysis and research. \tabref{tab:feature-descr} provides a description for all extracted features, available from the \verb!.xls! or \verb!.csv! file.

\begin{figure*}
\includegraphics[width=\textwidth]{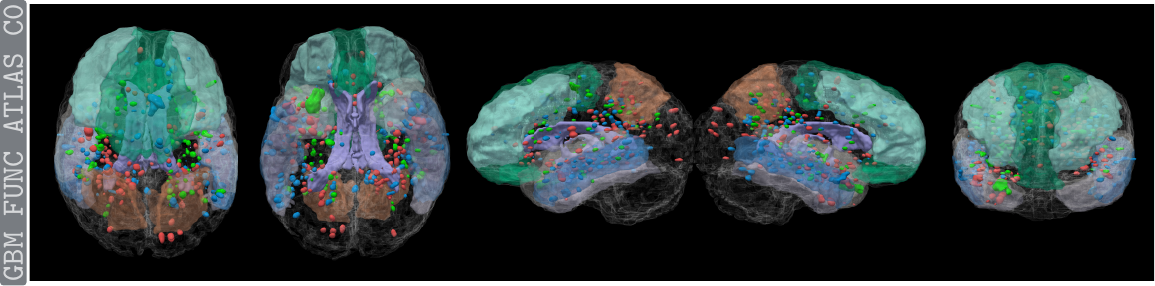}
\caption{\ipoint{Cohort GBM atlas for BratTS'18 survival data}. Tumor initiation of $161$ patients superimposed and displayed in the atlas space (short survivors in red; mid survivors in green; long survivors in blue). Frequently affected functional regions are highlighted (cf.\;\tabref{tab:func-atlas-freq}). Most tumor origins fall into an unlabeled part of the white matter, which is not highlighted in the figure.
\label{fig:gbm-atlas-3d}}
\end{figure*}

\begin{figure*}
\includegraphics[width=\textwidth]{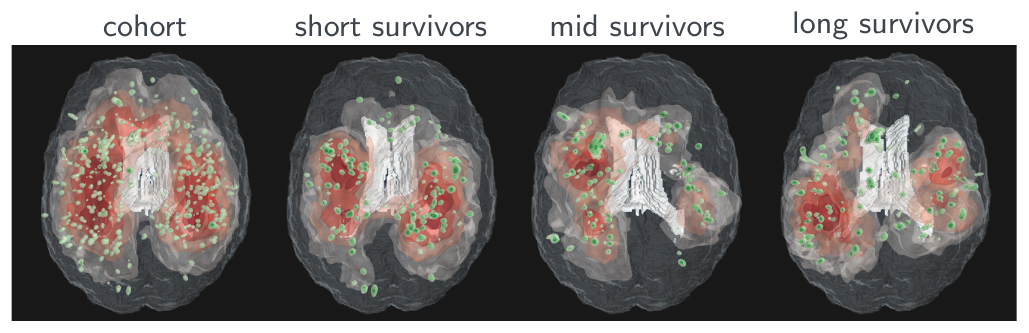}
\caption{\ipoint{Cohort GBM atlas for BratTS'18 survival data}. Distribution of observed tumor and tumor initiation of $161$ patients subdivided into short, mid and long survivors. The red contour levels indicate the $c(1)$ distribution normalized across the entire cohort; the green contour lines indicate the $c(0)$ distribution, respectively;
\label{fig:gbm-atlas-3d_v2}}
\end{figure*}

\begin{table*}[htb]
    \caption{\ipoint{Survival and tumor initiation}. We analyze correlation of patient survival and tumor initiation in functional regions of the brain. This study is based on an expert segmentation of functional regions of the normal SRI atlas brain~\cite{Rohlfing:2010sri}. Based on the probabilistic GBM atlas $a$ (cf.\;Fig. 1 main text), we compute tumor initiation frequencies $q^{\;c_0} \defeq \nf{\int_{\D{L}} a_X(\vect{x}) \d{\vect{x}}}{\int_{\Omega}a(\vect{x}) \d{\vect{x}}}$ per functional region $\D{L}$. $a_X$ for $X\in\{S,M,L\}$ denotes the probabilistic atlas of short, mid, and long survivors, respectively.
    For each region, we also report the rel.\,size, and overall rate of TIL occurrence $f(c_0)$, and OT occurrence $f(c_1)$.
    For comparison, we compute similar rates for $(i)$ the center of mass $\vect{x}_{cm}^{c_0}$ of TIL, $(ii)$ the center of mass $\vect{x}_{cm}^{TC}$ of the OT.
    When possible, an indication for region vs.\;survival class correlation is given based on $\ellTwoFreqShortIC$ (and $\hitFreqShortCMofIC$, $\hitFreqShortCMofTC$, and $\ellTwoFreqShortC$, respectively): superscripts $^{(+)}/^{(-)}$ depicts a strong, or weak indication, while ``$!$'' depicts disqualification of the given class for the considered label.
    }
    \label{tab:func-atlas-freq}
    \centering\scriptsize\setlength\tabcolsep{2.5pt}
\begin{tabular}{llLlllClllClllClllC}
\toprule
   \multicolumn{3}{c}{} & \textit{Short} & \textit{Mid} & \textit{Long} & \textit{C0}
                        & \textit{Short} & \textit{Mid} & \textit{Long} & \textit{CM(C0)}
                        & \textit{Short} & \textit{Mid} & \textit{Long} & \textit{CM(TC)}
                        & \textit{Short} & \textit{Mid} & \textit{Long} & \textit{C1} \\
   \textit{Name}   &  \textit{Size} &   $f(c_0)$
   &  $\ellTwoFreqShortIC$ & $\ellTwoFreqMidIC$ & $\ellTwoFreqLongIC$    & \textit{\detokenize{IND}}
   &  $\hitFreqShortCMofIC$ & $\hitFreqMidCMofIC$ & $\hitFreqLongCMofIC$ & \textit{\detokenize{IND}}
   &  $\hitFreqShortCMofTC$ & $\hitFreqMidCMofTC$ & $\hitFreqLongCMofTC$ & \textit{\detokenize{IND}}
   &  $\ellTwoFreqShortC$ & $\ellTwoFreqMidC$ & $\ellTwoFreqLongC$       & \textit{\detokenize{IND}}
   \\
\midrule
\textit{unlabelled } &        0.21 &       0.34 &        0.16 & 0.06 & 0.11 & $!M^{+}$              & 0.16 & 0.09 & 0.15 & ---                   & 0.18 & 0.09 & 0.15 & ---                   & 0.16 & 0.04 & 0.03 & $S^{-}$               \\
\textit{superior temporal gyrus  }   &        0.04 &       0.12 &      0.04 & 0.06 & 0.01 & $!M^{+}$              & 0.04 & 0.03 & 0.03 & ---                   & 0.04 & 0.03 & 0.02 & ---                   & 0.02 & 0.00 & 0.01 & $S^{-}$               \\
\textit{superior frontal gyrus   }   &        0.09 &       0.09 &      0.01 & 0.01 & 0.07 & $L^{-}$               & 0.02 & 0.01 & 0.03 & $!M^{+}$              & 0.03 & 0.01 & 0.02 & $!M^{+}$              & 0.00 & 0.00 & 0.00 & ---                   \\
\textit{middle temporal gyrus    }   &        0.03 &       0.07 &      0.03 & 0.00 & 0.04 & $L^{-}$               & 0.03 & 0.01 & 0.06 & $L^{-}$               & 0.03 & 0.00 & 0.08 & $L^{-}$               & 0.02 & 0.00 & 0.01 & $S^{-}$               \\
\textit{inferior temporal gyrus  }   &        0.03 &       0.05 &      0.02 & 0.00 & 0.03 & $L^{-}$               & 0.01 & 0.01 & 0.02 & $L^{-}$               & 0.01 & 0.01 & 0.01 & ---                   & 0.01 & 0.00 & 0.00 & $S^{-}$               \\
\textit{middle frontal gyrus     }   &        0.08 &       0.04 &      0.01 & 0.02 & 0.01 & ---                   & 0.01 & 0.02 & 0.01 & ---                   & 0.01 & 0.02 & 0.01 & ---                   & 0.00 & 0.00 & 0.00 & ---                   \\
\textit{superior parietal gyrus  }   &        0.04 &       0.04 &      0.01 & 0.01 & 0.01 & ---                   & 0.02 & 0.01 & 0.01 & $!M^{+}$              & 0.02 & 0.02 & 0.01 & ---                   & 0.01 & 0.00 & 0.00 & $S^{-}$               \\
\textit{ventricles                 } &        0.01 &       0.03 &        0.02 & 0.01 & 0.00 & $S^{-}$               & 0.02 & 0.00 & 0.00 & $S^{+}$               & 0.01 & 0.00 & 0.01 & $!M^{+}$              & 0.01 & 0.00 & 0.00 & $S^{-}$               \\
\textit{hippocampus                } &        0.01 &       0.02 &        0.02 & 0.00 & 0.00 & $S^{-}$               & 0.02 & 0.01 & 0.01 & $S^{-}$               & 0.02 & 0.01 & 0.00 & $S^{-}$               & 0.01 & 0.00 & 0.00 & $S^{-}$               \\
\textit{precuneus                  } &        0.01 &       0.02 &        0.01 & 0.00 & 0.01 & $S^{-}$               & 0.00 & 0.00 & 0.01 & $L^{+}$               & 0.01 & 0.00 & 0.01 & $!M^{+}$              & 0.00 & 0.00 & 0.00 & $S^{-}$               \\
\textit{precentral gyrus          }  &        0.04 &       0.02 &       0.00 & 0.01 & 0.00 & $M^{-}$               & 0.00 & 0.01 & 0.01 & $!S^{+}$              & 0.00 & 0.01 & 0.01 & $L^{-}$               & 0.00 & 0.00 & 0.00 & ---                   \\
\textit{insular cortex            }  &        0.01 &       0.02 &       0.01 & 0.00 & 0.00 & $S^{+}$               & 0.01 & 0.01 & 0.00 & $S^{-}$               & 0.01 & 0.01 & 0.00 & $M^{-}$               & 0.01 & 0.00 & 0.00 & $S^{-}$               \\
\textit{parahippocampal gyrus     }  &        0.01 &       0.02 &       0.01 & 0.01 & 0.00 & $!M^{+}$              & 0.01 & 0.00 & 0.01 & $S^{-}$               & 0.00 & 0.00 & 0.01 & $L^{+}$               & 0.01 & 0.00 & 0.00 & $S^{-}$               \\
\textit{superior occipital gyrus }   &        0.02 &       0.02 &      0.01 & 0.00 & 0.00 & $S^{+}$               & 0.01 & 0.01 & 0.00 & $!M^{+}$              & 0.01 & 0.01 & 0.01 & ---                   & 0.00 & 0.00 & 0.00 & $S^{-}$               \\
\textit{angular gyrus             }  &        0.02 &       0.01 &       0.00 & 0.01 & 0.01 & $!S^{+}$              & 0.00 & 0.01 & 0.00 & $M^{+}$               & 0.00 & 0.01 & 0.00 & $M^{+}$               & 0.00 & 0.00 & 0.00 & ---                   \\
\textit{cingulate gyrus           }  &        0.02 &       0.01 &       0.01 & 0.00 & 0.00 & $!M^{+}$              & 0.01 & 0.01 & 0.01 & ---                   & 0.01 & 0.01 & 0.01 & ---                   & 0.00 & 0.00 & 0.00 & $S^{-}$               \\
\textit{postcentral gyrus         }  &        0.03 &       0.01 &       0.01 & 0.00 & 0.00 & $S^{-}$               & 0.01 & 0.01 & 0.00 & $!M^{+}$              & 0.00 & 0.01 & 0.00 & $M^{+}$               & 0.00 & 0.00 & 0.00 & ---                   \\
\textit{fusiform gyrus            }  &        0.02 &       0.01 &       0.00 & 0.00 & 0.01 & $L^{-}$               & 0.01 & 0.00 & 0.00 & $S^{+}$               & 0.01 & 0.00 & 0.00 & $S^{+}$               & 0.00 & 0.00 & 0.00 & $!M^{+}$              \\
\textit{inferior frontal gyrus   }   &        0.03 &       0.01 &      0.00 & 0.00 & 0.00 & $L^{-}$               & 0.00 & 0.01 & 0.00 & $M^{+}$               & 0.00 & 0.01 & 0.00 & $M^{+}$               & 0.00 & 0.00 & 0.00 & $M^{-}$               \\
\textit{middle occipital gyrus   }   &        0.03 &       0.01 &      0.00 & 0.00 & 0.00 & $!M^{+}$              & 0.00 & 0.00 & 0.00 & ---                   & 0.01 & 0.00 & 0.00 & $S^{+}$               & 0.00 & 0.00 & 0.00 & $S^{+}$               \\
\textit{lingual gyrus             }  &        0.02 &       0.01 &       0.00 & 0.00 & 0.01 & $L^{-}$               & 0.00 & 0.00 & 0.01 & $L^{+}$               & 0.00 & 0.00 & 0.01 & $L^{+}$               & 0.00 & 0.00 & 0.00 & $S^{-}$               \\
\textit{caudate                    } &        0.01 &       0.01 &        0.00 & 0.01 & 0.00 & $M^{+}$               & 0.00 & 0.01 & 0.01 & $!S^{+}$              & 0.00 & 0.01 & 0.00 & $M^{+}$               & 0.00 & 0.00 & 0.00 & $S^{-}$               \\
\textit{supramarginal gyrus       }  &        0.02 &       0.01 &       0.00 & 0.00 & 0.00 & $L^{-}$               & 0.00 & 0.01 & 0.01 & $M^{-}$               & 0.00 & 0.01 & 0.01 & $!S^{+}$              & 0.00 & 0.00 & 0.00 & $L^{-}$               \\
\textit{putamen                    } &        0.01 &       0.01 &        0.00 & 0.01 & 0.00 & $M^{+}$               & 0.00 & 0.02 & 0.00 & $M^{+}$               & 0.00 & 0.01 & 0.00 & $M^{+}$               & 0.01 & 0.00 & 0.00 & $S^{-}$               \\
\textit{background                 } &        4.54 &       0.01 &        0.00 & 0.00 & 0.01 & $L^{+}$               & 0.00 & 0.00 & 0.00 & ---                   & 0.00 & 0.00 & 0.00 & ---                   & 0.00 & 0.00 & 0.00 & $!M^{+}$              \\
\textit{inferior occipital gyrus }   &        0.01 &       0.00 &      0.00 & 0.00 & 0.00 & $S^{+}$               & 0.01 & 0.00 & 0.00 & $S^{+}$               & 0.00 & 0.00 & 0.00 & ---                   & 0.00 & 0.00 & 0.00 & $L^{-}$               \\
\textit{cerebellum                 } &        0.12 &       0.00 &        0.00 & 0.00 & 0.00 & $L^{+}$               & 0.00 & 0.00 & 0.00 & ---                   & 0.00 & 0.00 & 0.00 & ---                   & 0.00 & 0.00 & 0.00 & $L^{+}$               \\
\textit{brainstem                  } &        0.01 &       0.00 &        0.00 & 0.00 & 0.00 & $M^{-}$               & 0.00 & 0.00 & 0.00 & ---                   & 0.00 & 0.00 & 0.00 & ---                   & 0.00 & 0.00 & 0.00 & $S^{-}$               \\
\textit{lateral orbitofrontal gyrus} &        0.01 &       0.00 &      0.00 & 0.00 & 0.00 & $S^{+}$               & 0.00 & 0.00 & 0.00 & ---                   & 0.00 & 0.00 & 0.00 & ---                   & 0.00 & 0.00 & 0.00 & $S^{-}$               \\
\textit{middle orbitofrontal gyrus}  &        0.01 &       0.00 &      0.00 & 0.00 & 0.00 & $L^{-}$               & 0.00 & 0.00 & 0.00 & ---                   & 0.00 & 0.00 & 0.00 & ---                   & 0.00 & 0.00 & 0.00 & $S^{-}$               \\
\textit{gyrus rectus              }  &        0.00 &       0.00 &       0.00 & 0.00 & 0.00 & $S^{+}$               & 0.00 & 0.00 & 0.00 & ---                   & 0.00 & 0.00 & 0.00 & ---                   & 0.00 & 0.00 & 0.00 & $S^{+}$               \\
\textit{cuneus                     } &        0.01 &       0.00 &        0.00 & 0.00 & 0.00 & $S^{+}$               & 0.00 & 0.00 & 0.00 & ---                   & 0.00 & 0.00 & 0.00 & ---                   & 0.00 & 0.00 & 0.00 & $S^{+}$               \\

\bottomrule
\end{tabular}
\end{table*}

\begin{table*}[htb]
    \caption{\ipoint{Survival and tumor initiation}. We analyze correlation of patient survival and tumor initiation in functional regions of the brain. This study is based on an expert segmentation of functional and structural regions of a normal brain (template obtained from the University of Pennsylvania). Based on the probabilistic GBM atlas $a$ (cf.\;Fig. 1 main text), we compute tumor initiation frequencies $q^{\;c_0} \defeq \nf{\int_{\D{L}} a_X(\vect{x}) \d{\vect{x}}}{\int_{\Omega}a(\vect{x}) \d{\vect{x}}}$ per functional region $\D{L}$. $a_X$ for $X\in\{S,M,L\}$ denotes the probabilistic atlas of short, mid, and long survivors, respectively.
    For each region, we also report the rel.\,size, and overall rate of TIL occurrence $f(c_0)$, and OT occurrence $f(c_1)$.
    For comparison, we compute similar rates for $(i)$ the center of mass $\vect{x}_{cm}^{c_0}$ of TIL, $(ii)$ the center of mass $\vect{x}_{cm}^{TC}$ of the OT.
    When possible, an indication for region vs.\;survival class correlation is given based on $\ellTwoFreqShortIC$ (and $\hitFreqShortCMofIC$, $\hitFreqShortCMofTC$, and $\ellTwoFreqShortC$, respectively): superscripts $^{(+)}/^{(-)}$ depicts a strong, or weak indication, while ``$!$'' depicts disqualification of the given class for the considered label.
    }
    \label{tab:func-atlas-freq-muse}
    \centering\tiny\setlength\tabcolsep{2.5pt}
\begin{tabular}{llLllllClllClllClllC}
\toprule
\multicolumn{4}{c}{} & \textit{Short} & \textit{Mid} & \textit{Long} & \textit{C0}
                     & \textit{Short} & \textit{Mid} & \textit{Long} & \textit{CM(C0)}
                     & \textit{Short} & \textit{Mid} & \textit{Long} & \textit{CM(TC)}
                     & \textit{Short} & \textit{Mid} & \textit{Long} & \textit{C1} \\
\textit{Name}   &  \textit{Size} &   $f(c_0)$ & $f(c_1)$
&  $\ellTwoFreqShortIC$ & $\ellTwoFreqMidIC$ & $\ellTwoFreqLongIC$    & \textit{\detokenize{IND}}
&  $\hitFreqShortCMofIC$ & $\hitFreqMidCMofIC$ & $\hitFreqLongCMofIC$ & \textit{\detokenize{IND}}
&  $\hitFreqShortCMofTC$ & $\hitFreqMidCMofTC$ & $\hitFreqLongCMofTC$ & \textit{\detokenize{IND}}
&  $\ellTwoFreqShortC$ & $\ellTwoFreqMidC$ & $\ellTwoFreqLongC$       & \textit{\detokenize{IND}}
\\
\midrule
\textit{temporal lobe WM (TL;L)                     }                      &        0.04 &       0.10 &       0.17 & 0.05 & 0.02 & 0.03 & ---                   & 0.05 & 0.02 & 0.04 & $!M^{+}$              & 0.07 & 0.03 & 0.04 & $!M^{+}$              & 0.04 & 0.01 & 0.01 & $S^{-}$               \\
\textit{temporal lobe WM (TL;R)                    }                      &        0.04 &       0.09 &       0.11 & 0.05 & 0.01 & 0.03 & $S^{-}$               & 0.06 & 0.01 & 0.02 & $S^{-}$               & 0.06 & 0.01 & 0.03 & $S^{-}$               & 0.04 & 0.00 & 0.01 & $S^{+}$               \\
\textit{frontal lobe WM (FL;L)                      }                      &        0.07 &       0.06 &       0.07 & 0.02 & 0.01 & 0.04 & $L^{-}$               & 0.03 & 0.02 & 0.03 & ---                   & 0.03 & 0.01 & 0.03 & $!M^{+}$              & 0.01 & 0.01 & 0.00 & $!M^{+}$              \\
\textit{parietal lobe WM (PL;R)                    }                      &        0.03 &       0.05 &       0.08 & 0.03 & 0.01 & 0.01 & $S^{-}$               & 0.04 & 0.02 & 0.01 & $S^{-}$               & 0.03 & 0.01 & 0.03 & $!M^{+}$              & 0.03 & 0.00 & 0.01 & $S^{-}$               \\
\textit{frontal lobe WM (FL;R)                     }                      &        0.07 &       0.05 &       0.04 & 0.02 & 0.02 & 0.01 & ---                   & 0.01 & 0.03 & 0.03 & $!S^{+}$              & 0.01 & 0.02 & 0.03 & ---                   & 0.01 & 0.00 & 0.00 & $S^{-}$               \\
\textit{temporal pole (TMP;L)                }                       &        0.01 &       0.05 &       0.01 & 0.01 & 0.03 & 0.00 & $M^{-}$               & 0.02 & 0.01 & 0.01 & $!M^{+}$              & 0.02 & 0.01 & 0.01 & ---                   & 0.01 & 0.00 & 0.00 & $S^{+}$               \\
\textit{parietal lobe WM (PL;L)                     }                      &        0.04 &       0.04 &       0.05 & 0.01 & 0.02 & 0.02 & $!S^{+}$              & 0.01 & 0.04 & 0.01 & $M^{-}$               & 0.01 & 0.02 & 0.00 & $M^{-}$               & 0.01 & 0.01 & 0.00 & $!M^{+}$              \\
\textit{supplementary motor cortex (SMC;L)    }                       &        0.01 &       0.04 &       0.00 & 0.00 & 0.00 & 0.03 & $L^{+}$               & 0.00 & 0.00 & 0.01 & $L^{+}$               & 0.00 & 0.00 & 0.01 & $L^{+}$               & 0.00 & 0.00 & 0.00 & ---                   \\
\textit{middle temporal gyrus (MTG;L)         }                       &        0.01 &       0.03 &       0.02 & 0.00 & 0.00 & 0.03 & $L^{+}$               & 0.01 & 0.00 & 0.02 & $L^{-}$               & 0.00 & 0.00 & 0.03 & $L^{+}$               & 0.00 & 0.00 & 0.00 & $!M^{+}$              \\
\textit{inferior temporal gyrus (ITG;L)       }                       &        0.01 &       0.02 &       0.01 & 0.02 & 0.00 & 0.01 & $S^{-}$               & 0.01 & 0.00 & 0.01 & $!M^{+}$              & 0.01 & 0.00 & 0.00 & $S^{+}$               & 0.00 & 0.00 & 0.00 & $L^{-}$               \\
\textit{lateral ventricle (LVE;L)                    }                      &        0.00 &       0.02 &       0.01 & 0.01 & 0.01 & 0.00 & ---                   & 0.01 & 0.00 & 0.01 & $!M^{+}$              & 0.00 & 0.00 & 0.01 & $L^{+}$               & 0.00 & 0.00 & 0.00 & $S^{-}$               \\
\textit{middle temporal gyrus (MTG;R)       }                       &        0.01 &       0.02 &       0.01 & 0.00 & 0.00 & 0.01 & $L^{-}$               & 0.00 & 0.01 & 0.03 & $L^{-}$               & 0.00 & 0.01 & 0.01 & $L^{-}$               & 0.00 & 0.00 & 0.00 & $S^{-}$               \\
\textit{precentral gyrus (PrG;R)            }                       &        0.01 &       0.02 &       0.00 & 0.00 & 0.01 & 0.00 & $M^{+}$               & 0.00 & 0.01 & 0.00 & $M^{+}$               & 0.00 & 0.01 & 0.00 & $M^{+}$               & 0.00 & 0.00 & 0.00 & $S^{-}$               \\
      \bottomrule
      \end{tabular}
 \end{table*}

\begin{landscape}
\begin{figure}
\centering
\includegraphics[scale=0.6, trim=60 0 0 0, clip]{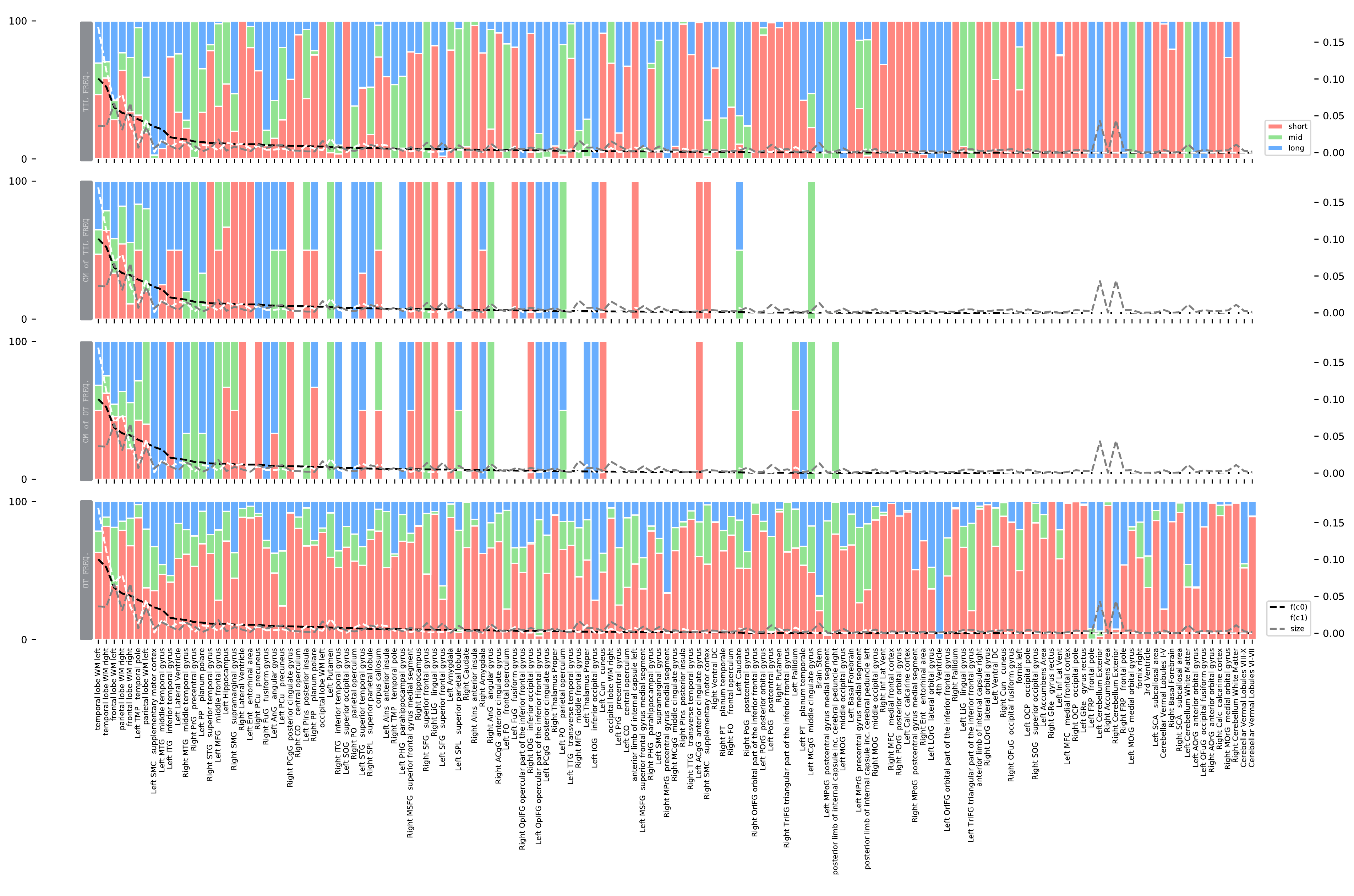}
\caption{\ipoint{Correlation of patient survival with TIL in certain functional regions}.
 Based on the probabilistic GBM atlas $a$ (Fig. 1 main text), we compute TIL and OT frequencies $\nf{\int_{\D{L}} a_X(\vect{x}) \d{\vect{x}}}{\int_{\Omega}a(\vect{x}) \d{\vect{x}}}$ per functional region $\D{L}$; $a_X$ for $X\in\{S,M,L\}$ denotes the probabilistic atlas of short, mid, and long survivors.
 For comparison, we compute similar rates for $(i)$ the center of mass $\vect{x}_{cm}^{c_0}$ of TIL, and $(ii)$ the center of mass $\vect{x}_{cm}^{TC}$ of the OT.
 Shown is the ratio of (TIL/OT/CM) frequencies for short (red), mid (green), and long (blue) survivors per label (imbalances indicate higher likelihood for survival class in this region), along with the rel.\,size and overall percentage (summed across short, mid, long) of TIL ($f(c_0)$) and OT ($f(c_1)$).
\label{fig:func-atlas-analysis-all-labels}
}
\end{figure}
\end{landscape}

\FloatBarrier

\begin{figure*}
\centering
\begin{minipage}{0.49\textwidth}
\includegraphics[width=1.\textwidth]{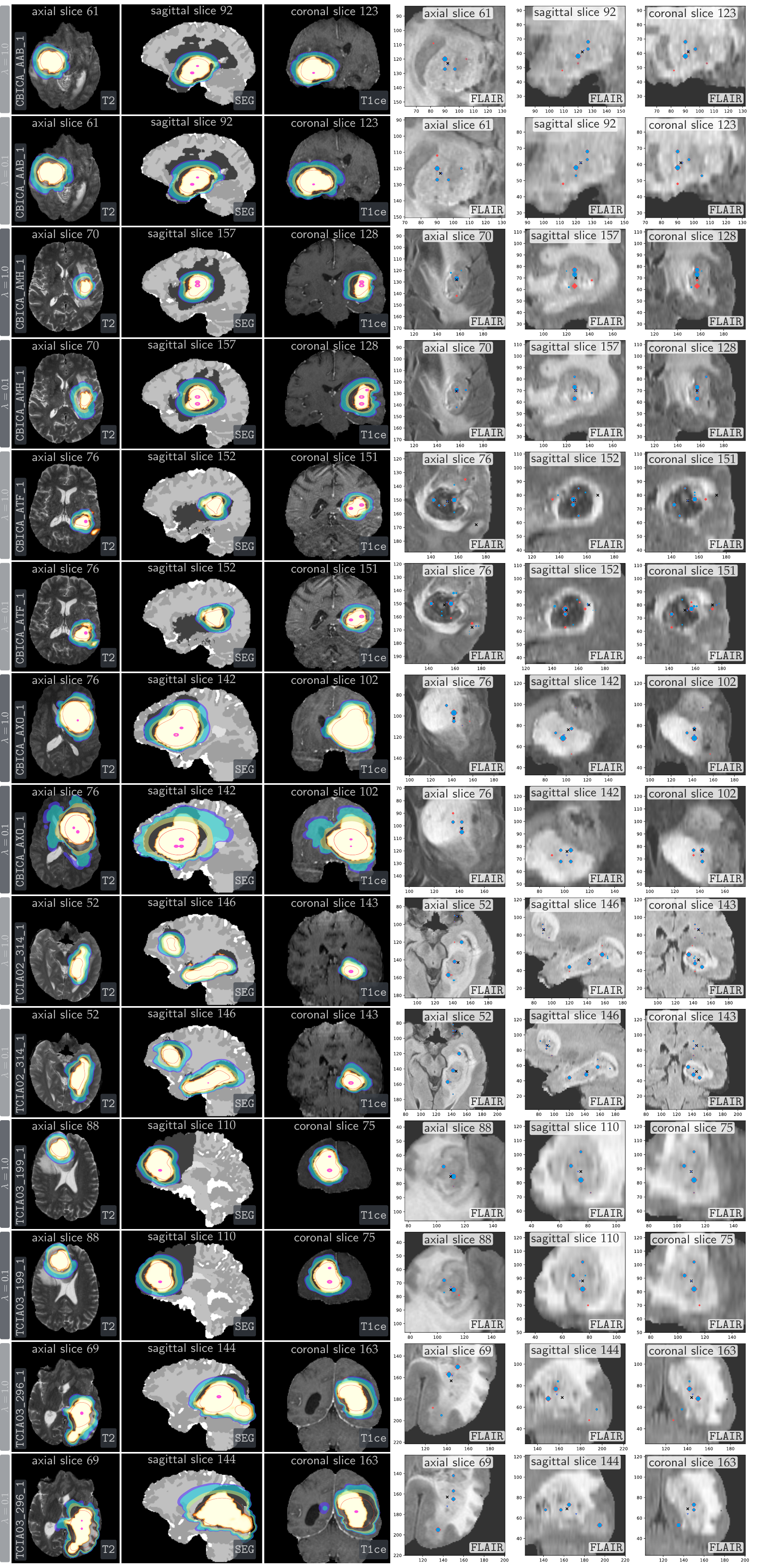}
\end{minipage}
\begin{minipage}{0.49\textwidth}
\includegraphics[width=1.\textwidth, valign=bottom]{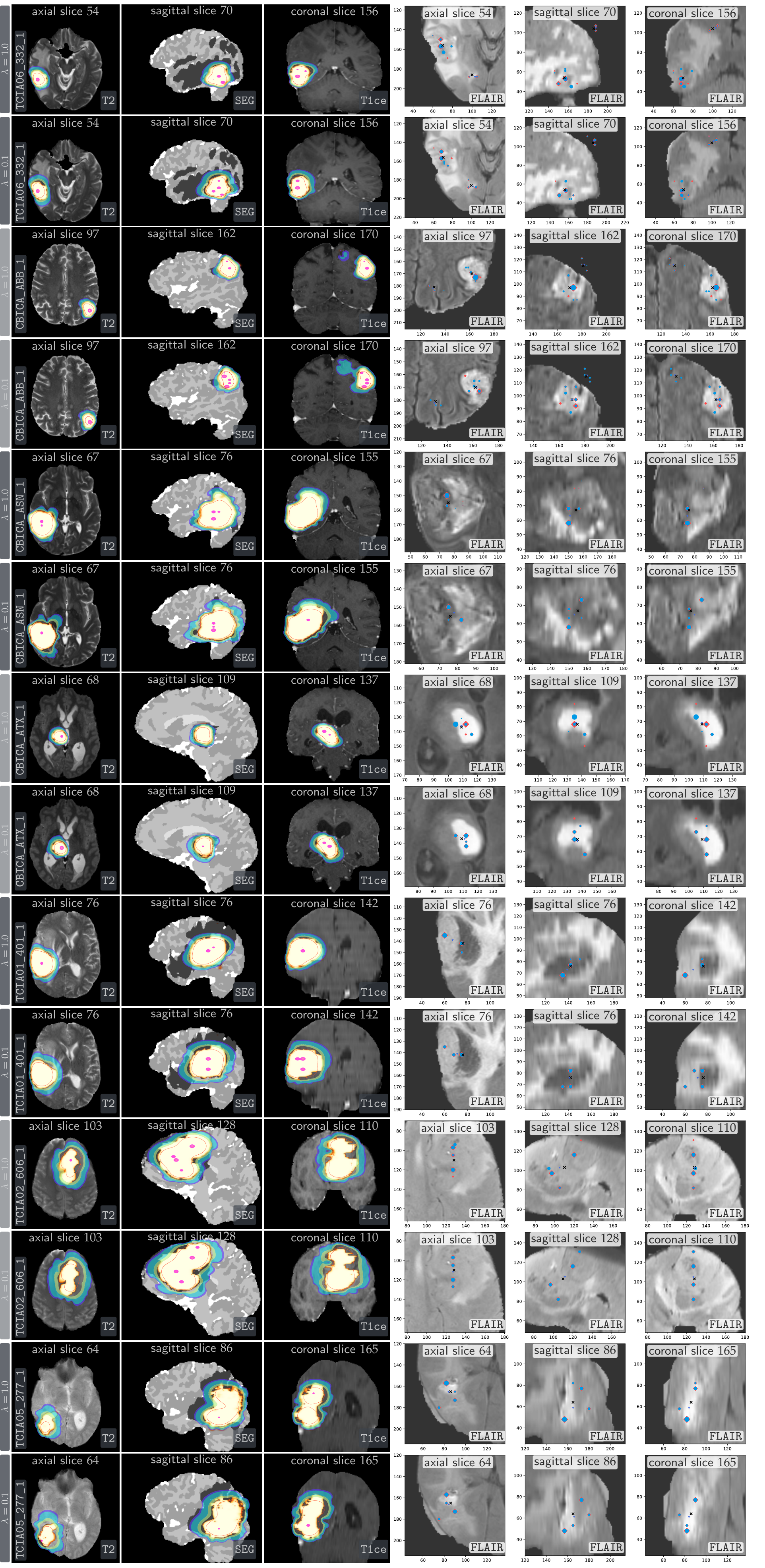}
\end{minipage}
\caption{\ipoint{Sensitivity to observation operator.}
Qualitative results for different choices of the observation operator for $\lambda \in \{1, 0.1\}$. Shown are axial, and saggital cuts of the 3D patient T2-MRI volume, overlaid with the $TC$ input data (yellow), the predicted tumor $c(1)$ from our calibrated model (contours from red to violet), and the tumor origin $c(0)$ (magenta). The red $c(1)$ contour line corresponds to $0.9 \leq c(1)$, and matches well with the input $TC$ data.
}
\label{fig:obs-operator}
\end{figure*}

\begin{table*}
    \centering\scriptsize\setlength\tabcolsep{4pt}
    \caption{
    \ipoint{Sensitivity to observation operator.}
    Calibrated parameters $\rho_w$ and $\kappa_w$, and corresponding data mismatches $\ellTwoObsMiss$ (at observation points), and $\ellTwoMiss$ (everywhere) for variations $\lambda \in \{1,0.1\}$ of the observation operator. We also report the integral measures $\int_{ED}c_1 \defeq \nf{\int_{ED}c(1)\dx}{\int_{\Omega}c(1)}$, and $\int_{B\setminus WT}c_1 \defeq \nf{\int_{B\setminus WT}c(1)\dx}{\int_{\Omega}c(1)}$ indicating the approximation quality of edema, and erroneous tumor growth in healthy tissue, respectively.
    }
    \label{tab:obs-operator}
\begin{tabular}{rLrrRRrr}
    \toprule
    BraTS ID                     & $\lambda$       & $\rho_w$        & $\kappa_w$      & $\ellTwoObsMiss$ & $\ellTwoMiss$    & $\int_{ED}c(1)$ & $\int_{B\setminus WT}c(1)$ \\
    \midrule
\texttt{CBICA\_AAB         } & $      1.0$ &           11.20 & \num{2.120e-02} & \num{2.213e-01} & \num{5.614e-01} & \num{3.790e-01} & \num{5.684e-02}    \\
& $      0.4$ &           11.20 & \num{2.200e-02} & \num{1.430e-01} & \num{5.999e-01} & \num{3.999e-01} & \num{5.973e-02}    \\
 & $      0.1$ &           10.60 & \num{3.730e-02} & \num{1.338e-01} & \num{8.138e-01} & \num{4.684e-01} & \num{1.128e-01}    \\
 & $      0.0$ &           10.20 & \num{4.520e-02} & \num{1.326e-01} & \num{9.052e-01} & \num{4.934e-01} & \num{1.320e-01}    \\
\cmidrule[0.05mm]{2-8}
\texttt{CBICA\_AMH         } & $      1.0$ &            9.92 & \num{1.260e-02} & \num{2.723e-01} & \num{6.796e-01} & \num{4.570e-01} & \num{6.356e-02}    \\
& $      0.4$ &            9.18 & \num{1.650e-02} & \num{2.015e-01} & \num{6.982e-01} & \num{4.539e-01} & \num{1.025e-01}    \\
 & $      0.1$ &            8.76 & \num{2.660e-02} & \num{1.678e-01} & \num{9.542e-01} & \num{5.397e-01} & \num{1.430e-01}    \\
 & $      0.0$ &            8.70 & \num{2.540e-02} & \num{1.612e-01} & \num{8.455e-01} & \num{5.068e-01} & \num{1.417e-01}    \\
\cmidrule[0.05mm]{2-8}
\texttt{CBICA\_ATF         } & $      1.0$ &            8.41 & \num{8.510e-03} & \num{3.285e-01} & \num{5.136e-01} & \num{3.280e-01} & \num{1.071e-01}    \\
 & $      0.4$ &            7.90 & \num{1.230e-02} & \num{2.543e-01} & \num{6.834e-01} & \num{3.630e-01} & \num{1.897e-01}    \\
 & $      0.1$ &            9.09 & \num{1.170e-02} & \num{1.631e-01} & \num{6.900e-01} & \num{3.430e-01} & \num{2.100e-01}    \\
 & $      0.0$ &            8.23 & \num{2.880e-02} & \num{1.810e-01} & \num{1.06e+00} & \num{3.392e-01} & \num{3.730e-01}    \\
\cmidrule[0.05mm]{2-8}
\texttt{CBICA\_AXO         } & $      1.0$ &           13.10 & \num{3.010e-02} & \num{2.447e-01} & \num{3.944e-01} & \num{1.831e-01} & \num{1.131e-01}    \\
& $      0.4$ &           13.50 & \num{2.620e-02} & \num{1.611e-01} & \num{3.663e-01} & \num{1.725e-01} & \num{9.821e-02}    \\
 & $      0.1$ &            9.97 & \num{2.590e-01} & \num{2.192e-01} & \num{7.242e-01} & \num{1.811e-01} & \num{4.180e-01}    \\
 & $      0.0$ &            9.84 & \num{2.310e-01} & \num{2.054e-01} & \num{7.182e-01} & \num{1.837e-01} & \num{4.033e-01}    \\
\cmidrule[0.05mm]{2-8}
\texttt{CBICA\_ABB         } & $      1.0$ &            9.11 & \num{7.720e-03} & \num{2.063e-01} & \num{3.038e-01} & \num{1.417e-01} & \num{1.690e-01}    \\
& $      0.4$ &            8.65 & \num{7.180e-03} & \num{1.559e-01} & \num{3.681e-01} & \num{1.505e-01} & \num{1.899e-01}    \\
 & $      0.1$ &            6.55 & \num{1.550e-02} & \num{2.018e-01} & \num{4.056e-01} & \num{1.390e-01} & \num{2.864e-01}    \\
 & $      0.0$ &            8.61 & \num{1.220e-02} & \num{1.216e-01} & \num{4.699e-01} & \num{1.632e-01} & \num{2.682e-01}    \\
\cmidrule[0.05mm]{2-8}
\texttt{CBICA\_ASN         } & $      1.0$ &           12.60 & \num{2.240e-02} & \num{2.501e-01} & \num{2.795e-01} & \num{4.358e-02} & \num{1.805e-01}    \\
& $      0.4$ &           13.10 & \num{2.820e-02} & \num{2.052e-01} & \num{3.971e-01} & \num{4.843e-02} & \num{2.332e-01}    \\
 & $      0.1$ &           12.10 & \num{4.730e-02} & \num{1.282e-01} & \num{5.685e-01} & \num{6.899e-02} & \num{3.402e-01}    \\
 & $      0.0$ &           12.30 & \num{7.080e-02} & \num{1.203e-01} & \num{6.758e-01} & \num{7.308e-02} & \num{4.052e-01}    \\
\cmidrule[0.05mm]{2-8}
\texttt{CBICA\_ATX         } & $      1.0$ &            9.78 & \num{6.070e-03} & \num{2.514e-01} & \num{3.412e-01} & \num{1.787e-01} & \num{1.007e-01}    \\
& $      0.4$ &            8.68 & \num{9.550e-03} & \num{1.880e-01} & \num{4.059e-01} & \num{1.946e-01} & \num{1.764e-01}    \\
 & $      0.1$ &            8.71 & \num{1.290e-02} & \num{1.359e-01} & \num{4.834e-01} & \num{1.993e-01} & \num{2.385e-01}    \\
 & $      0.0$ &            8.51 & \num{1.640e-02} & \num{1.346e-01} & \num{6.016e-01} & \num{1.933e-01} & \num{3.189e-01}    \\
\cmidrule[0.05mm]{2-8}
\texttt{TCIA02\_314        } & $      1.0$ &            9.71 & \num{1.070e-02} & \num{2.972e-01} & \num{3.513e-01} & \num{1.600e-01} & \num{1.041e-01}    \\
& $      0.4$ &            9.70 & \num{1.620e-02} & \num{2.008e-01} & \num{4.887e-01} & \num{2.064e-01} & \num{2.115e-01}    \\
 & $      0.1$ &            9.84 & \num{2.170e-02} & \num{1.359e-01} & \num{6.220e-01} & \num{2.212e-01} & \num{2.813e-01}    \\
 & $      0.0$ &            9.85 & \num{2.930e-02} & \num{1.304e-01} & \num{7.130e-01} & \num{2.180e-01} & \num{3.319e-01}    \\
\cmidrule[0.05mm]{2-8}
\texttt{TCIA03\_199        } & $      1.0$ &           11.40 & \num{1.290e-02} & \num{1.702e-01} & \num{5.590e-01} & \num{3.818e-01} & \num{4.402e-02}    \\
& $      0.4$ &           10.70 & \num{1.410e-02} & \num{1.300e-01} & \num{4.925e-01} & \num{3.544e-01} & \num{4.924e-02}    \\
 & $      0.1$ &           10.40 & \num{2.380e-02} & \num{1.321e-01} & \num{7.329e-01} & \num{4.602e-01} & \num{7.439e-02}    \\
 & $      0.0$ &           10.20 & \num{2.920e-02} & \num{1.310e-01} & \num{7.957e-01} & \num{4.725e-01} & \num{9.572e-02}    \\
\cmidrule[0.05mm]{2-8}
\texttt{TCIA03\_296        } & $      1.0$ &           11.70 & \num{2.760e-02} & \num{2.972e-01} & \num{5.061e-01} & \num{2.457e-01} & \num{1.302e-01}    \\
& $      1.0$ &           11.70 & \num{2.760e-02} & \num{2.972e-01} & \num{5.061e-01} & \num{2.457e-01} & \num{1.302e-01}    \\
 & $      0.1$ &           11.70 & \num{1.100e-01} & \num{1.539e-01} & \num{9.653e-01} & \num{2.849e-01} & \num{3.222e-01}    \\
 & $      0.0$ &            0.00 & \num{1.000e-04} & \num{9.986e-01} & \num{9.986e-01} & \num{2.631e-05} & \num{3.464e-06}    \\
\cmidrule[0.05mm]{2-8}
\texttt{TCIA06\_332        } & $      1.0$ &            9.42 & \num{5.330e-03} & \num{2.489e-01} & \num{3.287e-01} & \num{1.575e-01} & \num{8.008e-02}    \\
& $      0.4$ &            9.13 & \num{1.020e-02} & \num{1.764e-01} & \num{4.489e-01} & \num{2.200e-01} & \num{1.745e-01}    \\
& $      0.1$ &            9.06 & \num{1.550e-02} & \num{1.336e-01} & \num{5.499e-01} & \num{2.591e-01} & \num{2.179e-01}    \\
 & $      0.0$ &            8.87 & \num{1.460e-02} & \num{1.241e-01} & \num{5.722e-01} & \num{2.661e-01} & \num{2.209e-01}    \\
\cmidrule[0.05mm]{2-8}
\texttt{TCIA01\_401        } & $      1.0$ &           14.40 & \num{1.480e-02} & \num{2.117e-01} & \num{3.583e-01} & \num{1.835e-01} & \num{6.406e-02}    \\
 & $      0.1$ &           11.40 & \num{3.290e-02} & \num{1.155e-01} & \num{5.905e-01} & \num{3.243e-01} & \num{1.192e-01}    \\
 & $      0.4$ &           11.90 & \num{2.920e-02} & \num{1.546e-01} & \num{5.653e-01} & \num{3.130e-01} & \num{1.117e-01}    \\
 & $      0.0$ &           11.20 & \num{4.660e-02} & \num{1.160e-01} & \num{7.757e-01} & \num{3.940e-01} & \num{1.512e-01}    \\
\cmidrule[0.05mm]{2-8}
\texttt{TCIA02\_606        } & $      1.0$ &           11.60 & \num{2.680e-02} & \num{3.226e-01} & \num{5.059e-01} & \num{2.276e-01} & \num{1.517e-01}    \\
& $      0.4$ &           12.20 & \num{2.480e-02} & \num{1.975e-01} & \num{5.268e-01} & \num{2.292e-01} & \num{1.576e-01}    \\
 & $      0.1$ &           11.10 & \num{4.600e-02} & \num{1.379e-01} & \num{6.919e-01} & \num{2.403e-01} & \num{2.648e-01}    \\
 & $      0.0$ &           11.60 & \num{5.780e-02} & \num{1.295e-01} & \num{8.013e-01} & \num{2.479e-01} & \num{3.088e-01}    \\
\cmidrule[0.05mm]{2-8}
\texttt{TCIA05\_277        } & $      1.0$ &           10.90 & \num{2.490e-02} & \num{2.776e-01} & \num{5.631e-01} & \num{3.347e-01} & \num{1.089e-01}    \\
 & $      0.4$ &           11.10 & \num{2.550e-02} & \num{2.009e-01} & \num{5.832e-01} & \num{3.319e-01} & \num{1.103e-01}    \\
 & $      0.1$ &           10.50 & \num{4.270e-02} & \num{1.513e-01} & \num{7.361e-01} & \num{3.708e-01} & \num{1.668e-01}    \\
 & $      0.0$ &           10.50 & \num{4.820e-02} & \num{1.460e-01} & \num{7.909e-01} & \num{3.839e-01} & \num{1.793e-01}    \\
\bottomrule
\end{tabular}
\end{table*}

\begin{table*}
    \caption{
    \ipoint{Quantitative simulation results for the BraTS'18 data set}. For each patient, we report the estimated tumor proliferation and infiltration rate $\rho_w$, and $\kappa_w$, the relative $\ell_2$-data misfit at observation points $\ellTwoObsMiss$, the misfit $\scaledEllTwoDice$ in the tumor core only, after re-scaling $c(1)
    \in [0,1]$, the distance $\distCM$ (in voxels) between the center of mass of tumor initiation sites $c(0)$ and the center of mass of $TC$ (of the largest component), the distance $\distPi$ (in voxels) between the two most important tumor initiation sites, and integral measures $\int_{TC}c_1 \defeq \nf{\int_{TC}c(1)\dx}{\int_{\Omega}c(1)}$, and $\int_{ED}c_1 \defeq \nf{\int_{ED}c(1)\dx}{\int_{\Omega}c(1)}$ indicating the approximation quality of tumor core ($TC$), and edema ($ED$), respectively. $\int_{B\setminus WT}c_1 \defeq \nf{\int_{B\setminus WT}c(1)\dx}{\int_{\Omega}c(1)}$ indicates erroneous tumor growth in healthy tissue, predicted by our model. Further, we report the number of components with relative mass larger $0.1\%$, the patient's age (in years) and overall survival (in days).
    }
    \label{tab:brats-sim-results-ALL}
\centering\scriptsize\setlength\tabcolsep{5.5pt}
\begin{tabular}{lLLlLlcclllllcc}
    \toprule
    BraTS ID                     & $\rho_w$        & $\kappa_w$      & $\ellTwoObsMiss$ & $\scaledEllTwoDice$ &  $\ellTwoMiss$  & $\distCM$ & $\distPi$ & $\int_{TC}c_1$ & $\int_{ED}c_1$ & $\int_{B\setminus WT}c_1$  & $n_c$ & $age\,[y]$ & $srvl\,[d]$  \\
    \midrule
\texttt{2013\_10  } &           12.35 & \num{2.427e-02} & \num{2.326e-01} & \num{1.363e-01} & \num{3.093e-01} & $      6.8$ & $     10.8$ & \num{7.581e-01} & \num{1.077e-01} & \num{1.342e-01} & $        1$ &        --- &        ---  \\
\texttt{2013\_11  } &           11.38 & \num{3.919e-02} & \num{3.808e-01} & \num{3.211e-01} & \num{6.121e-01} & $      0.5$ & $     22.4$ & \num{5.597e-01} & \num{3.213e-01} & \num{1.190e-01} & $        1$ & $      29$ & $      150$  \\
\texttt{2013\_12  } &           12.70 & \num{2.184e-02} & \num{1.703e-01} & \num{1.278e-01} & \num{3.931e-01} & $      2.6$ & $     16.0$ & \num{7.104e-01} & \num{2.259e-01} & \num{6.373e-02} & $        1$ &        --- &        ---  \\
\texttt{2013\_13  } &            7.10 & \num{4.149e-03} & \num{2.680e-01} & \num{2.046e-01} & \num{4.518e-01} & $      0.6$ & $     18.3$ & \num{5.730e-01} & \num{2.951e-01} & \num{1.319e-01} & $        1$ &        --- &        ---  \\
\texttt{2013\_14  } &           15.69 & \num{1.189e-02} & \num{1.554e-01} & \num{1.343e-01} & \num{3.338e-01} & $      5.2$ & $     11.3$ & \num{7.700e-01} & \num{2.003e-01} & \num{2.971e-02} & $        1$ &        --- &        ---  \\
\texttt{2013\_17  } &           12.55 & \num{2.567e-02} & \num{2.571e-01} & \num{1.469e-01} & \num{4.642e-01} & $      2.6$ & $     16.0$ & \num{6.416e-01} & \num{2.602e-01} & \num{9.819e-02} & $        1$ &        --- &        ---  \\
\texttt{2013\_18  } &           10.16 & \num{1.552e-02} & \num{1.971e-01} & \num{1.418e-01} & \num{6.133e-01} & $      2.2$ & $     26.5$ & \num{5.009e-01} & \num{4.505e-01} & \num{4.862e-02} & $        1$ &        --- &        ---  \\
\texttt{2013\_19  } &           12.75 & \num{1.185e-02} & \num{2.717e-01} & \num{1.589e-01} & \num{4.479e-01} & $      2.7$ & $     24.0$ & \num{6.800e-01} & \num{2.117e-01} & \num{1.082e-01} & $        1$ &        --- &        ---  \\
\texttt{2013\_20  } &           14.36 & \num{2.006e-02} & \num{2.978e-01} & \num{2.551e-01} & \num{4.017e-01} & $      5.7$ & $      8.0$ & \num{7.535e-01} & \num{1.799e-01} & \num{6.658e-02} & $        1$ &        --- &        ---  \\
\texttt{2013\_21  } &           12.52 & \num{1.989e-02} & \num{2.606e-01} & \num{2.277e-01} & \num{3.546e-01} & $      5.3$ & $     16.0$ & \num{7.545e-01} & \num{1.766e-01} & \num{6.889e-02} & $        1$ &        --- &        ---  \\
\texttt{2013\_22  } &           13.24 & \num{2.359e-02} & \num{1.863e-01} & \num{1.259e-01} & \num{4.104e-01} & $     10.6$ & $     13.9$ & \num{7.007e-01} & \num{2.375e-01} & \num{6.183e-02} & $        2$ &        --- &        ---  \\
\texttt{2013\_23  } &           13.35 & \num{2.597e-02} & \num{2.575e-01} & \num{1.844e-01} & \num{3.134e-01} & $      5.6$ & $     16.0$ & \num{8.067e-01} & \num{7.924e-02} & \num{1.141e-01} & $        1$ &        --- &        ---  \\
\texttt{2013\_25  } &            8.02 & \num{5.627e-03} & \num{2.496e-01} & \num{2.128e-01} & \num{4.400e-01} & $      1.0$ & $     15.0$ & \num{6.140e-01} & \num{3.268e-01} & \num{5.924e-02} & $        1$ &        --- &        ---  \\
\texttt{2013\_26  } &            7.25 & \num{3.570e-03} & \num{2.469e-01} & \num{1.958e-01} & \num{4.014e-01} & $      0.4$ & $     11.3$ & \num{6.436e-01} & \num{2.838e-01} & \num{7.261e-02} & $        1$ &        --- &        ---  \\
\texttt{2013\_27  } &           12.73 & \num{1.115e-02} & \num{1.739e-01} & \num{1.418e-01} & \num{3.086e-01} & $      8.2$ & $     17.9$ & \num{7.484e-01} & \num{1.996e-01} & \num{5.201e-02} & $        1$ & $      68$ & $      120$  \\
\texttt{2013\_2   } &           10.15 & \num{8.872e-03} & \num{4.104e-01} & \num{3.525e-01} & \num{4.919e-01} & $      9.0$ & $      8.0$ & \num{6.574e-01} & \num{1.991e-01} & \num{1.434e-01} & $        1$ &        --- &        ---  \\
\texttt{2013\_3   } &           11.74 & \num{1.083e-02} & \num{2.144e-01} & \num{1.453e-01} & \num{4.082e-01} & $      3.6$ & $     20.4$ & \num{6.934e-01} & \num{2.404e-01} & \num{6.625e-02} & $        1$ &        --- &        ---  \\
\texttt{2013\_4   } &           10.53 & \num{1.365e-02} & \num{2.244e-01} & \num{1.601e-01} & \num{4.907e-01} & $      5.0$ & $     18.3$ & \num{5.734e-01} & \num{3.406e-01} & \num{8.608e-02} & $        2$ &        --- &        ---  \\
\texttt{2013\_5   } &            7.72 & \num{7.990e-03} & \num{3.013e-01} & \num{2.577e-01} & \num{8.082e-01} & $      3.5$ & $     58.2$ & \num{3.825e-01} & \num{5.747e-01} & \num{4.280e-02} & $        1$ &        --- &        ---  \\
\texttt{2013\_7   } &           11.82 & \num{2.077e-02} & \num{2.304e-01} & \num{1.476e-01} & \num{5.496e-01} & $      2.7$ & $     16.0$ & \num{5.844e-01} & \num{3.548e-01} & \num{6.084e-02} & $        1$ &        --- &        ---  \\
\texttt{CBICA\_AAB} &           11.24 & \num{2.122e-02} & \num{2.213e-01} & \num{1.493e-01} & \num{5.614e-01} & $      1.2$ & $     17.9$ & \num{5.642e-01} & \num{3.789e-01} & \num{5.683e-02} & $        1$ & $      60$ & $      289$  \\
\texttt{CBICA\_AAG} &           11.23 & \num{7.278e-03} & \num{1.709e-01} & \num{1.500e-01} & \num{4.019e-01} & $      5.0$ & $     17.5$ & \num{6.625e-01} & \num{2.987e-01} & \num{3.877e-02} & $        1$ & $      52$ & $      616$  \\
\texttt{CBICA\_AAL} &            6.43 & \num{4.103e-03} & \num{2.558e-01} & \num{1.749e-01} & \num{3.750e-01} & $      1.8$ & $      8.0$ & \num{6.268e-01} & \num{1.756e-01} & \num{1.976e-01} & $        1$ & $      54$ & $      464$  \\
\texttt{CBICA\_AAP} &           11.57 & \num{1.308e-02} & \num{2.684e-01} & \num{2.092e-01} & \num{5.991e-01} & $      3.9$ & $      8.0$ & \num{5.795e-01} & \num{3.434e-01} & \num{7.709e-02} & $        1$ & $      39$ & $      788$  \\
\texttt{CBICA\_ABB} &            9.11 & \num{7.722e-03} & \num{2.063e-01} & \num{1.336e-01} & \num{3.038e-01} & $      2.3$ & $     17.0$ & \num{6.894e-01} & \num{1.417e-01} & \num{1.690e-01} & $        1$ & $      68$ & $      465$  \\
\texttt{CBICA\_ABE} &           10.73 & \num{2.461e-02} & \num{2.698e-01} & \num{1.595e-01} & \num{3.946e-01} & $      1.8$ & $     16.0$ & \num{6.674e-01} & \num{2.076e-01} & \num{1.250e-01} & $        1$ & $      67$ & $      269$  \\
\texttt{CBICA\_ABM} &            7.63 & \num{7.112e-03} & \num{2.089e-01} & \num{1.538e-01} & \num{4.310e-01} & $      1.3$ & $     18.3$ & \num{5.586e-01} & \num{3.673e-01} & \num{7.407e-02} & $        2$ & $      69$ & $      503$  \\
\texttt{CBICA\_ABN} &            4.03 & \num{1.528e-03} & \num{2.713e-01} & \num{2.494e-01} & \num{3.902e-01} & $      0.4$ & $      8.0$ & \num{5.666e-01} & \num{3.488e-01} & \num{8.464e-02} & $        2$ & $      68$ & $     1278$  \\
\texttt{CBICA\_ABO} &           13.37 & \num{1.345e-02} & \num{2.842e-01} & \num{2.696e-01} & \num{3.545e-01} & $      6.8$ & $     17.9$ & \num{8.125e-01} & \num{1.324e-01} & \num{5.517e-02} & $        1$ & $      56$ & $     1155$  \\
\texttt{CBICA\_ABY} &            7.57 & \num{4.211e-03} & \num{2.935e-01} & \num{2.057e-01} & \num{4.268e-01} & $      1.4$ & $     16.5$ & \num{6.166e-01} & \num{2.443e-01} & \num{1.392e-01} & $        1$ & $      48$ & $      515$  \\
\texttt{CBICA\_ALN} &            7.05 & \num{5.709e-03} & \num{1.984e-01} & \num{1.606e-01} & \num{5.030e-01} & $      1.3$ & $     16.0$ & \num{5.235e-01} & \num{4.171e-01} & \num{5.938e-02} & $        1$ & $      60$ & $      421$  \\
\texttt{CBICA\_ALU} &            7.36 & \num{4.833e-03} & \num{1.547e-01} & \num{1.376e-01} & \num{4.149e-01} & $      0.8$ & $     12.0$ & \num{6.134e-01} & \num{3.551e-01} & \num{3.150e-02} & $        3$ & $      65$ & $      495$  \\
\texttt{CBICA\_ALX} &            9.70 & \num{8.817e-03} & \num{2.414e-01} & \num{1.750e-01} & \num{4.359e-01} & $      3.9$ & $     11.3$ & \num{6.210e-01} & \num{2.792e-01} & \num{9.980e-02} & $        1$ & $      59$ & $      698$  \\
\texttt{CBICA\_AME} &           10.13 & \num{1.040e-02} & \num{2.624e-01} & \num{1.930e-01} & \num{4.791e-01} & $      3.6$ & $     19.6$ & \num{6.196e-01} & \num{2.791e-01} & \num{1.013e-01} & $        3$ & $      51$ & $      359$  \\
\texttt{CBICA\_AMH} &            9.92 & \num{1.258e-02} & \num{2.723e-01} & \num{2.138e-01} & \num{6.796e-01} & $      6.2$ & $      8.0$ & \num{4.794e-01} & \num{4.570e-01} & \num{6.356e-02} & $        1$ & $      62$ & $      169$  \\
\texttt{CBICA\_ANG} &           11.03 & \num{1.782e-02} & \num{2.596e-01} & \num{1.770e-01} & \num{4.136e-01} & $      5.6$ & $     16.0$ & \num{6.635e-01} & \num{2.223e-01} & \num{1.141e-01} & $        1$ & $      55$ & $      368$  \\
\texttt{CBICA\_ANP} &            4.55 & \num{2.260e-03} & \num{4.187e-01} & \num{3.074e-01} & \num{4.513e-01} & $      4.4$ & $     16.5$ & \num{5.652e-01} & \num{1.195e-01} & \num{3.152e-01} & $        1$ & $      61$ & $      486$  \\
\texttt{CBICA\_ANZ} &            9.33 & \num{1.115e-02} & \num{2.740e-01} & \num{2.321e-01} & \num{4.504e-01} & $      4.1$ & $     16.0$ & \num{6.551e-01} & \num{2.623e-01} & \num{8.264e-02} & $        1$ & $      68$ & $      287$  \\
\texttt{CBICA\_AOD} &           14.68 & \num{2.384e-02} & \num{2.621e-01} & \num{1.988e-01} & \num{3.130e-01} & $      7.0$ & $     19.6$ & \num{8.137e-01} & \num{7.068e-02} & \num{1.156e-01} & $        1$ & $      60$ & $       55$  \\
\texttt{CBICA\_AOH} &            5.10 & \num{3.421e-03} & \num{3.785e-01} & \num{2.672e-01} & \num{4.124e-01} & $      0.6$ & $      8.0$ & \num{5.744e-01} & \num{1.508e-01} & \num{2.747e-01} & $        1$ & $      56$ & $      576$  \\
\texttt{CBICA\_AOO} &           12.09 & \num{8.075e-03} & \num{2.179e-01} & \num{1.786e-01} & \num{3.500e-01} & $      3.2$ & $     25.3$ & \num{7.335e-01} & \num{1.895e-01} & \num{7.695e-02} & $        1$ & $      44$ & $      350$  \\
\texttt{CBICA\_AOP} &            9.30 & \num{7.879e-03} & \num{2.955e-01} & \num{2.196e-01} & \num{3.620e-01} & $      0.7$ & $     11.3$ & \num{6.898e-01} & \num{1.623e-01} & \num{1.479e-01} & $        1$ & $      67$ & $      332$  \\
\texttt{CBICA\_AOZ} &            9.49 & \num{9.996e-03} & \num{2.585e-01} & \num{1.915e-01} & \num{4.751e-01} & $      4.0$ & $     19.6$ & \num{5.853e-01} & \num{2.982e-01} & \num{1.165e-01} & $        3$ & $      46$ & $      331$  \\
\texttt{CBICA\_APR} &           10.48 & \num{1.965e-02} & \num{3.319e-01} & \num{1.699e-01} & \num{5.261e-01} & $      3.8$ & $     25.3$ & \num{5.704e-01} & \num{2.509e-01} & \num{1.787e-01} & $        1$ & $      62$ & $       23$  \\
\texttt{CBICA\_APY} &            8.48 & \num{9.149e-03} & \num{2.934e-01} & \num{2.140e-01} & \num{5.438e-01} & $      1.2$ & $     39.2$ & \num{5.646e-01} & \num{3.242e-01} & \num{1.113e-01} & $        3$ & $      45$ & $      203$  \\
\texttt{CBICA\_APZ} &           10.21 & \num{1.170e-02} & \num{3.513e-01} & \num{2.653e-01} & \num{3.936e-01} & $      6.5$ & $     17.9$ & \num{6.879e-01} & \num{1.419e-01} & \num{1.702e-01} & $        1$ & $      60$ & $      336$  \\
\texttt{CBICA\_AQA} &            8.88 & \num{9.306e-03} & \num{2.265e-01} & \num{1.565e-01} & \num{2.580e-01} & $      4.7$ & $      8.0$ & \num{7.176e-01} & \num{1.276e-01} & \num{1.548e-01} & $        2$ & $      76$ & $      106$  \\
\texttt{CBICA\_AQD} &            8.33 & \num{4.416e-03} & \num{1.544e-01} & \num{1.236e-01} & \num{3.528e-01} & $      1.1$ & $     12.6$ & \num{6.629e-01} & \num{2.717e-01} & \num{6.535e-02} & $        1$ & $      73$ & $       32$  \\
\texttt{CBICA\_AQG} &            5.84 & \num{2.956e-03} & \num{3.190e-01} & \num{2.396e-01} & \num{3.770e-01} & $      1.0$ & $     41.2$ & \num{5.917e-01} & \num{1.899e-01} & \num{2.184e-01} & $        4$ & $      53$ & $      466$  \\
\texttt{CBICA\_AQJ} &           16.55 & \num{1.809e-02} & \num{2.030e-01} & \num{1.407e-01} & \num{3.009e-01} & $      6.8$ & $      8.0$ & \num{8.184e-01} & \num{1.109e-01} & \num{7.063e-02} & $        1$ & $      66$ & $      170$  \\
\texttt{CBICA\_AQN} &            8.86 & \num{9.831e-03} & \num{3.312e-01} & \num{2.278e-01} & \num{3.581e-01} & $      4.9$ & $     32.0$ & \num{6.961e-01} & \num{1.020e-01} & \num{2.019e-01} & $        1$ & $      63$ & $      488$  \\
\texttt{CBICA\_AQO} &           11.26 & \num{1.050e-02} & \num{1.882e-01} & \num{1.303e-01} & \num{4.436e-01} & $      4.3$ & $     16.0$ & \num{6.285e-01} & \num{3.190e-01} & \num{5.250e-02} & $        1$ & $      67$ & $      473$  \\
\texttt{CBICA\_AQP} &           13.12 & \num{9.301e-03} & \num{1.497e-01} & \num{1.192e-01} & \num{2.742e-01} & $      2.8$ & $     11.3$ & \num{8.019e-01} & \num{1.371e-01} & \num{6.105e-02} & $        1$ & $      46$ & $     1283$  \\
\texttt{CBICA\_AQQ} &           10.55 & \num{9.181e-02} & \num{5.342e-01} & \num{3.714e-01} & \num{5.812e-01} & $     14.9$ & $     10.2$ & \num{5.266e-01} & \num{1.438e-01} & \num{3.297e-01} & $        1$ & $      69$ & $       33$  \\
\texttt{CBICA\_AQR} &            6.23 & \num{6.490e-03} & \num{3.693e-01} & \num{2.746e-01} & \num{4.772e-01} & $      1.4$ & $     16.5$ & \num{5.543e-01} & \num{2.416e-01} & \num{2.040e-01} & $        1$ & $      71$ & $       89$  \\
\texttt{CBICA\_AQT} &            9.95 & \num{1.426e-02} & \num{1.437e-01} & \num{1.359e-01} & \num{5.445e-01} & $      1.1$ & $     11.3$ & \num{5.496e-01} & \num{4.343e-01} & \num{1.609e-02} & $        1$ & $      75$ & $      172$  \\
\texttt{CBICA\_AQU} &            9.66 & \num{1.138e-02} & \num{2.372e-01} & \num{1.882e-01} & \num{4.831e-01} & $      1.4$ & $     25.3$ & \num{5.907e-01} & \num{3.311e-01} & \num{7.824e-02} & $        2$ & $      72$ & $       30$  \\
\texttt{CBICA\_AQV} &           10.60 & \num{9.412e-03} & \num{2.732e-01} & \num{2.132e-01} & \num{5.885e-01} & $      7.2$ & $     26.9$ & \num{5.576e-01} & \num{3.518e-01} & \num{9.062e-02} & $        1$ & $      53$ & $       84$  \\
\texttt{CBICA\_AQY} &           11.71 & \num{1.025e-02} & \num{2.141e-01} & \num{1.984e-01} & \num{2.763e-01} & $      1.6$ & $     19.6$ & \num{8.166e-01} & \num{1.400e-01} & \num{4.338e-02} & $        2$ & $      57$ & $      229$  \\
\bottomrule
\end{tabular}
\end{table*}
\begin{table*}
\ContinuedFloat
\centering\scriptsize\setlength\tabcolsep{5.5pt}
\begin{tabular}{lLLlLlcclllllcc}
    \toprule
    BraTS ID                     & $\rho_w$        & $\kappa_w$      & $\ellTwoObsMiss$ & $\scaledEllTwoDice$ &  $\ellTwoMiss$  & $\distCM$ & $\distPi$ & $\int_{TC}c_1$ & $\int_{ED}c_1$ & $\int_{B\setminus WT}c_1$  & $n_c$ & $age\,[y]$ & $srvl\,[d]$  \\
    \midrule
\texttt{CBICA\_AQZ} &           12.15 & \num{1.867e-02} & \num{2.625e-01} & \num{1.715e-01} & \num{4.003e-01} & $      5.9$ & $     16.0$ & \num{7.033e-01} & \num{1.648e-01} & \num{1.319e-01} & $        1$ & $      63$ & $      286$  \\
\texttt{CBICA\_ARF} &           10.52 & \num{1.070e-02} & \num{2.628e-01} & \num{2.246e-01} & \num{2.868e-01} & $      4.7$ & $     24.0$ & \num{7.981e-01} & \num{7.793e-02} & \num{1.240e-01} & $        1$ & $      75$ & $      726$  \\
\texttt{CBICA\_ARW} &            6.97 & \num{4.686e-03} & \num{1.897e-01} & \num{1.503e-01} & \num{2.885e-01} & $      1.5$ & $     11.3$ & \num{6.603e-01} & \num{2.217e-01} & \num{1.180e-01} & $        3$ & $      44$ & $      495$  \\
\texttt{CBICA\_ARZ} &            2.93 & \num{9.257e-04} & \num{3.539e-01} & \num{3.268e-01} & \num{4.029e-01} & $      1.5$ & $      6.0$ & \num{5.882e-01} & \num{2.852e-01} & \num{1.266e-01} & $        7$ & $      54$ & $      871$  \\
\texttt{CBICA\_ASA} &           11.34 & \num{1.414e-02} & \num{3.922e-01} & \num{3.615e-01} & \num{4.364e-01} & $      6.3$ & $      9.2$ & \num{7.645e-01} & \num{1.450e-01} & \num{9.052e-02} & $        1$ & $      63$ & $      210$  \\
\texttt{CBICA\_ASE} &            3.01 & \num{1.601e-03} & \num{4.201e-01} & \num{4.075e-01} & \num{5.284e-01} & $      0.9$ & $      6.9$ & \num{4.879e-01} & \num{4.134e-01} & \num{9.877e-02} & $        4$ & $      46$ & $      318$  \\
\texttt{CBICA\_ASG} &           11.16 & \num{1.526e-02} & \num{2.258e-01} & \num{1.734e-01} & \num{4.490e-01} & $      2.8$ & $     19.6$ & \num{6.537e-01} & \num{2.686e-01} & \num{7.771e-02} & $        2$ & $      57$ & $      208$  \\
\texttt{CBICA\_ASH} &            6.46 & \num{3.685e-03} & \num{1.821e-01} & \num{1.463e-01} & \num{4.178e-01} & $      0.5$ & $      8.9$ & \num{5.661e-01} & \num{3.924e-01} & \num{4.143e-02} & $        1$ & $      46$ & $      660$  \\
\texttt{CBICA\_ASK} &            7.84 & \num{5.324e-03} & \num{2.125e-01} & \num{1.655e-01} & \num{5.047e-01} & $      1.6$ & $     11.3$ & \num{5.428e-01} & \num{3.929e-01} & \num{6.430e-02} & $        1$ & $      77$ & $      522$  \\
\texttt{CBICA\_ASN} &           12.22 & \num{1.826e-02} & \num{2.306e-01} & \num{1.617e-01} & \num{2.468e-01} & $      3.8$ & $     17.9$ & \num{8.193e-01} & \num{3.367e-02} & \num{1.471e-01} & $        1$ & $      39$ & $      407$  \\
\texttt{CBICA\_ASO} &           10.10 & \num{1.178e-02} & \num{2.427e-01} & \num{1.888e-01} & \num{5.598e-01} & $      1.7$ & $     17.9$ & \num{5.389e-01} & \num{4.045e-01} & \num{5.652e-02} & $        1$ & $      52$ & $      265$  \\
\texttt{CBICA\_ASU} &            9.68 & \num{9.926e-03} & \num{2.174e-01} & \num{1.867e-01} & \num{2.978e-01} & $      2.1$ & $      8.0$ & \num{7.328e-01} & \num{1.869e-01} & \num{8.032e-02} & $        1$ & $      81$ & $       85$  \\
\texttt{CBICA\_ASV} &            9.98 & \num{8.347e-03} & \num{1.817e-01} & \num{1.490e-01} & \num{5.397e-01} & $      2.4$ & $     16.0$ & \num{5.968e-01} & \num{3.638e-01} & \num{3.943e-02} & $        1$ & $      54$ & $      597$  \\
\texttt{CBICA\_ASW} &           11.23 & \num{1.506e-02} & \num{3.398e-01} & \num{2.516e-01} & \num{5.038e-01} & $      4.8$ & $     17.9$ & \num{6.328e-01} & \num{2.374e-01} & \num{1.298e-01} & $        1$ & $      68$ & $      239$  \\
\texttt{CBICA\_ASY} &            7.77 & \num{6.026e-03} & \num{1.564e-01} & \num{1.294e-01} & \num{4.694e-01} & $      1.0$ & $     15.0$ & \num{5.591e-01} & \num{4.054e-01} & \num{3.551e-02} & $        4$ & $      66$ & $      610$  \\
\texttt{CBICA\_ATB} &            9.27 & \num{1.590e-02} & \num{2.935e-01} & \num{1.780e-01} & \num{3.471e-01} & $      3.9$ & $     17.9$ & \num{6.539e-01} & \num{1.278e-01} & \num{2.183e-01} & $        2$ & $      71$ & $      208$  \\
\texttt{CBICA\_ATD} &            9.77 & \num{1.144e-02} & \num{2.462e-01} & \num{2.075e-01} & \num{9.635e-01} & $      4.8$ & $     18.5$ & \num{3.643e-01} & \num{6.033e-01} & \num{3.240e-02} & $        1$ & $      69$ & $      355$  \\
\texttt{CBICA\_ATF} &            8.54 & \num{7.209e-03} & \num{3.177e-01} & \num{2.643e-01} & \num{5.339e-01} & $      3.3$ & $     17.9$ & \num{5.643e-01} & \num{3.397e-01} & \num{9.600e-02} & $        1$ & $      68$ & $      152$  \\
\texttt{CBICA\_ATP} &            6.03 & \num{2.205e-03} & \num{1.623e-01} & \num{1.339e-01} & \num{3.087e-01} & $      0.8$ & $      9.8$ & \num{6.198e-01} & \num{3.307e-01} & \num{4.949e-02} & $        1$ & $      51$ & $      385$  \\
\texttt{CBICA\_ATV} &           10.67 & \num{1.881e-02} & \num{2.296e-01} & \num{1.729e-01} & \num{4.440e-01} & $      3.8$ & $     17.9$ & \num{6.020e-01} & \num{3.150e-01} & \num{8.298e-02} & $        1$ & $      62$ & $      453$  \\
\texttt{CBICA\_ATX} &            9.78 & \num{6.070e-03} & \num{2.514e-01} & \num{1.936e-01} & \num{3.412e-01} & $      2.9$ & $     24.7$ & \num{7.206e-01} & \num{1.787e-01} & \num{1.007e-01} & $        1$ & $      36$ & $     1592$  \\
\texttt{CBICA\_AUN} &           13.31 & \num{9.958e-03} & \num{1.718e-01} & \num{1.493e-01} & \num{3.270e-01} & $      1.7$ & $     17.9$ & \num{7.620e-01} & \num{2.044e-01} & \num{3.361e-02} & $        1$ & $      68$ & $      376$  \\
\texttt{CBICA\_AUQ} &            6.28 & \num{8.181e-03} & \num{3.673e-01} & \num{2.651e-01} & \num{5.392e-01} & $      3.0$ & $     11.3$ & \num{4.369e-01} & \num{3.938e-01} & \num{1.692e-01} & $        1$ & $      60$ & $     1337$  \\
\texttt{CBICA\_AVG} &            6.98 & \num{3.796e-03} & \num{1.468e-01} & \num{1.204e-01} & \num{3.934e-01} & $      2.2$ & $     13.3$ & \num{6.001e-01} & \num{3.410e-01} & \num{5.889e-02} & $        1$ & $      63$ & $      579$  \\
\texttt{CBICA\_AVJ} &            9.25 & \num{5.098e-03} & \num{1.840e-01} & \num{1.490e-01} & \num{4.030e-01} & $      3.8$ & $     11.3$ & \num{6.361e-01} & \num{3.028e-01} & \num{6.115e-02} & $        2$ & $      45$ & $      614$  \\
\texttt{CBICA\_AVV} &           10.76 & \num{2.463e-02} & \num{2.802e-01} & \num{1.803e-01} & \num{4.104e-01} & $      2.4$ & $     16.0$ & \num{6.703e-01} & \num{1.972e-01} & \num{1.325e-01} & $        1$ & $      72$ & $      387$  \\
\texttt{CBICA\_AWG} &           10.71 & \num{8.258e-03} & \num{3.372e-01} & \num{3.115e-01} & \num{3.828e-01} & $      2.5$ & $     16.0$ & \num{7.959e-01} & \num{1.224e-01} & \num{8.169e-02} & $        1$ & $      55$ & $      180$  \\
\texttt{CBICA\_AWH} &           10.48 & \num{1.722e-02} & \num{2.876e-01} & \num{1.803e-01} & \num{4.052e-01} & $      3.2$ & $     11.3$ & \num{6.320e-01} & \num{2.005e-01} & \num{1.674e-01} & $        1$ & $      52$ & $      139$  \\
\texttt{CBICA\_AWI} &            8.65 & \num{1.148e-02} & \num{3.252e-01} & \num{2.463e-01} & \num{4.105e-01} & $      1.8$ & $     34.4$ & \num{6.582e-01} & \num{1.853e-01} & \num{1.565e-01} & $        2$ & $      46$ & $      375$  \\
\texttt{CBICA\_AXJ} &            8.33 & \num{1.204e-02} & \num{2.446e-01} & \num{2.059e-01} & \num{6.580e-01} & $      1.4$ & $     24.0$ & \num{4.584e-01} & \num{5.023e-01} & \num{3.932e-02} & $        1$ & $      27$ & $     1767$  \\
\texttt{CBICA\_AXL} &           11.12 & \num{1.736e-02} & \num{2.917e-01} & \num{1.831e-01} & \num{3.169e-01} & $      1.1$ & $     17.9$ & \num{7.453e-01} & \num{6.551e-02} & \num{1.892e-01} & $        1$ & $      74$ & $      168$  \\
\texttt{CBICA\_AXM} &            8.98 & \num{7.333e-03} & \num{2.408e-01} & \num{2.000e-01} & \num{3.873e-01} & $      3.0$ & $     13.9$ & \num{6.613e-01} & \num{2.608e-01} & \num{7.788e-02} & $        1$ & $      66$ & $      438$  \\
\texttt{CBICA\_AXN} &           10.24 & \num{1.705e-02} & \num{3.221e-01} & \num{2.542e-01} & \num{3.480e-01} & $      3.0$ & $     11.3$ & \num{7.571e-01} & \num{7.363e-02} & \num{1.693e-01} & $        1$ & $      85$ & $      345$  \\
\texttt{CBICA\_AXO} &           13.10 & \num{3.006e-02} & \num{2.447e-01} & \num{1.328e-01} & \num{3.944e-01} & $      9.3$ & $     17.9$ & \num{7.038e-01} & \num{1.831e-01} & \num{1.131e-01} & $        1$ & $      56$ & $      394$  \\
\texttt{CBICA\_AXQ} &            6.78 & \num{8.366e-03} & \num{3.528e-01} & \num{2.305e-01} & \num{3.850e-01} & $      1.6$ & $     32.5$ & \num{6.435e-01} & \num{1.193e-01} & \num{2.371e-01} & $        2$ & $      66$ & $      114$  \\
\texttt{CBICA\_AXW} &            8.68 & \num{6.786e-03} & \num{1.959e-01} & \num{1.617e-01} & \num{4.310e-01} & $      2.8$ & $     17.9$ & \num{6.288e-01} & \num{3.206e-01} & \num{5.061e-02} & $        1$ & $      79$ & $      191$  \\
\texttt{CBICA\_AYA} &            7.55 & \num{3.558e-03} & \num{2.516e-01} & \num{1.887e-01} & \num{3.510e-01} & $      2.2$ & $     25.4$ & \num{6.637e-01} & \num{2.207e-01} & \num{1.157e-01} & $        1$ & $      74$ & $       50$  \\
\texttt{CBICA\_AYI} &            5.54 & \num{3.029e-03} & \num{3.878e-01} & \num{2.890e-01} & \num{4.822e-01} & $      1.9$ & $     15.0$ & \num{5.767e-01} & \num{2.446e-01} & \num{1.788e-01} & $        1$ & $      65$ & $      387$  \\
\texttt{CBICA\_AYU} &            8.86 & \num{8.683e-03} & \num{2.180e-01} & \num{1.704e-01} & \num{3.932e-01} & $      4.5$ & $     11.3$ & \num{6.314e-01} & \num{2.698e-01} & \num{9.880e-02} & $        1$ & $      63$ & $       58$  \\
\texttt{CBICA\_AYW} &           10.12 & \num{3.960e-03} & \num{3.174e-01} & \num{3.056e-01} & \num{3.667e-01} & $      2.0$ & $     17.9$ & \num{8.312e-01} & \num{1.259e-01} & \num{4.292e-02} & $        1$ & $      49$ & $      734$  \\
\texttt{CBICA\_AZD} &            9.30 & \num{7.879e-03} & \num{2.955e-01} & \num{2.196e-01} & \num{3.620e-01} & $      0.7$ & $     11.3$ & \num{6.898e-01} & \num{1.623e-01} & \num{1.479e-01} & $        1$ & $      46$ & $      448$  \\
\texttt{CBICA\_AZH} &            8.98 & \num{5.525e-03} & \num{1.657e-01} & \num{1.417e-01} & \num{3.948e-01} & $      1.8$ & $      8.0$ & \num{6.288e-01} & \num{3.373e-01} & \num{3.389e-02} & $        1$ & $      54$ & $      401$  \\
\texttt{CBICA\_BFB} &            9.98 & \num{2.198e-02} & \num{3.340e-01} & \num{2.581e-01} & \num{5.374e-01} & $      4.2$ & $     10.8$ & \num{5.832e-01} & \num{2.871e-01} & \num{1.297e-01} & $        1$ &        --- &        ---  \\
\texttt{CBICA\_BFP} &            8.50 & \num{3.690e-03} & \num{2.121e-01} & \num{1.758e-01} & \num{3.638e-01} & $      1.8$ & $     12.6$ & \num{6.797e-01} & \num{2.444e-01} & \num{7.592e-02} & $        1$ &        --- &        ---  \\
\texttt{CBICA\_BHB} &            2.75 & \num{1.043e-03} & \num{1.174e-01} & \num{1.429e-01} & \num{2.894e-01} & $      0.1$ & $      8.9$ & \num{4.983e-01} & \num{4.843e-01} & \num{1.747e-02} & $        1$ & $      55$ & $      510$  \\
\texttt{CBICA\_BHK} &            8.07 & \num{6.546e-03} & \num{2.014e-01} & \num{1.641e-01} & \num{5.016e-01} & $      1.2$ & $     15.0$ & \num{5.584e-01} & \num{3.756e-01} & \num{6.607e-02} & $        1$ &        --- &        ---  \\
\texttt{CBICA\_BHM} &            9.17 & \num{9.308e-03} & \num{2.388e-01} & \num{1.666e-01} & \num{3.721e-01} & $      5.5$ & $     16.0$ & \num{6.598e-01} & \num{1.971e-01} & \num{1.432e-01} & $        1$ & $      62$ & $      436$  \\
\texttt{TCIA01\_131} &            8.90 & \num{5.764e-03} & \num{2.254e-01} & \num{1.718e-01} & \num{3.244e-01} & $      5.4$ & $     13.9$ & \num{6.870e-01} & \num{2.005e-01} & \num{1.124e-01} & $        2$&        --- &        ---  \\
\texttt{TCIA01\_147} &            9.88 & \num{1.869e-02} & \num{3.357e-01} & \num{2.648e-01} & \num{4.869e-01} & $      4.8$ & $     33.9$ & \num{6.281e-01} & \num{2.652e-01} & \num{1.068e-01} & $        1$& $      61$ & $      209$  \\
\texttt{TCIA01\_150} &           12.36 & \num{2.054e-02} & \num{2.598e-01} & \num{1.534e-01} & \num{4.826e-01} & $      3.9$ & $     11.3$ & \num{6.511e-01} & \num{2.406e-01} & \num{1.083e-01} & $        1$& $      51$ & $     1489$  \\
\texttt{TCIA01\_180} &           12.77 & \num{1.392e-02} & \num{1.945e-01} & \num{1.371e-01} & \num{4.088e-01} & $      6.7$ & $     17.9$ & \num{6.907e-01} & \num{2.506e-01} & \num{5.863e-02} & $        1$&        --- &        ---  \\
\texttt{TCIA01\_186} &            8.76 & \num{8.883e-03} & \num{2.540e-01} & \num{2.078e-01} & \num{4.361e-01} & $      1.3$ & $     17.9$ & \num{6.218e-01} & \num{3.060e-01} & \num{7.223e-02} & $        1$& $      33$ & $      370$  \\
\texttt{TCIA01\_190} &            8.97 & \num{6.167e-03} & \num{2.517e-01} & \num{1.851e-01} & \num{3.907e-01} & $      2.1$ & $     12.6$ & \num{6.325e-01} & \num{2.582e-01} & \num{1.093e-01} & $        1$& $      61$ & $      322$  \\
\texttt{TCIA01\_201} &           12.97 & \num{1.435e-02} & \num{2.014e-01} & \num{1.846e-01} & \num{4.152e-01} & $      2.0$ & $     13.9$ & \num{7.222e-01} & \num{2.473e-01} & \num{3.055e-02} & $        1$& $      60$ & $      430$  \\
\texttt{TCIA01\_203} &           13.39 & \num{1.743e-02} & \num{2.137e-01} & \num{1.786e-01} & \num{3.203e-01} & $      3.9$ & $     19.6$ & \num{7.595e-01} & \num{1.859e-01} & \num{5.455e-02} & $        4$& $      45$ & $      268$  \\
\texttt{TCIA01\_221} &           12.06 & \num{1.642e-02} & \num{3.337e-01} & \num{2.772e-01} & \num{5.582e-01} & $      9.5$ & $     24.0$ & \num{6.362e-01} & \num{2.838e-01} & \num{7.995e-02} & $        1$&        --- &        ---  \\
\texttt{TCIA01\_231} &           10.40 & \num{1.364e-02} & \num{3.074e-01} & \num{2.285e-01} & \num{6.206e-01} & $      6.9$ & $      9.2$ & \num{5.470e-01} & \num{3.666e-01} & \num{8.644e-02} & $        1$& $      63$ & $     1561$  \\
\texttt{TCIA01\_235} &           11.65 & \num{1.051e-02} & \num{1.631e-01} & \num{1.359e-01} & \num{5.073e-01} & $      6.5$ & $     26.8$ & \num{5.957e-01} & \num{3.733e-01} & \num{3.106e-02} & $        1$& $      57$ & $      804$  \\
\texttt{TCIA01\_335} &           11.75 & \num{2.334e-02} & \num{2.221e-01} & \num{1.469e-01} & \num{3.395e-01} & $      1.9$ & $      8.0$ & \num{7.475e-01} & \num{1.485e-01} & \num{1.040e-01} & $        1$& $      54$ & $      355$  \\
\texttt{TCIA01\_378} &           11.88 & \num{1.897e-02} & \num{2.448e-01} & \num{1.245e-01} & \num{4.695e-01} & $      2.7$ & $     25.3$ & \num{6.450e-01} & \num{2.523e-01} & \num{1.026e-01} & $        2$& $      74$ & $      110$  \\
\texttt{TCIA01\_390} &           11.87 & \num{1.958e-02} & \num{2.331e-01} & \num{1.270e-01} & \num{6.740e-01} & $      5.0$ & $      8.0$ & \num{5.224e-01} & \num{3.819e-01} & \num{9.565e-02} & $        1$& $      63$ & $      634$  \\
\texttt{TCIA01\_401} &           14.38 & \num{1.483e-02} & \num{2.117e-01} & \num{1.604e-01} & \num{3.583e-01} & $     10.3$ & $     24.0$ & \num{7.526e-01} & \num{1.834e-01} & \num{6.403e-02} & $        1$& $      78$ & $      448$  \\
\texttt{TCIA01\_411} &            4.34 & \num{1.813e-03} & \num{1.871e-01} & \num{1.474e-01} & \num{3.580e-01} & $      0.3$ & $      8.0$ & \num{5.109e-01} & \num{4.381e-01} & \num{5.104e-02} & $        1$& $      42$ & $      822$  \\
\texttt{TCIA01\_412} &           11.57 & \num{2.724e-02} & \num{3.630e-01} & \num{3.083e-01} & \num{4.506e-01} & $      4.1$ & $      8.0$ & \num{7.262e-01} & \num{1.649e-01} & \num{1.089e-01} & $        1$& $      68$ & $      291$  \\
\texttt{TCIA01\_425} &            2.43 & \num{2.215e-03} & \num{5.094e-01} & \num{3.508e-01} & \num{5.151e-01} & $      0.9$ & $     17.0$ & \num{7.349e-01} & \num{2.082e-01} & \num{5.692e-02} & $        1$& $      56$ & $      558$  \\
\texttt{TCIA01\_429} &           10.39 & \num{2.153e-02} & \num{2.998e-01} & \num{1.563e-01} & \num{5.125e-01} & $      1.4$ & $     17.9$ & \num{5.829e-01} & \num{2.743e-01} & \num{1.428e-01} & $        1$& $      54$ & $       86$  \\
\texttt{TCIA01\_448} &            9.58 & \num{1.431e-02} & \num{2.529e-01} & \num{1.832e-01} & \num{5.152e-01} & $      7.3$ & $      8.0$ & \num{5.639e-01} & \num{3.444e-01} & \num{9.176e-02} & $        1$& $      44$ & $      199$  \\
\texttt{TCIA01\_460} &           11.55 & \num{1.867e-02} & \num{2.465e-01} & \num{1.404e-01} & \num{3.767e-01} & $      3.4$ & $     25.3$ & \num{6.926e-01} & \num{1.776e-01} & \num{1.298e-01} & $        3$& $      18$ & $      630$  \\
\texttt{TCIA01\_499} &           10.46 & \num{1.641e-02} & \num{2.915e-01} & \num{2.206e-01} & \num{4.264e-01} & $      5.7$ & $      8.0$ & \num{6.728e-01} & \num{1.866e-01} & \num{1.405e-01} & $        1$& $      50$ & $      600$  \\
\texttt{TCIA02\_117} &           10.05 & \num{1.460e-02} & \num{2.125e-01} & \num{1.517e-01} & \num{3.169e-01} & $      0.7$ & $     16.0$ & \num{7.089e-01} & \num{1.710e-01} & \num{1.201e-01} & $        1$&        --- &        ---  \\
\texttt{TCIA02\_118} &           12.52 & \num{4.055e-02} & \num{3.115e-01} & \num{1.892e-01} & \num{4.255e-01} & $      3.7$ & $      8.0$ & \num{6.951e-01} & \num{1.626e-01} & \num{1.423e-01} & $        1$& $      47$ & $      104$  \\
\texttt{TCIA02\_135} &           10.62 & \num{2.726e-02} & \num{2.380e-01} & \num{1.654e-01} & \num{5.421e-01} & $      3.1$ & $      8.0$ & \num{5.699e-01} & \num{3.191e-01} & \num{1.110e-01} & $        1$& $      69$ & $      828$  \\
\texttt{TCIA02\_151} &           12.33 & \num{1.466e-02} & \num{2.001e-01} & \num{1.029e-01} & \num{4.158e-01} & $      3.3$ & $     24.0$ & \num{6.770e-01} & \num{2.499e-01} & \num{7.308e-02} & $        1$& $      47$ & $     1731$  \\
\texttt{TCIA02\_168} &           13.41 & \num{2.202e-02} & \num{2.414e-01} & \num{1.236e-01} & \num{6.733e-01} & $      6.3$ & $     26.5$ & \num{5.451e-01} & \num{3.850e-01} & \num{6.987e-02} & $        1$&        --- &        ---  \\
\bottomrule
\end{tabular}
\end{table*}
\begin{table*}
\ContinuedFloat
\centering\scriptsize\setlength\tabcolsep{5.5pt}
\begin{tabular}{lLLlLlcclllllcc}
    \toprule
    BraTS ID                     & $\rho_w$        & $\kappa_w$      & $\ellTwoObsMiss$ & $\scaledEllTwoDice$ &  $\ellTwoMiss$  & $\distCM$ & $\distPi$ & $\int_{TC}c_1$ & $\int_{ED}c_1$ & $\int_{B\setminus WT}c_1$  & $n_c$ & $age\,[y]$ & $srvl\,[d]$  \\
    \midrule
\texttt{TCIA02\_171} &            6.47 & \num{4.861e-03} & \num{1.827e-01} & \num{1.742e-01} & \num{5.783e-01} & $      0.3$ & $      8.0$ & \num{4.810e-01} & \num{5.012e-01} & \num{1.783e-02} & $        2$&        --- &        ---  \\
\texttt{TCIA02\_179} &            7.86 & \num{8.612e-03} & \num{1.795e-01} & \num{1.376e-01} & \num{4.166e-01} & $      0.8$ & $     17.9$ & \num{5.880e-01} & \num{3.514e-01} & \num{6.062e-02} & $        1$& $      46$ & $      405$  \\
\texttt{TCIA02\_198} &            9.56 & \num{4.448e-03} & \num{1.963e-01} & \num{1.709e-01} & \num{4.176e-01} & $     11.4$ & $     19.6$ & \num{6.412e-01} & \num{3.146e-01} & \num{4.424e-02} & $        1$& $      54$ & $      394$  \\
\texttt{TCIA02\_208} &            6.79 & \num{5.243e-03} & \num{2.248e-01} & \num{1.616e-01} & \num{5.425e-01} & $      0.8$ & $      9.8$ & \num{5.196e-01} & \num{3.920e-01} & \num{8.836e-02} & $        1$&        --- &        ---  \\
\texttt{TCIA02\_222} &           11.22 & \num{2.999e-02} & \num{3.080e-01} & \num{2.097e-01} & \num{4.961e-01} & $      1.7$ & $      8.0$ & \num{5.931e-01} & \num{2.896e-01} & \num{1.173e-01} & $        1$&        --- &        ---  \\
\texttt{TCIA02\_226} &            7.92 & \num{5.153e-03} & \num{2.174e-01} & \num{1.681e-01} & \num{4.513e-01} & $      2.4$ & $     11.3$ & \num{5.980e-01} & \num{3.512e-01} & \num{5.083e-02} & $        1$& $      73$ & $      329$  \\
\texttt{TCIA02\_274} &           12.05 & \num{1.598e-02} & \num{3.046e-01} & \num{2.590e-01} & \num{4.681e-01} & $      2.6$ & $     16.0$ & \num{6.903e-01} & \num{2.277e-01} & \num{8.196e-02} & $        1$& $      54$ & $      357$  \\
\texttt{TCIA02\_283} &            7.04 & \num{9.443e-03} & \num{3.768e-01} & \num{1.896e-01} & \num{4.812e-01} & $      2.3$ & $     18.3$ & \num{5.227e-01} & \num{1.957e-01} & \num{2.816e-01} & $        1$& $      74$ & $      262$  \\
\texttt{TCIA02\_290} &           12.11 & \num{1.257e-02} & \num{2.403e-01} & \num{1.656e-01} & \num{3.687e-01} & $      7.3$ & $     24.0$ & \num{7.389e-01} & \num{1.561e-01} & \num{1.050e-01} & $        2$& $      43$ & $      737$  \\
\texttt{TCIA02\_300} &           16.42 & \num{1.386e-02} & \num{2.054e-01} & \num{1.357e-01} & \num{2.976e-01} & $      7.8$ & $      8.0$ & \num{8.247e-01} & \num{1.060e-01} & \num{6.926e-02} & $        1$& $      64$ & $      127$  \\
\texttt{TCIA02\_309} &           10.36 & \num{1.212e-02} & \num{2.341e-01} & \num{1.445e-01} & \num{4.262e-01} & $      4.4$ & $     25.3$ & \num{6.353e-01} & \num{2.328e-01} & \num{1.320e-01} & $        3$&        --- &        ---  \\
\texttt{TCIA02\_314} &            9.71 & \num{1.069e-02} & \num{2.972e-01} & \num{2.617e-01} & \num{3.513e-01} & $      4.1$ & $     49.3$ & \num{7.359e-01} & \num{1.600e-01} & \num{1.041e-01} & $        1$& $      40$ & $      362$  \\
\texttt{TCIA02\_321} &           13.46 & \num{2.161e-02} & \num{2.647e-01} & \num{1.884e-01} & \num{3.531e-01} & $      4.7$ & $      8.0$ & \num{7.514e-01} & \num{1.253e-01} & \num{1.234e-01} & $        1$& $      81$ & $       67$  \\
\texttt{TCIA02\_322} &            9.06 & \num{5.196e-03} & \num{2.255e-01} & \num{1.697e-01} & \num{3.919e-01} & $      3.1$ & $     20.0$ & \num{6.766e-01} & \num{2.407e-01} & \num{8.268e-02} & $        1$& $      57$ & $      621$  \\
\texttt{TCIA02\_331} &            8.95 & \num{8.174e-03} & \num{2.701e-01} & \num{1.884e-01} & \num{3.868e-01} & $      1.0$ & $     25.3$ & \num{6.928e-01} & \num{1.722e-01} & \num{1.351e-01} & $        1$& $      84$ & $      187$  \\
\texttt{TCIA02\_368} &           11.71 & \num{2.483e-02} & \num{2.762e-01} & \num{1.674e-01} & \num{6.624e-01} & $      5.6$ & $     17.9$ & \num{5.131e-01} & \num{4.140e-01} & \num{7.288e-02} & $        1$& $      62$ & $      317$  \\
\texttt{TCIA02\_370} &           11.20 & \num{2.674e-02} & \num{2.688e-01} & \num{1.869e-01} & \num{5.078e-01} & $      3.2$ & $     25.3$ & \num{6.310e-01} & \num{2.776e-01} & \num{9.145e-02} & $        1$&        --- &        ---  \\
\texttt{TCIA02\_374} &           15.69 & \num{1.189e-02} & \num{1.554e-01} & \num{1.343e-01} & \num{3.338e-01} & $      5.2$ & $     11.3$ & \num{7.700e-01} & \num{2.003e-01} & \num{2.971e-02} & $        1$&        --- &        ---  \\
\texttt{TCIA02\_377} &           12.39 & \num{1.581e-02} & \num{2.075e-01} & \num{1.773e-01} & \num{4.189e-01} & $      6.1$ & $     19.6$ & \num{7.214e-01} & \num{2.185e-01} & \num{6.016e-02} & $        1$& $      63$ & $      812$  \\
\texttt{TCIA02\_394} &           12.68 & \num{1.240e-02} & \num{2.056e-01} & \num{1.780e-01} & \num{2.648e-01} & $      3.9$ & $      8.0$ & \num{8.071e-01} & \num{1.047e-01} & \num{8.817e-02} & $        1$& $      64$ & $      616$  \\
\texttt{TCIA02\_430} &           11.61 & \num{2.226e-02} & \num{2.034e-01} & \num{1.505e-01} & \num{3.206e-01} & $      5.0$ & $      8.0$ & \num{7.458e-01} & \num{1.769e-01} & \num{7.724e-02} & $        1$& $      53$ & $       71$  \\
\texttt{TCIA02\_455} &           10.79 & \num{1.393e-02} & \num{1.652e-01} & \num{1.278e-01} & \num{4.479e-01} & $      8.0$ & $     19.6$ & \num{6.311e-01} & \num{3.144e-01} & \num{5.451e-02} & $        1$& $      54$ & $      424$  \\
\texttt{TCIA02\_471} &           12.87 & \num{2.540e-02} & \num{3.012e-01} & \num{1.650e-01} & \num{4.643e-01} & $     18.1$ & $     25.3$ & \num{6.532e-01} & \num{2.136e-01} & \num{1.332e-01} & $        1$& $      76$ & $      111$  \\
\texttt{TCIA02\_473} &            6.58 & \num{6.496e-03} & \num{2.741e-01} & \num{1.741e-01} & \num{4.444e-01} & $      2.4$ & $     16.0$ & \num{5.466e-01} & \num{3.047e-01} & \num{1.488e-01} & $        1$& $      61$ & $      175$  \\
\texttt{TCIA02\_491} &           13.75 & \num{1.522e-02} & \num{2.425e-01} & \num{2.067e-01} & \num{2.977e-01} & $     10.4$ & $     25.3$ & \num{8.301e-01} & \num{9.948e-02} & \num{7.045e-02} & $        1$& $      81$ & $       82$  \\
\texttt{TCIA02\_605} &           10.53 & \num{1.688e-02} & \num{3.568e-01} & \num{2.744e-01} & \num{5.671e-01} & $      6.4$ & $     22.6$ & \num{5.335e-01} & \num{3.459e-01} & \num{1.206e-01} & $        1$&        --- &        ---  \\
\texttt{TCIA02\_606} &           11.64 & \num{2.683e-02} & \num{3.226e-01} & \num{1.741e-01} & \num{5.059e-01} & $      5.4$ & $     40.0$ & \num{6.206e-01} & \num{2.276e-01} & \num{1.517e-01} & $        1$&        --- &        ---  \\
\texttt{TCIA02\_607} &           13.49 & \num{1.248e-02} & \num{2.675e-01} & \num{1.964e-01} & \num{3.684e-01} & $      6.2$ & $     17.9$ & \num{7.542e-01} & \num{1.353e-01} & \num{1.105e-01} & $        1$&        --- &        ---  \\
\texttt{TCIA02\_608} &           12.12 & \num{1.903e-02} & \num{3.375e-01} & \num{1.745e-01} & \num{6.132e-01} & $      4.8$ & $     19.6$ & \num{5.793e-01} & \num{3.004e-01} & \num{1.203e-01} & $        1$&        --- &        ---  \\
\texttt{TCIA03\_121} &           13.32 & \num{1.165e-02} & \num{2.087e-01} & \num{1.595e-01} & \num{3.680e-01} & $     12.8$ & $     24.0$ & \num{7.347e-01} & \num{1.873e-01} & \num{7.798e-02} & $        1$& $      30$ & $      747$  \\
\texttt{TCIA03\_133} &            8.27 & \num{5.313e-03} & \num{2.704e-01} & \num{2.053e-01} & \num{4.344e-01} & $      2.8$ & $     17.9$ & \num{6.348e-01} & \num{2.711e-01} & \num{9.417e-02} & $        1$& $      63$ & $      382$  \\
\texttt{TCIA03\_138} &           12.63 & \num{3.431e-02} & \num{3.106e-01} & \num{1.951e-01} & \num{3.964e-01} & $     12.2$ & $     17.9$ & \num{7.285e-01} & \num{1.249e-01} & \num{1.465e-01} & $        1$& $      71$ & $       82$  \\
\texttt{TCIA03\_199} &           11.40 & \num{1.294e-02} & \num{1.702e-01} & \num{1.185e-01} & \num{5.590e-01} & $      2.1$ & $     19.6$ & \num{5.742e-01} & \num{3.818e-01} & \num{4.401e-02} & $        1$& $      48$ & $     1282$  \\
\texttt{TCIA03\_257} &            9.25 & \num{1.113e-02} & \num{2.524e-01} & \num{2.190e-01} & \num{4.890e-01} & $      4.2$ & $     17.9$ & \num{6.128e-01} & \num{3.252e-01} & \num{6.209e-02} & $        2$& $      69$ & $      425$  \\
\texttt{TCIA03\_265} &           13.81 & \num{1.050e-02} & \num{2.247e-01} & \num{1.603e-01} & \num{3.674e-01} & $      2.2$ & $      8.0$ & \num{7.456e-01} & \num{1.888e-01} & \num{6.553e-02} & $        1$& $      59$ & $      103$  \\
\texttt{TCIA03\_296} &           11.69 & \num{2.764e-02} & \num{2.972e-01} & \num{1.800e-01} & \num{5.061e-01} & $      7.2$ & $     19.6$ & \num{6.241e-01} & \num{2.457e-01} & \num{1.302e-01} & $        1$& $      60$ & $       22$  \\
\texttt{TCIA03\_338} &            8.46 & \num{1.007e-02} & \num{3.030e-01} & \num{2.221e-01} & \num{4.514e-01} & $      2.8$ & $     17.4$ & \num{6.158e-01} & \num{2.635e-01} & \num{1.206e-01} & $        1$& $      76$ & $      468$  \\
\texttt{TCIA03\_375} &           12.20 & \num{2.662e-02} & \num{2.487e-01} & \num{1.517e-01} & \num{3.343e-01} & $      5.2$ & $     11.3$ & \num{7.352e-01} & \num{1.309e-01} & \num{1.339e-01} & $        1$& $      60$ & $      946$  \\
\texttt{TCIA03\_419} &           12.20 & \num{3.746e-02} & \num{2.761e-01} & \num{1.565e-01} & \num{4.043e-01} & $      2.0$ & $     18.0$ & \num{7.023e-01} & \num{1.612e-01} & \num{1.366e-01} & $        1$& $      70$ & $      327$  \\
\texttt{TCIA03\_474} &           13.07 & \num{2.457e-02} & \num{2.630e-01} & \num{2.289e-01} & \num{3.464e-01} & $      3.7$ & $     25.3$ & \num{7.923e-01} & \num{1.345e-01} & \num{7.325e-02} & $        1$& $      61$ & $      635$  \\
\texttt{TCIA03\_498} &           11.10 & \num{8.243e-03} & \num{1.799e-01} & \num{1.706e-01} & \num{3.847e-01} & $      2.1$ & $     16.0$ & \num{6.906e-01} & \num{2.880e-01} & \num{2.144e-02} & $        1$& $      59$ & $      467$  \\
\texttt{TCIA04\_111} &            9.15 & \num{7.361e-03} & \num{1.757e-01} & \num{1.497e-01} & \num{4.941e-01} & $      0.8$ & $     19.6$ & \num{5.907e-01} & \num{3.663e-01} & \num{4.299e-02} & $        1$& $      75$ & $      626$  \\
\texttt{TCIA04\_149} &           12.59 & \num{1.287e-02} & \num{2.662e-01} & \num{1.474e-01} & \num{3.982e-01} & $      5.6$ & $     19.6$ & \num{7.065e-01} & \num{1.637e-01} & \num{1.298e-01} & $        1$& $      36$ & $      448$  \\
\texttt{TCIA04\_192} &            9.60 & \num{5.759e-03} & \num{2.195e-01} & \num{1.942e-01} & \num{3.559e-01} & $      9.1$ & $     25.3$ & \num{7.077e-01} & \num{2.432e-01} & \num{4.908e-02} & $        1$& $      75$ & $      121$  \\
\texttt{TCIA04\_328} &            8.84 & \num{9.029e-03} & \num{2.315e-01} & \num{1.830e-01} & \num{5.564e-01} & $      0.8$ & $     16.0$ & \num{5.252e-01} & \num{3.990e-01} & \num{7.578e-02} & $        1$&        --- &        ---  \\
\texttt{TCIA04\_343} &            8.32 & \num{5.101e-03} & \num{1.997e-01} & \num{1.736e-01} & \num{4.687e-01} & $      3.6$ & $      8.0$ & \num{5.951e-01} & \num{3.442e-01} & \num{6.074e-02} & $        1$& $      52$ & $      296$  \\
\texttt{TCIA04\_361} &           14.39 & \num{1.266e-02} & \num{1.903e-01} & \num{1.557e-01} & \num{5.136e-01} & $      4.8$ & $     11.3$ & \num{6.501e-01} & \num{3.096e-01} & \num{4.032e-02} & $        1$& $      75$ & $      476$  \\
\texttt{TCIA04\_437} &            6.04 & \num{2.512e-03} & \num{1.534e-01} & \num{1.332e-01} & \num{3.653e-01} & $      0.5$ & $     12.0$ & \num{5.860e-01} & \num{3.646e-01} & \num{4.947e-02} & $        1$& $      46$ & $      333$  \\
\texttt{TCIA04\_479} &           11.50 & \num{1.347e-02} & \num{1.978e-01} & \num{1.477e-01} & \num{4.562e-01} & $      1.8$ & $     16.0$ & \num{6.552e-01} & \num{2.858e-01} & \num{5.895e-02} & $        1$& $      56$ & $      372$  \\
\texttt{TCIA05\_277} &           10.94 & \num{2.485e-02} & \num{2.776e-01} & \num{1.636e-01} & \num{5.631e-01} & $      8.1$ & $     51.2$ & \num{5.565e-01} & \num{3.347e-01} & \num{1.089e-01} & $        1$& $      70$ & $      232$  \\
\texttt{TCIA05\_396} &            9.72 & \num{1.371e-02} & \num{1.919e-01} & \num{1.558e-01} & \num{7.926e-01} & $      2.3$ & $     22.6$ & \num{4.251e-01} & \num{5.473e-01} & \num{2.763e-02} & $        1$&        --- &        ---  \\
\texttt{TCIA05\_444} &           14.12 & \num{2.767e-02} & \num{3.337e-01} & \num{1.492e-01} & \num{5.251e-01} & $      6.8$ & $     11.3$ & \num{6.544e-01} & \num{2.052e-01} & \num{1.403e-01} & $        1$&        --- &        ---  \\
\texttt{TCIA05\_478} &           12.47 & \num{2.746e-02} & \num{2.887e-01} & \num{2.528e-01} & \num{4.043e-01} & $      3.7$ & $     17.9$ & \num{7.386e-01} & \num{1.899e-01} & \num{7.143e-02} & $        1$& $      59$ & $       30$  \\
\texttt{TCIA06\_165} &           11.07 & \num{2.287e-02} & \num{2.808e-01} & \num{2.272e-01} & \num{6.062e-01} & $      2.9$ & $     11.3$ & \num{5.600e-01} & \num{3.675e-01} & \num{7.243e-02} & $        1$& $      51$ & $        5$  \\
\texttt{TCIA06\_184} &           13.50 & \num{2.956e-02} & \num{3.168e-01} & \num{1.856e-01} & \num{4.903e-01} & $      6.2$ & $     17.9$ & \num{6.720e-01} & \num{2.055e-01} & \num{1.225e-01} & $        1$& $      61$ & $      434$  \\
\texttt{TCIA06\_211} &           12.56 & \num{1.440e-02} & \num{3.557e-01} & \num{3.326e-01} & \num{4.346e-01} & $      8.8$ & $     54.3$ & \num{7.828e-01} & \num{1.544e-01} & \num{6.280e-02} & $        1$&        --- &        ---  \\
\texttt{TCIA06\_247} &           11.76 & \num{1.625e-02} & \num{2.738e-01} & \num{1.574e-01} & \num{4.054e-01} & $      7.4$ & $     17.9$ & \num{6.819e-01} & \num{1.568e-01} & \num{1.613e-01} & $        1$& $      76$ & $      244$  \\
\texttt{TCIA06\_332} &            9.42 & \num{5.333e-03} & \num{2.489e-01} & \num{2.281e-01} & \num{3.287e-01} & $      4.0$ & $     15.7$ & \num{7.624e-01} & \num{1.575e-01} & \num{8.009e-02} & $        1$&        --- &        ---  \\
\texttt{TCIA06\_372} &           10.80 & \num{2.410e-02} & \num{2.964e-01} & \num{2.353e-01} & \num{4.856e-01} & $      5.2$ & $     46.0$ & \num{6.598e-01} & \num{2.496e-01} & \num{9.062e-02} & $        1$& $      74$ & $      213$  \\
\texttt{TCIA06\_409} &            8.71 & \num{1.190e-02} & \num{2.760e-01} & \num{1.912e-01} & \num{5.791e-01} & $      1.7$ & $     13.9$ & \num{5.272e-01} & \num{3.744e-01} & \num{9.839e-02} & $        1$& $      69$ & $       99$  \\
\texttt{TCIA06\_603} &           11.43 & \num{2.267e-02} & \num{2.841e-01} & \num{1.657e-01} & \num{3.719e-01} & $      9.6$ & $      8.0$ & \num{7.150e-01} & \num{1.548e-01} & \num{1.302e-01} & $        1$&        --- &        ---  \\
\texttt{TCIA08\_105} &           11.99 & \num{3.972e-02} & \num{2.919e-01} & \num{1.972e-01} & \num{4.621e-01} & $      5.5$ & $     16.0$ & \num{6.785e-01} & \num{2.085e-01} & \num{1.130e-01} & $        1$& $      66$ & $       77$  \\
\texttt{TCIA08\_113} &            8.97 & \num{1.044e-02} & \num{2.400e-01} & \num{2.030e-01} & \num{5.878e-01} & $      1.7$ & $     20.4$ & \num{4.955e-01} & \num{4.319e-01} & \num{7.260e-02} & $        1$&        --- &        ---  \\
\texttt{TCIA08\_162} &           12.77 & \num{9.774e-03} & \num{3.272e-01} & \num{3.214e-01} & \num{3.393e-01} & $      2.0$ & $     25.3$ & \num{9.049e-01} & \num{5.179e-02} & \num{4.331e-02} & $        1$&        --- &        ---  \\
\texttt{TCIA08\_167} &           13.54 & \num{2.131e-02} & \num{2.641e-01} & \num{2.408e-01} & \num{2.984e-01} & $      5.1$ & $      8.0$ & \num{8.474e-01} & \num{9.358e-02} & \num{5.901e-02} & $        1$& $      74$ & $      153$  \\
\texttt{TCIA08\_205} &           16.55 & \num{1.809e-02} & \num{2.030e-01} & \num{1.407e-01} & \num{3.009e-01} & $      6.8$ & $      8.0$ & \num{8.184e-01} & \num{1.109e-01} & \num{7.063e-02} & $        1$&        --- &        ---  \\
\texttt{TCIA08\_218} &           11.55 & \num{1.230e-02} & \num{1.931e-01} & \num{1.289e-01} & \num{6.264e-01} & $      5.6$ & $     16.0$ & \num{5.453e-01} & \num{3.968e-01} & \num{5.784e-02} & $        1$& $      57$ & $      346$  \\
\texttt{TCIA08\_234} &            8.98 & \num{6.871e-03} & \num{1.837e-01} & \num{1.494e-01} & \num{5.266e-01} & $      4.5$ & $     16.5$ & \num{5.642e-01} & \num{3.940e-01} & \num{4.175e-02} & $        1$&        --- &        ---  \\
\texttt{TCIA08\_242} &            9.42 & \num{5.333e-03} & \num{2.489e-01} & \num{2.281e-01} & \num{3.287e-01} & $      4.0$ & $     15.7$ & \num{7.624e-01} & \num{1.575e-01} & \num{8.009e-02} & $        1$& $      66$ & $      147$  \\
\texttt{TCIA08\_278} &           13.50 & \num{1.510e-02} & \num{2.809e-01} & \num{1.232e-01} & \num{4.167e-01} & $     16.9$ & $     41.6$ & \num{7.056e-01} & \num{1.547e-01} & \num{1.397e-01} & $        1$& $      50$ & $     1458$  \\
\texttt{TCIA08\_280} &           10.33 & \num{7.563e-03} & \num{2.026e-01} & \num{1.870e-01} & \num{3.899e-01} & $      1.6$ & $     14.4$ & \num{7.075e-01} & \num{2.619e-01} & \num{3.064e-02} & $        1$& $      57$ & $      508$  \\
\texttt{TCIA08\_319} &           10.99 & \num{1.020e-02} & \num{2.342e-01} & \num{1.699e-01} & \num{3.816e-01} & $      3.3$ & $      8.0$ & \num{7.026e-01} & \num{2.146e-01} & \num{8.275e-02} & $        1$& $      64$ & $      254$  \\
\texttt{TCIA08\_406} &            5.14 & \num{4.175e-03} & \num{2.204e-01} & \num{1.834e-01} & \num{5.659e-01} & $      1.8$ & $      4.0$ & \num{4.324e-01} & \num{4.983e-01} & \num{6.927e-02} & $        2$& $      78$ & $       82$  \\
\texttt{TCIA08\_436} &           15.55 & \num{1.376e-02} & \num{1.308e-01} & \num{1.067e-01} & \num{3.251e-01} & $      4.4$ & $     16.0$ & \num{7.667e-01} & \num{1.998e-01} & \num{3.352e-02} & $        1$&        --- &        ---  \\
\texttt{TCIA08\_469} &           15.85 & \num{1.652e-02} & \num{1.807e-01} & \num{1.368e-01} & \num{5.284e-01} & $      5.6$ & $     11.3$ & \num{6.519e-01} & \num{3.114e-01} & \num{3.672e-02} & $        1$& $      63$ & $      519$  \\
\bottomrule
\end{tabular}
\end{table*}

\begin{figure*}
\centering
\includegraphics[width=\textwidth]{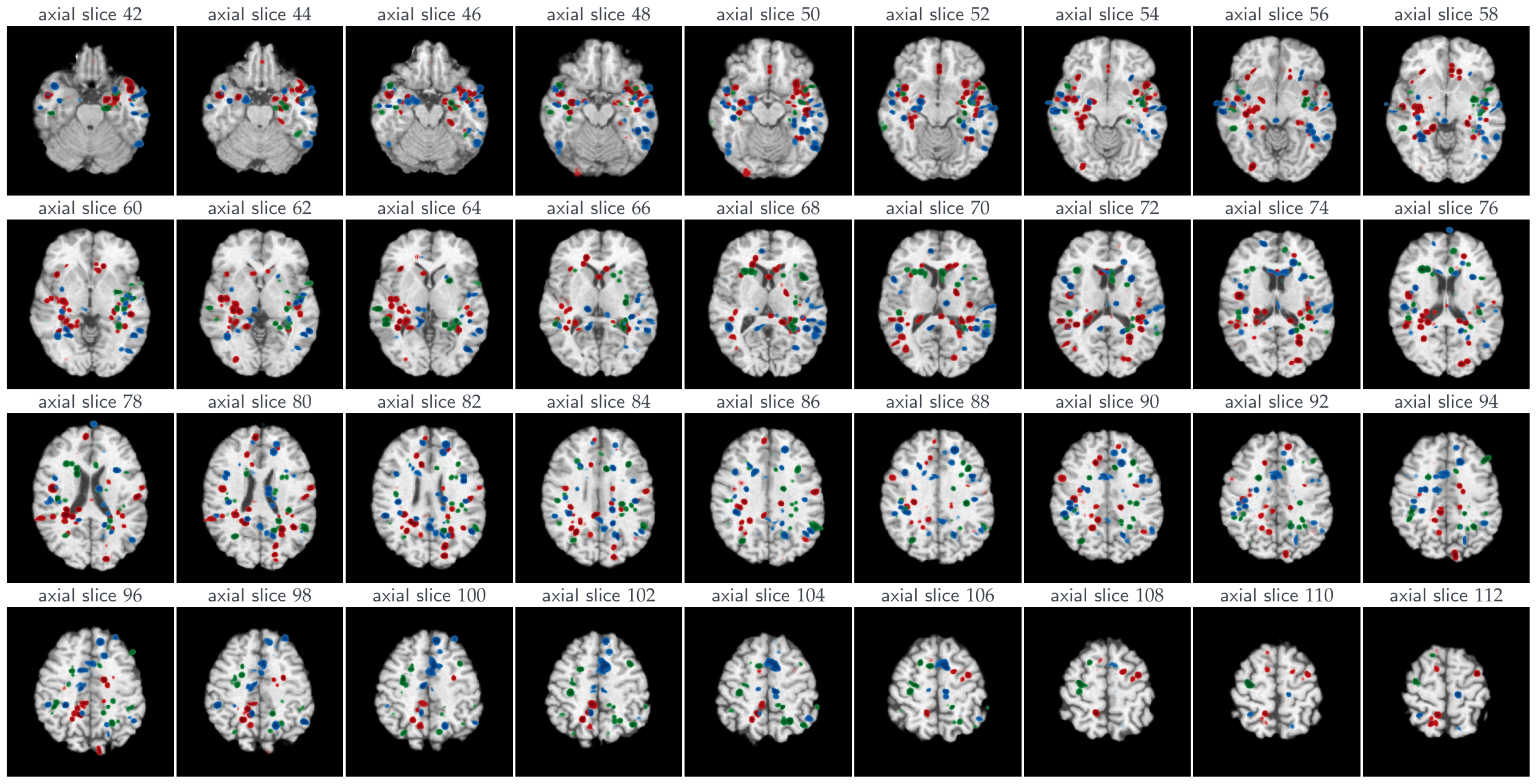}
\caption{\ipoint{Cohort GBM TIL atlas (of the reconstructed tumor initiation $c(0)$) for BratTS'18 survival data}. Shown are axial slices $42$ through $112$ of a 3D volume. Location of tumor initiation of $165$ patients are superimposed and displayed in the atlas space, and color coded by survival classes: short survivors in red; mid survivors in green; long survivors.}
\label{fig:brats18-gbm-atlas-c0-4x9}
\end{figure*}

\begin{figure*}
\centering
\includegraphics[width=\textwidth]{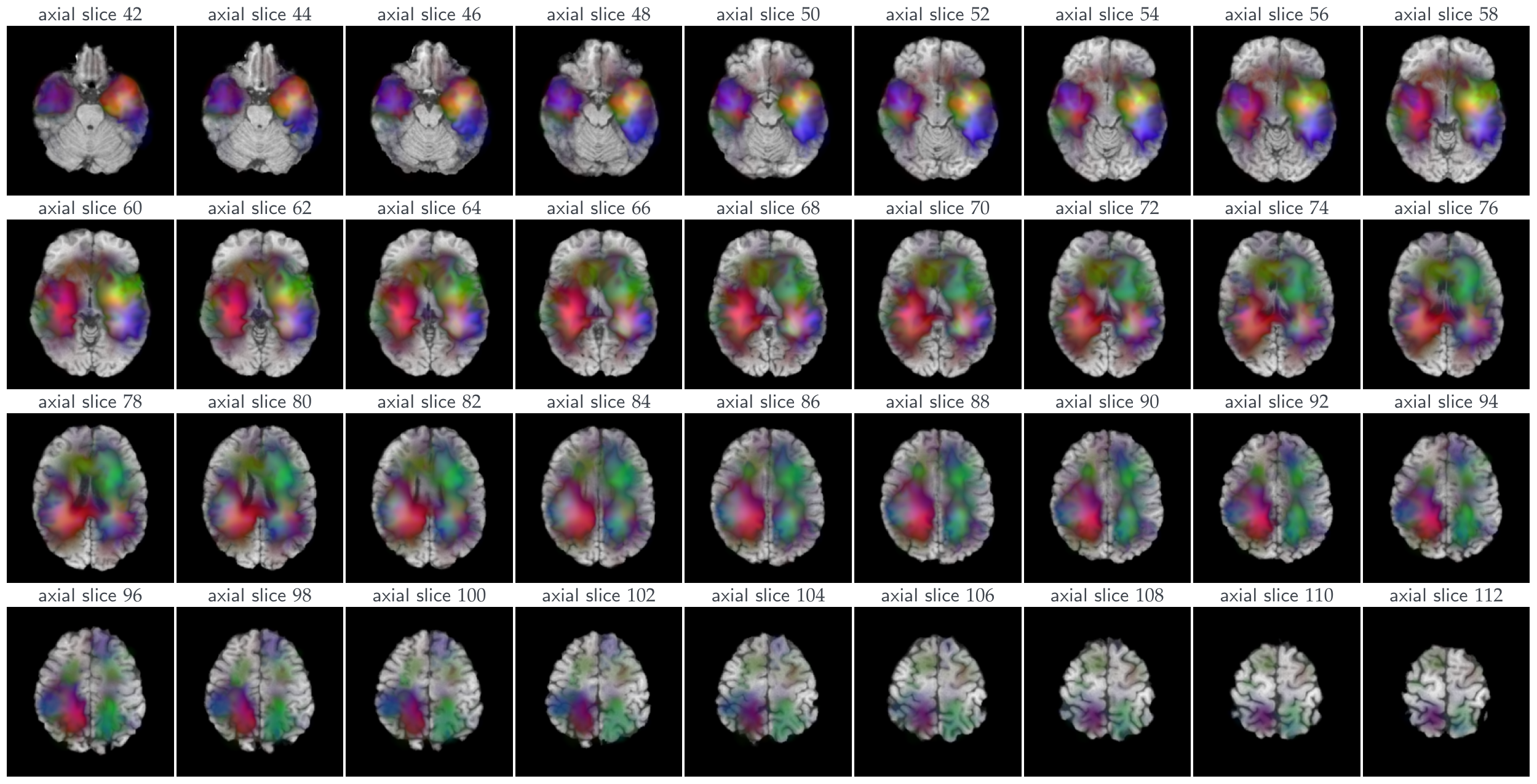}
\caption{\ipoint{Cohort GBM OT atlas (of simulated $c(1)$ tumor concentration) for the BratTS'18 survival data}. Shown are axial slices $42$ through $112$ of a 3D volume. $c(1)$ concentrations of $165$ patients are superimposed and displayed in the atlas space, and color coded by survival classes: short survivors: red channel; mid survivors: green channel; long survivors: blue channel. The channels are blended together, and displayed as RGB image.}
\label{fig:brats18-gbm-atlas-c1-4x9}
\end{figure*}

\begin{figure*}
\centering
\includegraphics[width=\textwidth]{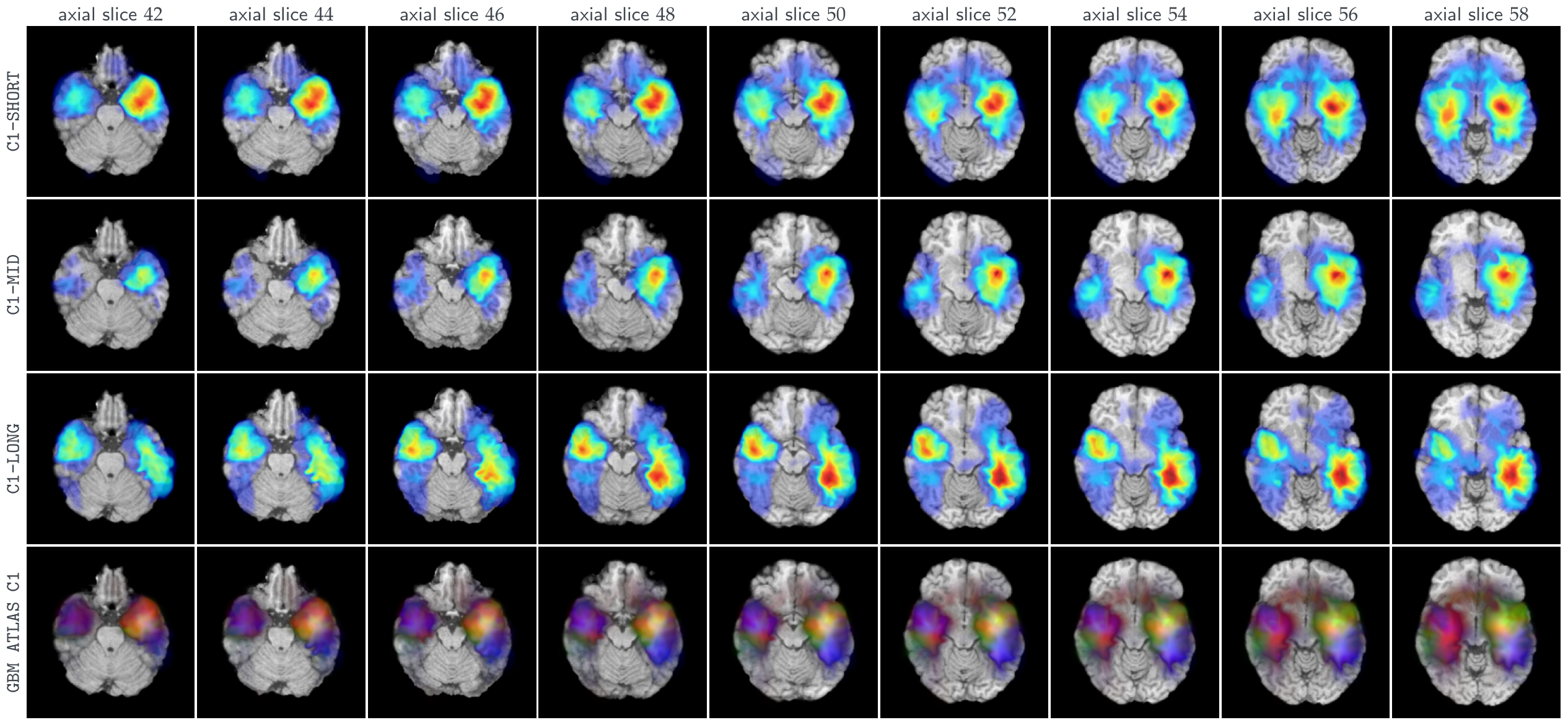}
\includegraphics[width=\textwidth]{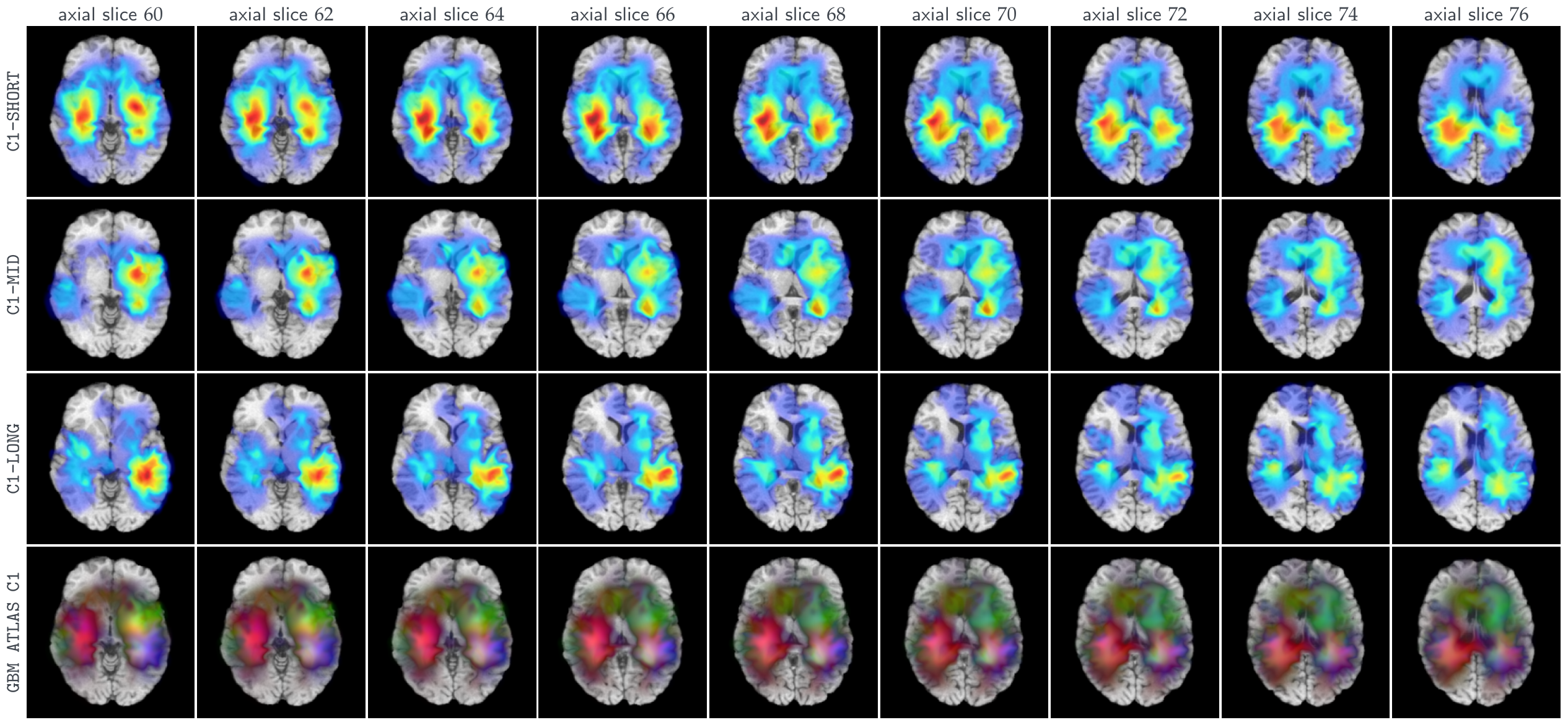}
\caption{\ipoint{Cohort GBM OT atlas (of simulated $c(1)$ tumor concentration) for BratTS'18 survival data}. Shown are axial slices $42$ through $76$ of a 3D volume. $c(1)$ concentrations of $165$ patients are superimposed in the atlas space, and color coded by survival classes: short survivors: red channel; mid survivors: green channel; long survivors: blue channel. The channels are blended together, and displayed as RGB image. Spatial distributions of $c(1)$ for each class individually are displayed in rows 1 through 3.}
\label{fig:brats18-gbm-atlas-c1-sml-1}
\end{figure*}

\begin{figure*}
\centering
\includegraphics[width=\textwidth]{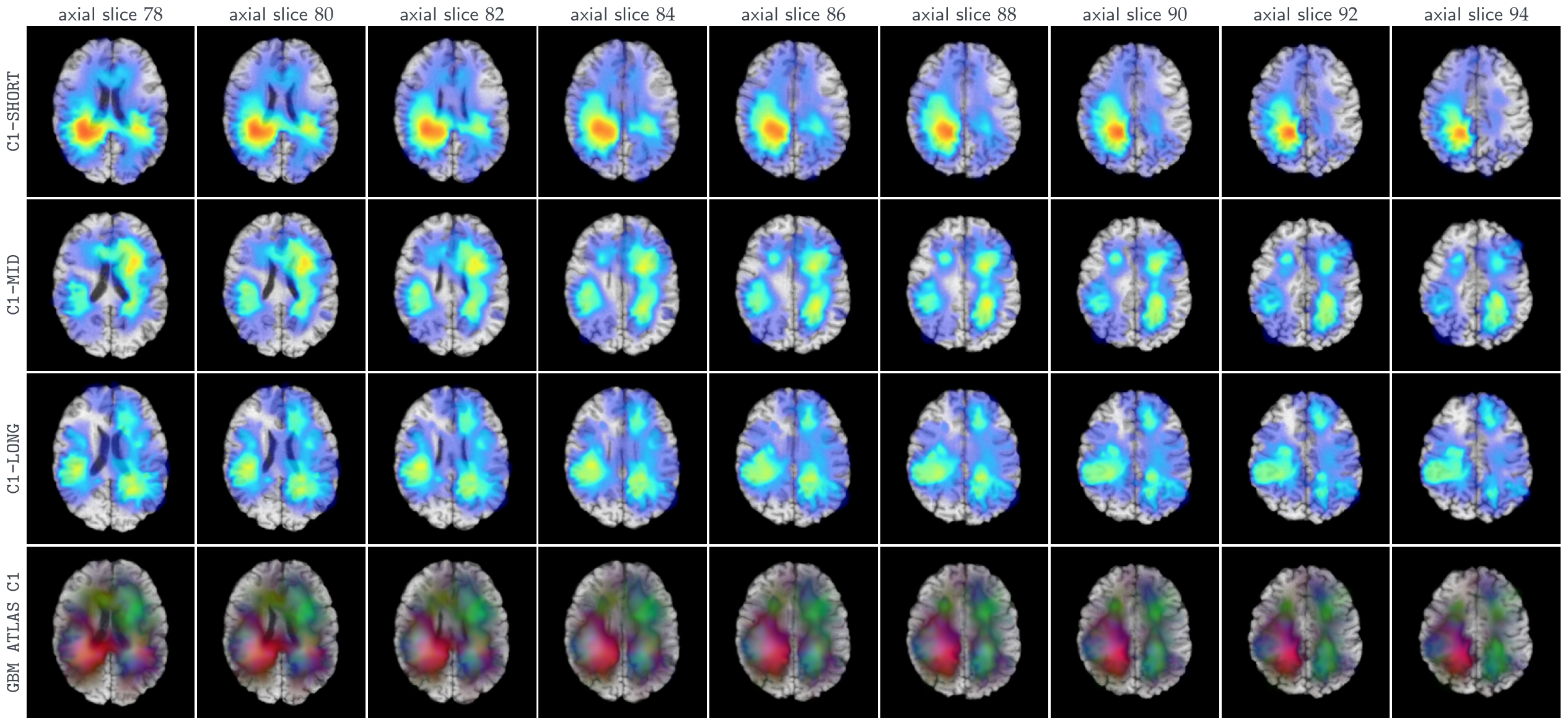}
\includegraphics[width=\textwidth]{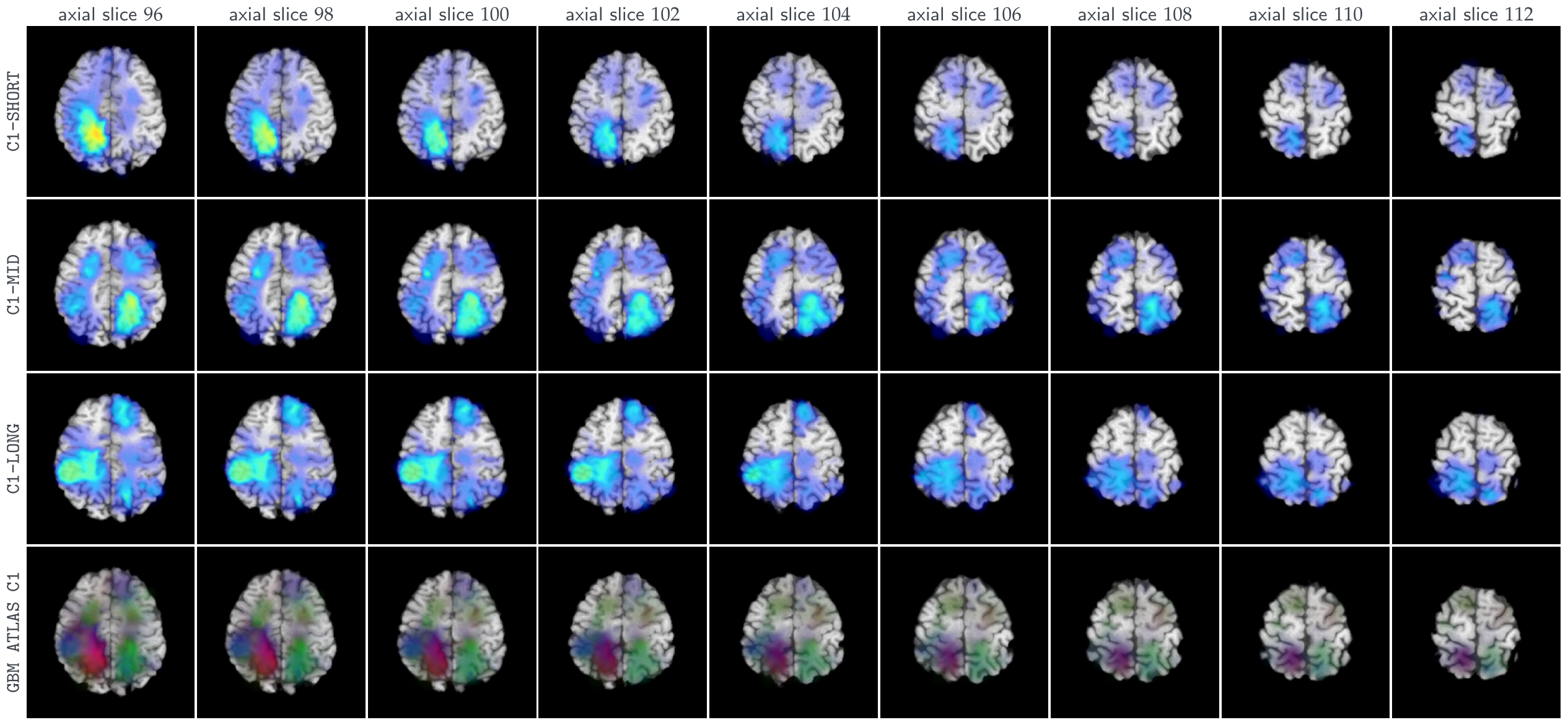}
\caption{\ipoint{Cohort GBM OT atlas (of simulated $c(1)$ tumor concentration) for BratTS'18 survival data}. Shown are axial slices $78$ through $112$ of a 3D volume. $c(1)$ concentrations of $165$ patients are superimposed in the atlas space, and color coded by survival classes: short survivors: red channel; mid survivors: green channel; long survivors: blue channel. The channels are blended together, and displayed as RGB image. Spatial distributions of $c(1)$ for each class individually are displayed in rows 1 through 3.}
\label{fig:brats18-gbm-atlas-c1-sml-2}
\end{figure*}

\begin{figure*}
\centering
\begin{minipage}{0.49\textwidth}
\includegraphics[width=1.\textwidth]{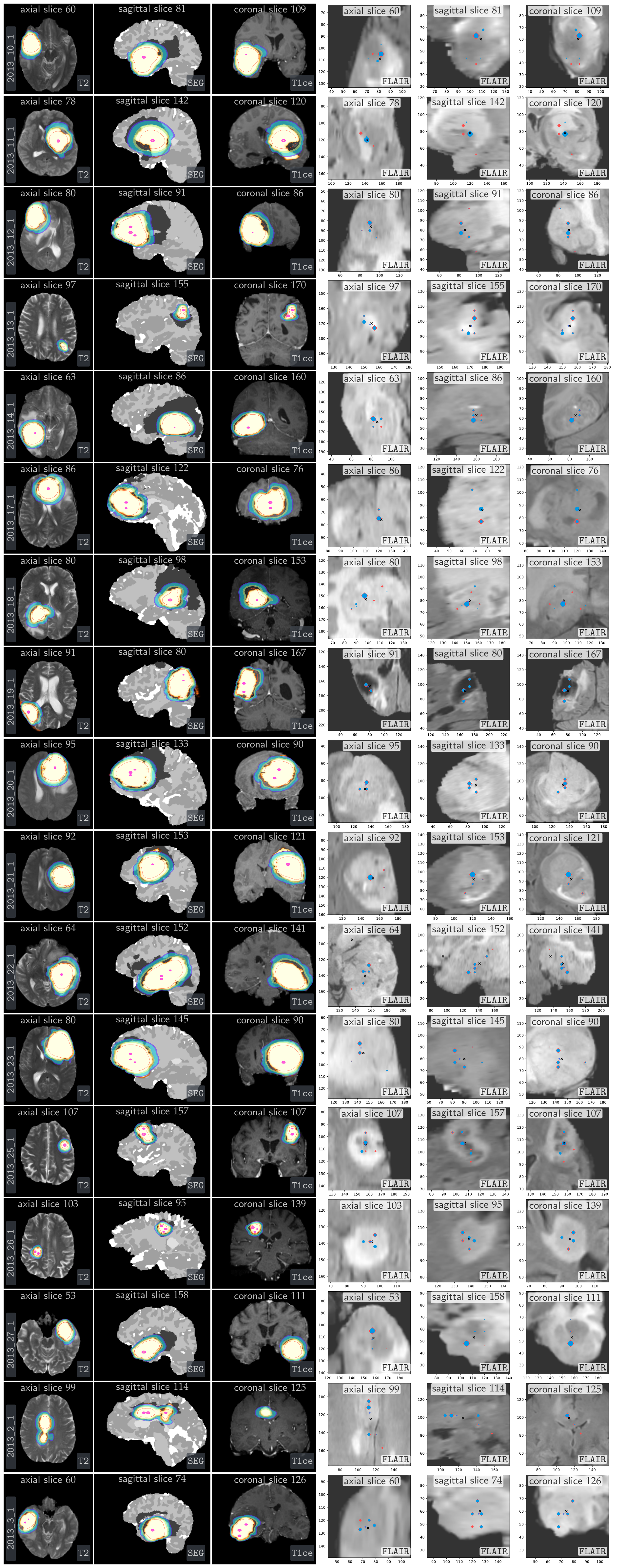}
\end{minipage}
\begin{minipage}{0.49\textwidth}
\includegraphics[width=1.\textwidth]{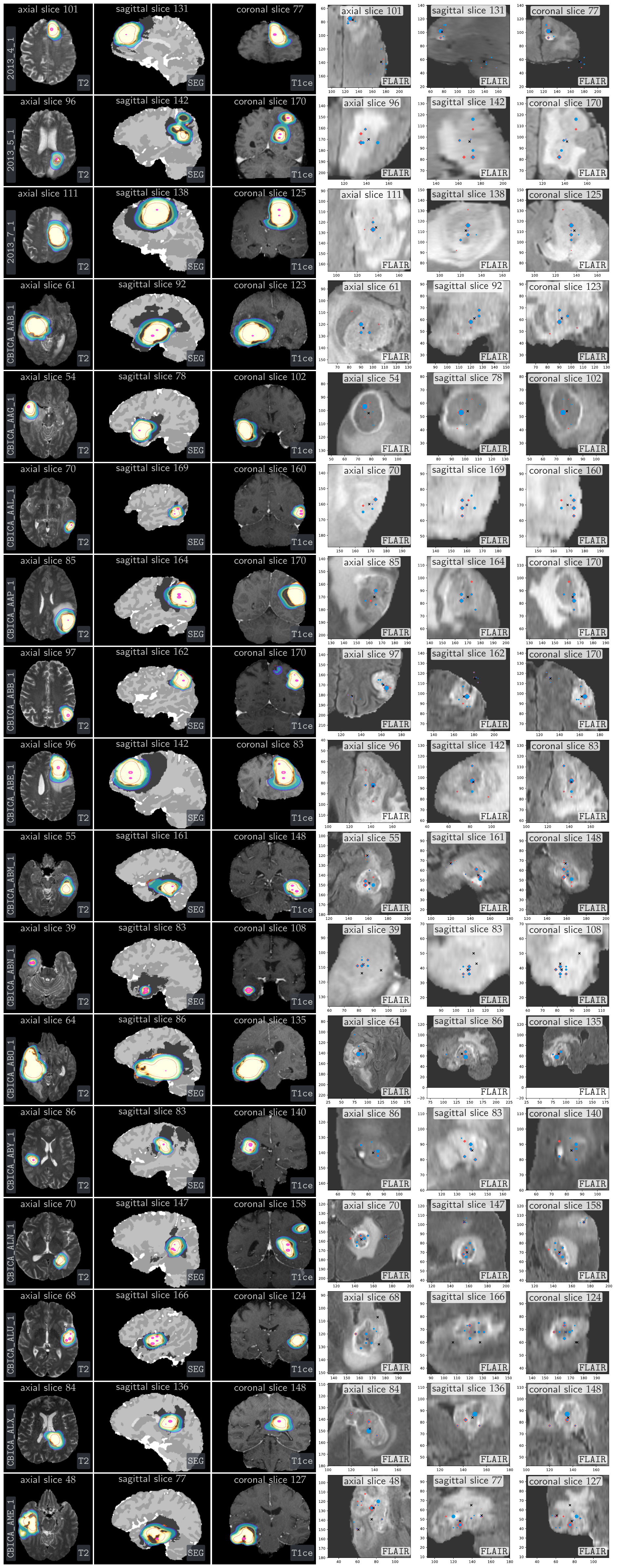}
\end{minipage}
\caption{\ipoint{Qualitative Results of BraTS'18 Cohort}. Left: Axial, coronal, and saggital cuts of the 3D patient T2-MRI volume, overlaid with the $TC$ input data (yellow), the predicted tumor $c(1)$ from our calibrated model (contours from red to violet), and the tumor origin $c(0)$ (magenta). The red $c(1)$ contour line corresponds to $0.9 \leq c(1)$, and matches well with the input $TC$ data (cf.\;\figref{fig:obs-operator} for details). Right: Projection of estimated tumor initiations $\vect{p}$ onto axial, coronal, and saggital plane (red corresp. to $128^3$ solution; blue corresp. to final $256^3$ solution). The size of the markers indicates activation weight. The center of mass of $TC$ components are indicated with a black cross.}
\label{fig:brats18-sim-1}
\end{figure*}

\begin{figure*}
\centering
\begin{minipage}{0.49\textwidth}
\includegraphics[width=1.\textwidth]{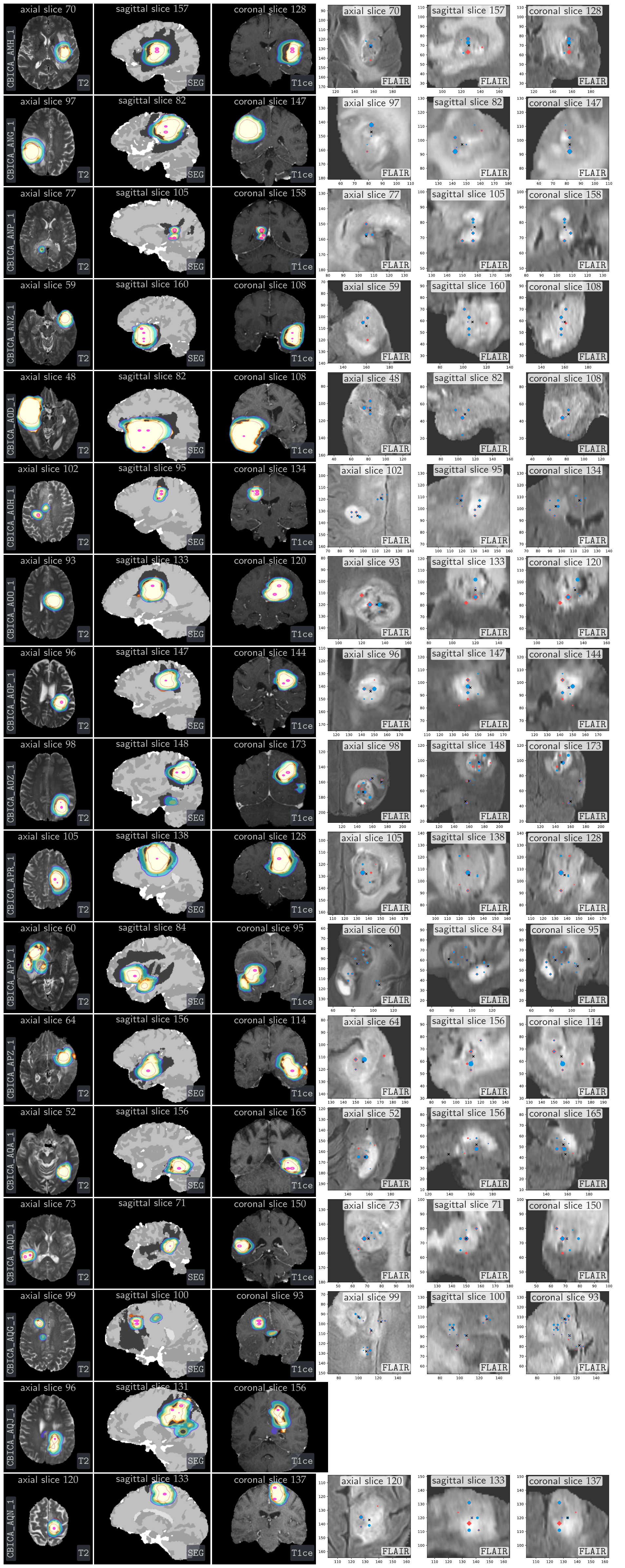}
\end{minipage}
\begin{minipage}{0.49\textwidth}
\includegraphics[width=1.\textwidth]{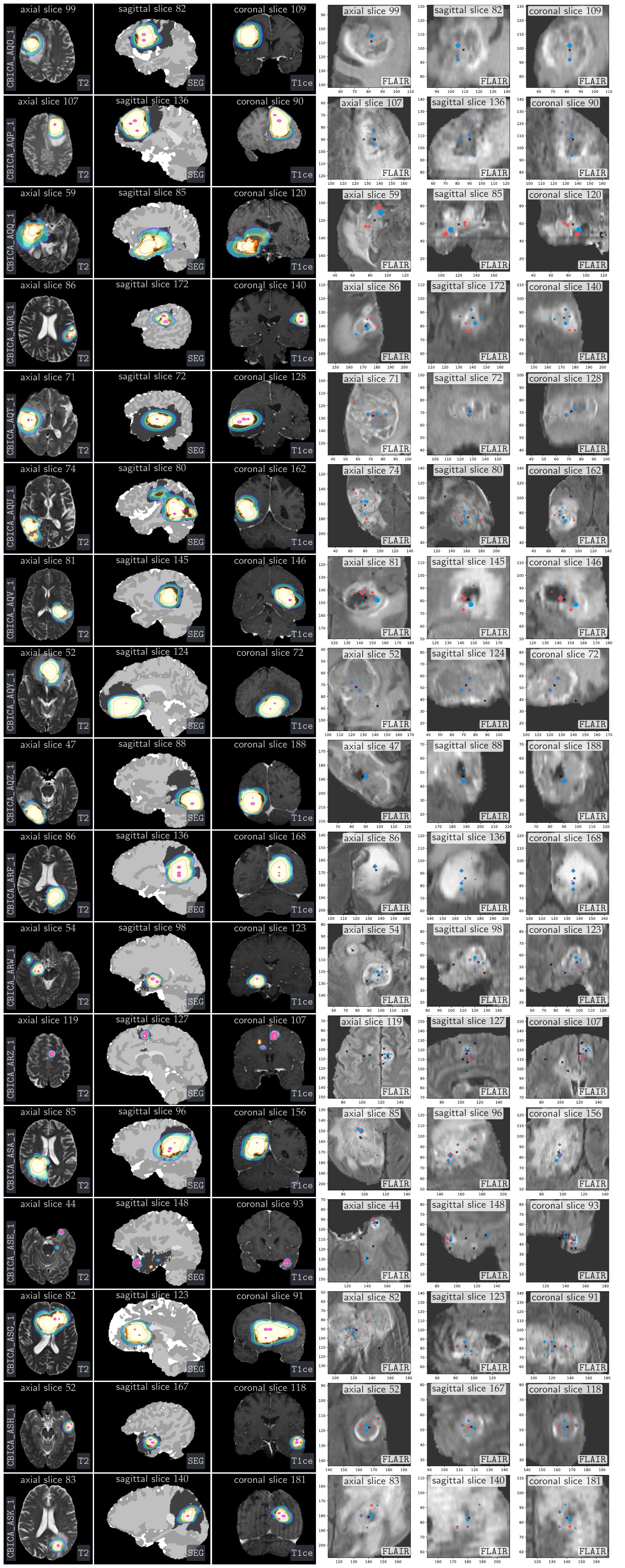}
\end{minipage}
\caption{\ipoint{Qualitative Results of BraTS'18 Cohort}. Left: Axial, coronal, and saggital cuts of the 3D patient T2-MRI volume, overlaid with the $TC$ input data (yellow), the predicted tumor $c(1)$ from our calibrated model (contours from red to violet), and the tumor origin $c(0)$ (magenta). The red $c(1)$ contour line corresponds to $0.9 \leq c(1)$, and matches well with the input $TC$ data (cf.\;\figref{fig:obs-operator} for details). Right: Projection of estimated tumor initiations $\vect{p}$ onto axial, coronal, and saggital plane (red corresp. to $128^3$ solution; blue corresp. to final $256^3$ solution). The size of the markers indicates activation weight. The center of mass of $TC$ components are indicated with a black cross.}
\label{fig:brats18-sim-2}
\end{figure*}

\begin{figure*}
\centering
\begin{minipage}{0.49\textwidth}
\includegraphics[width=1.\textwidth]{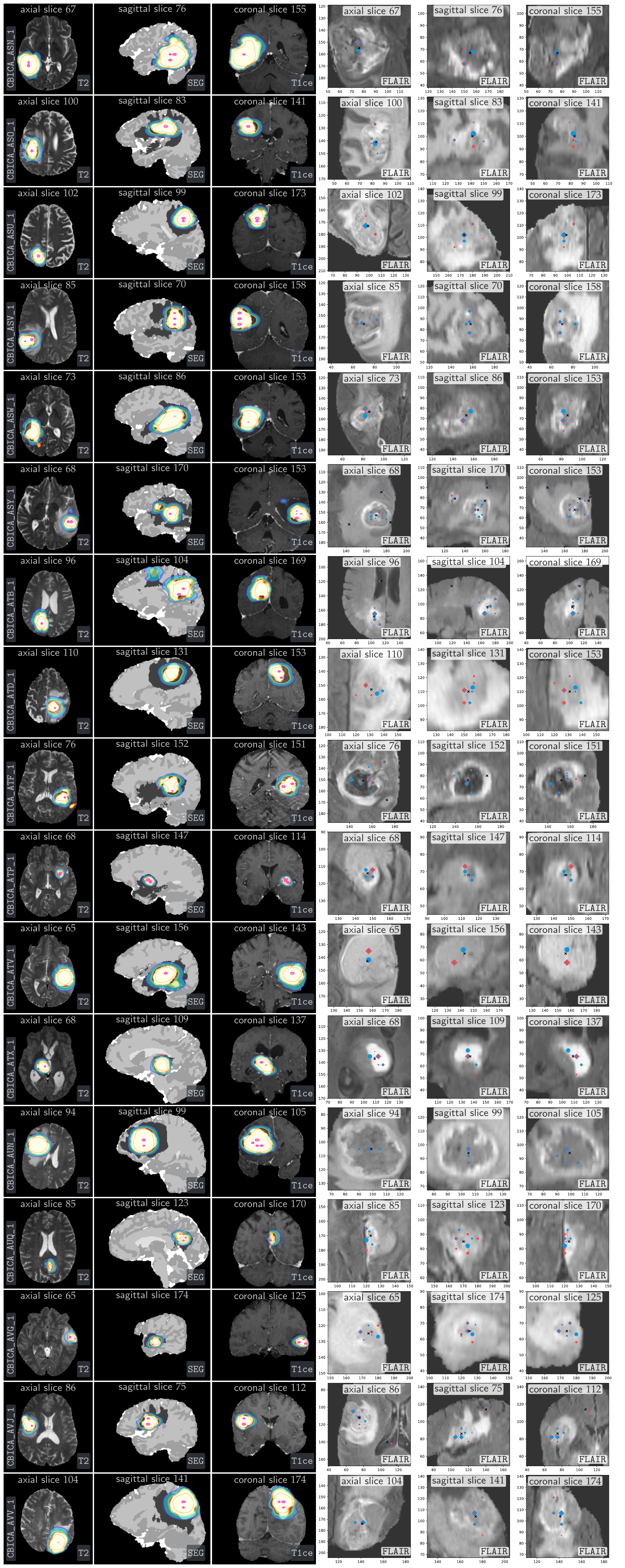}
\end{minipage}
\begin{minipage}{0.49\textwidth}
\includegraphics[width=1.\textwidth]{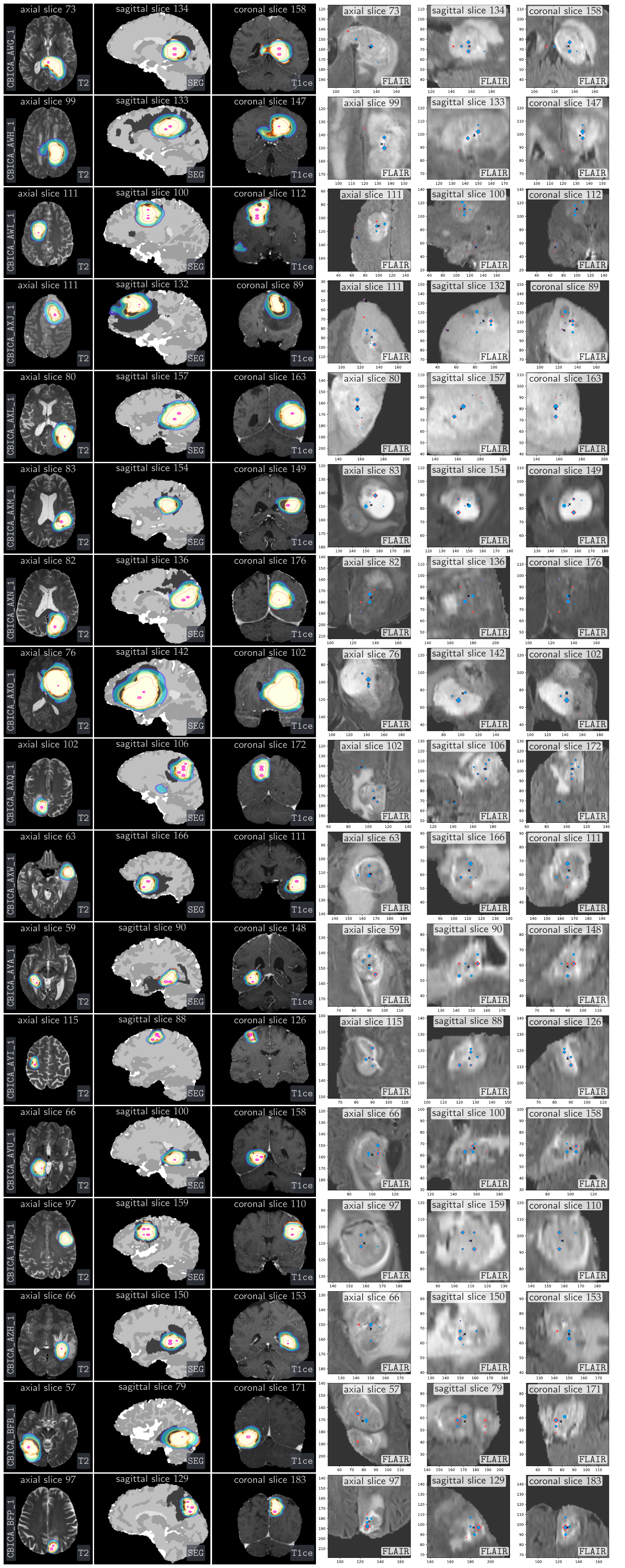}
\end{minipage}
\caption{\ipoint{Qualitative Results of BraTS'18 Cohort}. Left: Axial, coronal, and saggital cuts of the 3D patient T2-MRI volume, overlaid with the $TC$ input data (yellow), the predicted tumor $c(1)$ from our calibrated model (contours from red to violet), and the tumor origin $c(0)$ (magenta). The red $c(1)$ contour line corresponds to $0.9 \leq c(1)$, and matches well with the input $TC$ data (cf.\;\figref{fig:obs-operator} for details). Right: Projection of estimated tumor initiations $\vect{p}$ onto axial, coronal, and saggital plane (red corresp. to $128^3$ solution; blue corresp. to final $256^3$ solution). The size of the markers indicates activation weight. The center of mass of $TC$ components are indicated with a black cross.}
\label{fig:brats18-sim-3}
\end{figure*}

\begin{figure*}
\centering
\begin{minipage}{0.49\textwidth}
\includegraphics[width=1.\textwidth]{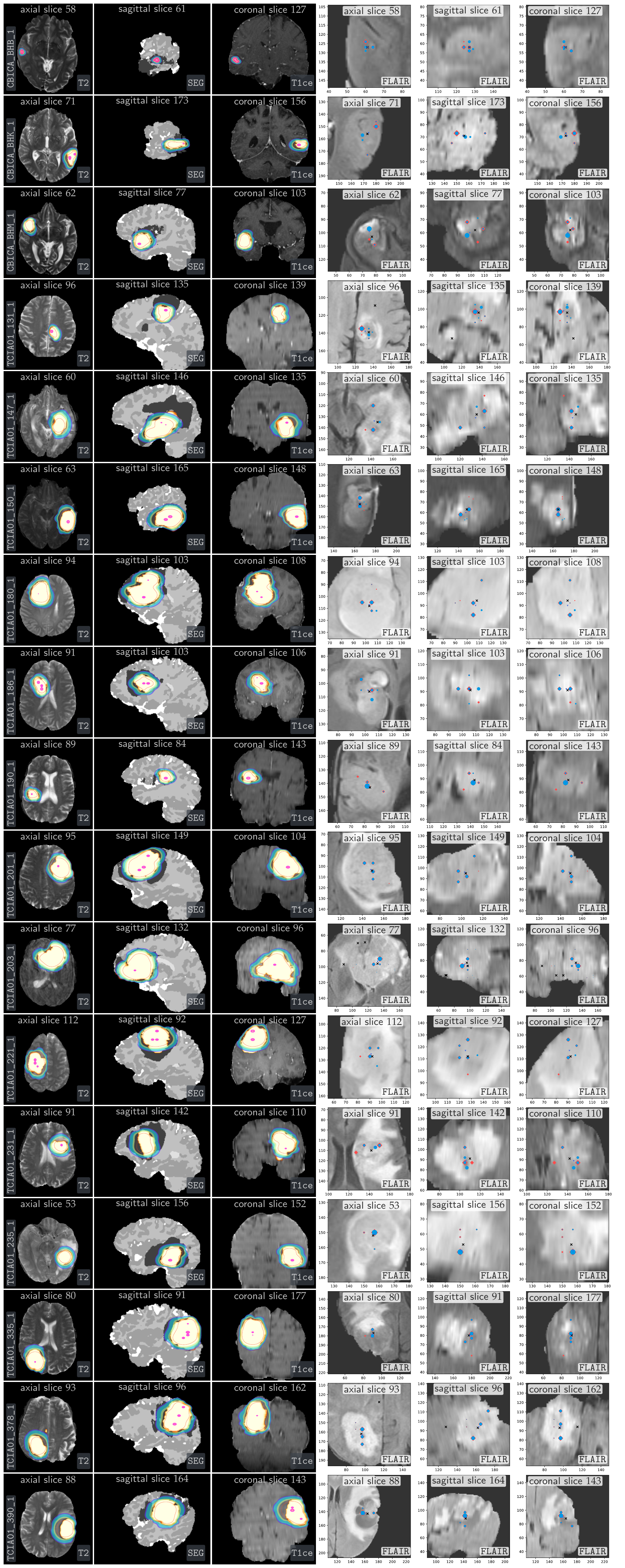}
\end{minipage}
\begin{minipage}{0.49\textwidth}
\includegraphics[width=1.\textwidth]{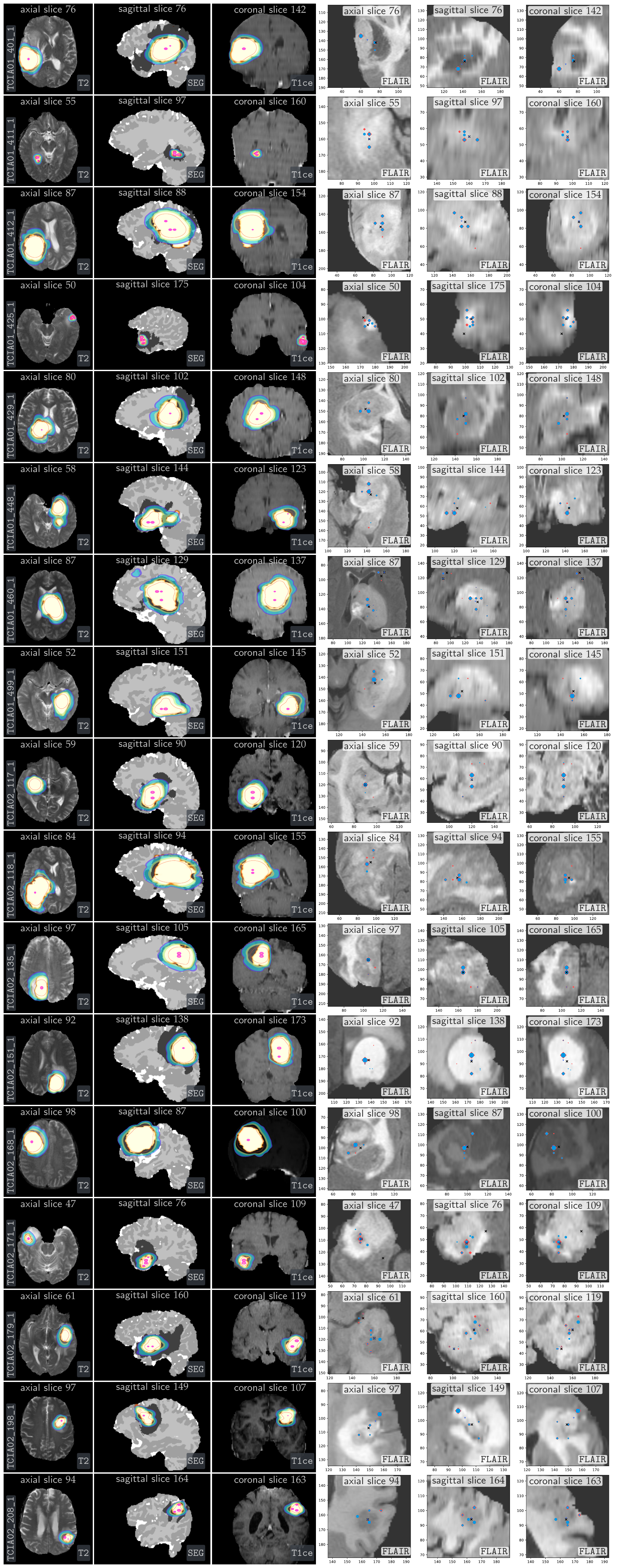}
\end{minipage}
\caption{\ipoint{Qualitative Results of BraTS'18 Cohort}. Left: Axial, coronal, and saggital cuts of the 3D patient T2-MRI volume, overlaid with the $TC$ input data (yellow), the predicted tumor $c(1)$ from our calibrated model (contours from red to violet), and the tumor origin $c(0)$ (magenta). The red $c(1)$ contour line corresponds to $0.9 \leq c(1)$, and matches well with the input $TC$ data (cf.\;\figref{fig:obs-operator} for details). Right: Projection of estimated tumor initiations $\vect{p}$ onto axial, coronal, and saggital plane (red corresp. to $128^3$ solution; blue corresp. to final $256^3$ solution). The size of the markers indicates activation weight. The center of mass of $TC$ components are indicated with a black cross.}
\label{fig:brats18-sim-4}
\end{figure*}

\begin{figure*}
\centering
\begin{minipage}{0.49\textwidth}
\includegraphics[width=1.\textwidth]{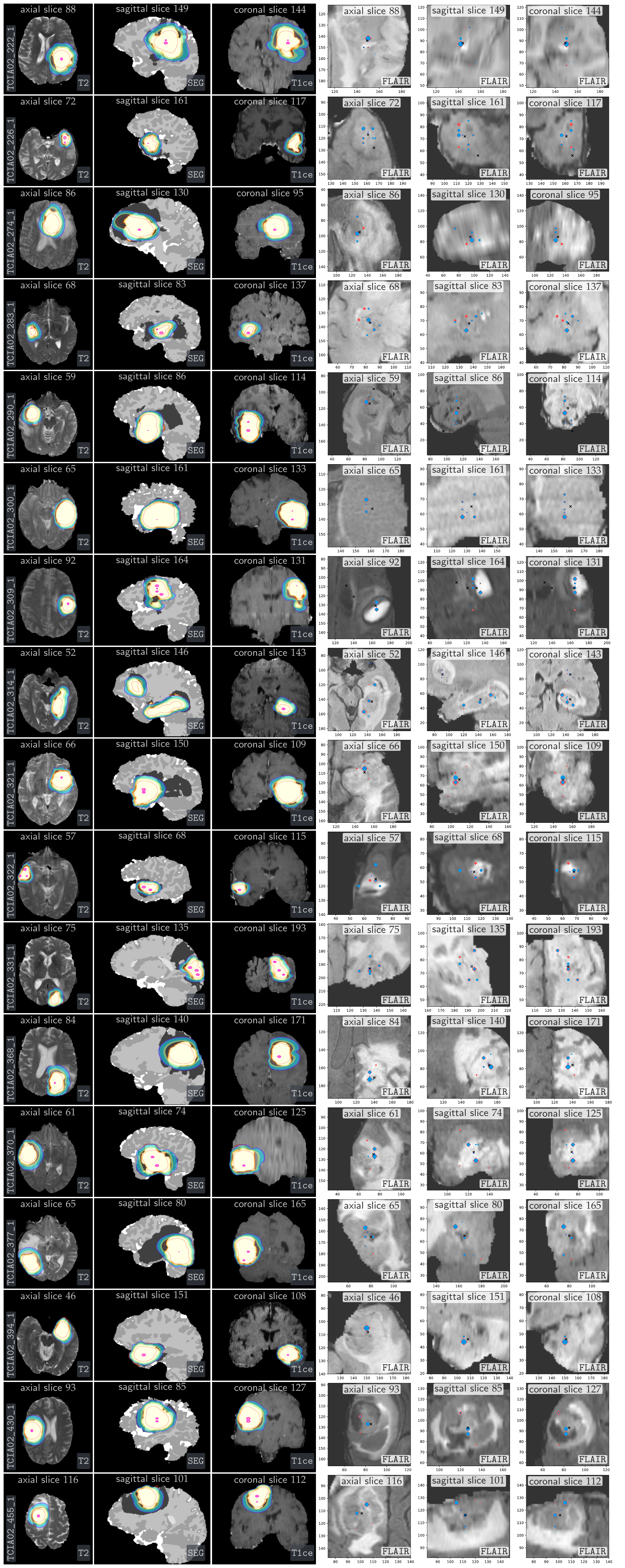}
\end{minipage}
\begin{minipage}{0.49\textwidth}
\includegraphics[width=1.\textwidth]{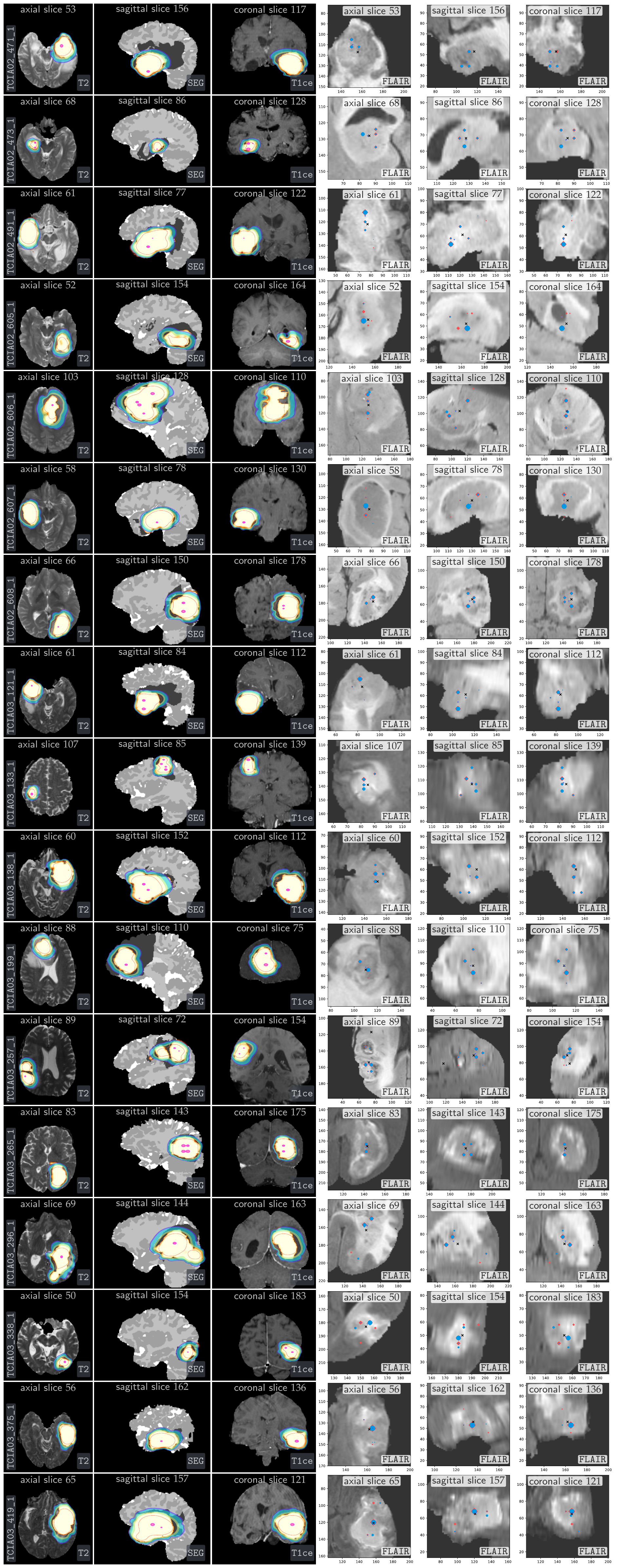}
\end{minipage}
\caption{\ipoint{Qualitative Results of BraTS'18 Cohort}. Left: Axial, coronal, and saggital cuts of the 3D patient T2-MRI volume, overlaid with the $TC$ input data (yellow), the predicted tumor $c(1)$ from our calibrated model (contours from red to violet), and the tumor origin $c(0)$ (magenta). The red $c(1)$ contour line corresponds to $0.9 \leq c(1)$, and matches well with the input $TC$ data (cf.\;\figref{fig:obs-operator} for details). Right: Projection of estimated tumor initiations $\vect{p}$ onto axial, coronal, and saggital plane (red corresp. to $128^3$ solution; blue corresp. to final $256^3$ solution). The size of the markers indicates activation weight. The center of mass of $TC$ components are indicated with a black cross.}
\label{fig:brats18-sim-5}
\end{figure*}

\begin{figure*}
\centering
\begin{minipage}{0.49\textwidth}
\includegraphics[width=1.\textwidth]{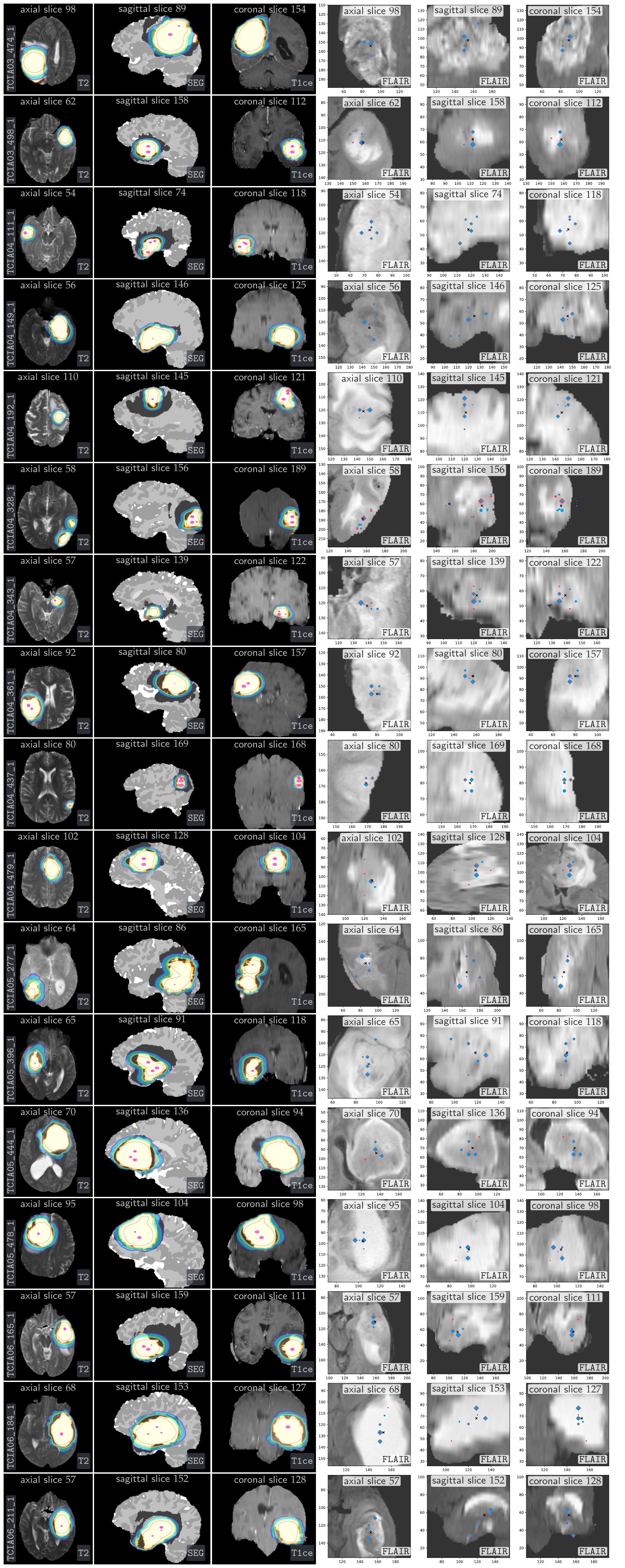}
\end{minipage}
\begin{minipage}{0.49\textwidth}
\includegraphics[width=1.\textwidth]{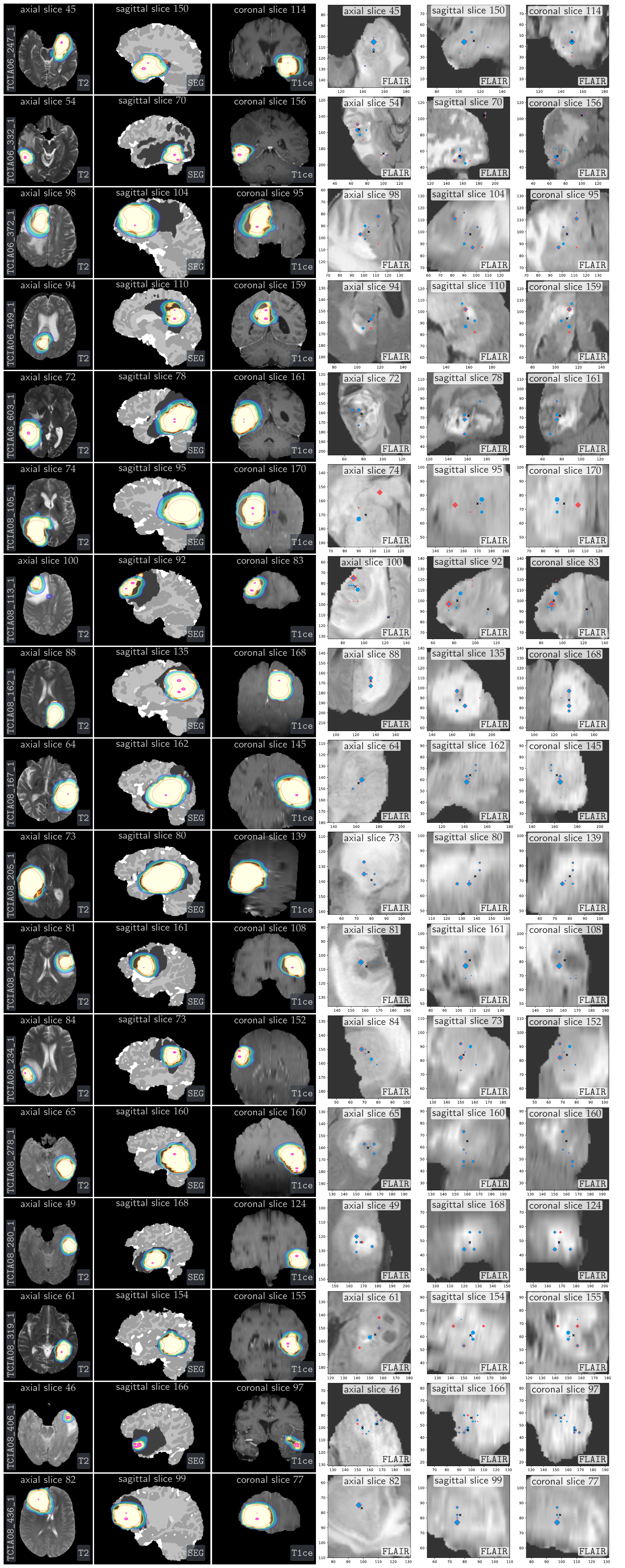}
\end{minipage}
\caption{\ipoint{Qualitative Results of BraTS'18 Cohort}. Left: Axial, coronal, and saggital cuts of the 3D patient T2-MRI volume, overlaid with the $TC$ input data (yellow), the predicted tumor $c(1)$ from our calibrated model (contours from red to violet), and the tumor origin $c(0)$ (magenta). The red $c(1)$ contour line corresponds to $0.9 \leq c(1)$, and matches well with the input $TC$ data (cf.\;\figref{fig:obs-operator} for details). Right: Projection of estimated tumor initiations $\vect{p}$ onto axial, coronal, and saggital plane (red corresp. to $128^3$ solution; blue corresp. to final $256^3$ solution). The size of the markers indicates activation weight. The center of mass of $TC$ components are indicated with a black cross.}
\label{fig:brats18-sim-6}
\end{figure*}

\addtocounter{table}{2}

\begin{table*}\centering\scriptsize\setlength\tabcolsep{5.5pt}
    \caption{\ipoint{Description of all extracted features, available in the supplementary material.}\label{tab:feature-descr}}
\begin{tabular}{ll}
    \toprule
    \ipoint{Feature}                          & \ipoint{Description} \\
\midrule
\textit{BID}                                  & Patient ID \\
\textit{age}                                  & Patient age \\
\textit{survival [d]}                         & Patient overall survival after diagnosis [in days] \\
\textit{survival\_class}                      & Patient survival class \\
\textit{resection\_status}                    & Patient resection status \\
\textit{n\_comps($TC$)}                       & Number of components of patient tumor core \\
\textit{kappa}                                & Net migration rate \\
\textit{rho}                                  & Net proliferation rate \\
\textit{rho/kappa}                            & Ratio proliferation rate to migration rate \\
\textit{kappa/rho [shape]}                    & Ratio migration rate to proliferation rate (determines shape of traveling wave) \\
\textit{2sqrt(rho$\times$kappa) [velocity]}   & Product of migration rate and proliferation rate (determines velocity of traveling wave)\\
\textit{l2[Oc(1)-TC]}                         & Data-misfit at observation points as defined in~\eqref{eq:ell2-obs-misfit} \\
\textit{l2[c(1)\textbar\_TC-TC,scaled]\_r}    & Scaled data-misfit in TC region only, as defined in~\eqref{eq:scaled-ell2-dice} \\
\textit{l2[c(1)-TC]	cm(c(0)) (aspace)}        & Data-misfit in the entire brain, as defined in~\eqref{eq:sell2-misfit}  \\
\textit{cm(TC) (aspace)}                      & Center of mass (CM) of tumor core (NEC + EN) in atlas space (Jacob atlas)  \\
\textit{cm(NEC) (aspace)}                     & CM of necrotic tumor core in atlas space (Jacob atlas)  \\
\textit{dist[c(0), BG] (aspace)}              & (Shortest) distance (in voxels) of $c(0)$ to the skull in atlas space (Jacob atlas)  \\
\textit{dist[c(0), VE] (aspace)}              & (Shortest) distance (in voxels) of $c(0)$ to the ventricles in atlas space (Jacob atlas) \\
\textit{dist[max1(p) - max2(p)]}              &  Distance (in voxels) between the two largest initial tumor activations in patient space.  \\
\textit{dist[wcm(p\textbar\_\#c) - cm(TC\textbar\_\#c)]\_(c=0)} & Distance (in voxels) between the weighted CM of $c(0)$, and the CM of tumor core data (for largest component)\\
\textit{vol(c(0))\_r}	                      & Volume of initial tumor $c(0)$ (relative to the brain volume) \\
\textit{vol(TC)\_r}                           & Volume of the tumor core (TC) (relative to the brain volume) \\
\textit{vol(ED)\_r}                           & Volume of the peritumoral edema (ED) (relative to the brain volume) \\
\textit{vol(EN)\_r}                           & Volume of the enhancing tumor (EN) (relative to the brain volume) \\
\textit{vol(NEC)\_r}                          & Volume of the necrotic tumor core (NEC) (relative to the brain volume) \\
\textit{vol(WT)\_r}                           & Volume of the whole tumor (WT) composed of (TC), and (ED) (relative to the brain volume) \\
\textit{vol(EN)/vol(TC)}                      & Volume of enhancing tumor, relative to the tumor core \\
\textit{vol(ED)/kappa}                        & Volume of edema, relative to the tumor core \\
\textit{vol(TC)/rho}                          & Ratio of (absolute) volume of the tumor core to the estimated proliferation rate \\
\textit{vol(TC)/kappa}                        & Ratio of (absolute) volume of the tumor core to the estimated migration rate \\
\textit{area(ED)\_a}                          & Surface area of peritumoral edema \\
\textit{area(NEC)\_a}                         & Surface area of the necrotic tumor core \\
\textit{area(TC)\_a}                          & Surface area of the tumor core \\
\textit{int\{c(1)\textbar\_TC\}\_r}           & Relative measure quantifying recon. quality of TC, defined in~\eqref{eq:integral-measures} \\
\textit{int\{c(1)\textbar\_ED\}\_r}           & Relative measure quantifying recon. quality of ED, defined in~\eqref{eq:integral-measures} \\
\textit{int\{c(1)\textbar\_B-WT\}\_r}         & Relative measure quantifying erroneous tumor prediction, defined in~\eqref{eq:integral-measures} \\
\textit{int\{c(0)\}/int\{c(1)\}}              & Relative measure relating mass of $c(0)$ to mass of $c(1)$  \\
\textit{int\{c(1)\textbar\_TC\}}              & Absolute measure quantifying recon. quality of TC  \\
\textit{int\{c(1)\textbar\_ED\}}              & Absolute measure quantifying recon. quality of ED  \\
\textit{int\{c(1)\textbar\_B-WT\}}            & Absolute measure quantifying erroneous tumor prediction.  \\
\textit{int\{c(1)\}}                          & Integral over predicted tumor with respect the entire brain \\
\textit{labels(descr)\_func\_atlas(c0)}       & Functional regions in the SRI24 atlas brain, overlapped by the estimated initial tumor $c(0)$ \\
\textit{labels\_percentage\_func\_atlas(c0)}  & Measure of size of overlap of the respective functional region (L2-norm of overlap relative to L2-norm of $c(0)$)  \\
\textit{labels(descr)\_func\_atlas(cm(c0))}   & Functional region in the SRI24 atlas, which the CM of the initial tumor $c(0)$ falls into \\
\textit{labels(descr)\_func\_atlas(cm(TC))}   & Functional region in the SRI24 atlas, which the CM of the tumor core falls into \\
\midrule
\textit{cm(NEC\textbar\_\#c) (\#c=i,apsace)}  & CM of the necrotic tumor core in (if applicable) three largest components $i\in\{0,1,2\}$ in atlas space \\
\textit{cm(NEC\textbar\_\#c) (\#c=i,pspace)}  & CM of the necrotic tumor core in (if applicable) three largest components $i\in\{0,1,2\}$ in patient space \\
\textit{cm(TC\textbar\_\#c) (\#c=i,aspace)}   & CM of the tumor core in (if applicable) three largest components $i\in\{0,1,2\}$ in atlas space \\
\textit{cm(TC\textbar\_\#c) (\#c=i,pspace)}   & CM of the tumor core in (if applicable) three largest components $i\in\{0,1,2\}$ in patient space \\
\textit{cm(c(0)\textbar\_\#c) (\#c=i,aspace)} & CM of the initial tumor $c(0)$ in (if applicable) three largest components $i\in\{0,1,2\}$ in atlas space \\
\textit{cm(c(0)\textbar\_\#c) (\#c=i,pspace)} & CM of the initial tumor $c(0)$ in (if applicable) three largest components $i\in\{0,1,2\}$ in patient space \\
\textit{dist[wcm(p\textbar\_\#c) - cm(TC\textbar\_\#c)]\_(c=i)} & Distance (in voxels) between the weighted CM of $c(0)$, and the CM of tumor core data (for largest three comps.)\\
\textit{l2[TC\textbar\_\#c]\_a(\#c=i)}	      & L2-norm of tumor core for largest three components $i\in\{0,1,2\}$ \\
\textit{l2[c(0)\textbar\_\#c]\_a(\#c=i)}      & L2-norm of initial tumor $c(0)$ for largest three components $i\in\{0,1,2\}$ \\
\textit{l2[c(0)\textbar\_\#c]\_r(\#c=i)}      & L2-norm  of initial tumor for largest three components $i\in\{0,1,2\}$ (rel. to L2-norm of $c(0)$) \\
\textit{l2[c(1)\textbar\_\#c - TC\textbar\_\#c]\_a(\#c=i)}             & L2-misfit of tumor prediction $c(1)$ and TC target data for largest three components $i\in\{0,1,2\}$ \\
\textit{vol(TC\textbar\_\#c)\_a(\#c=i)}        & Volume of tumor core subdivided into largest three components $i\in\{0,1,2\}$ \\
\textit{vol(TC\textbar\_\#c)\_r(\#c=i)}	      &  Relative size of largest three components  $i\in\{0,1,2\}$ \\
\textit{vol(c(0))\_a}	                      & Volume of the initial tumor $c(0)$ \\
\textit{vol(EN)\_a}	                          & Volume of enhancing tumor \\
\textit{vol(ED)\_a}                           & Volume of peritumoral edema \\
\textit{vol(NEC)\_a}                          & Volume of necrotic tumor core \\
\textit{vol(TC)\_a}                           & Volume of tumor core \\
\textit{vol(WT)\_a}                           & Surface area of whole tumor, relative to surface area of a sphere occupying the same volume \\
\textit{area(ED)\_r}                          & Surface area of edema, relative to surface area of a sphere occupying the same volume \\
\textit{area(TC)\_r}                          & Surface area of tumor core, relative to surface area of a sphere occupying the same volume \\
\textit{area(NEC)\_r}                         & Surface area of necrotic core, relative to surface area of a sphere occupying the same volume \\
\bottomrule
\end{tabular}
\end{table*}

\ifCLASSOPTIONcaptionsoff
  \newpage
\fi


\FloatBarrier

\bibliographystyle{bib/IEEEtran}
\bibliography{bib/IEEEabrv,bib/literature.bib}